\documentclass[a4paper,fleqn,usenatbib]{mnras}

\usepackage{newtxtext,newtxmath}

\usepackage[T1]{fontenc}
\usepackage{ae,aecompl}

\usepackage{graphicx}	
\usepackage{amsmath}	
\usepackage{amssymb}	
\usepackage{times}
\usepackage{graphics, subfigure}
\usepackage{epsfig}
\usepackage{multirow}
\usepackage{ulem}
\usepackage{pdfpages}

\def\simlt{\mathrel{\hbox{\rlap{\hbox{\lower4pt\hbox{$\sim$}}}\hbox{$<$}}}}
\def\simgt{\mathrel{\hbox{\rlap{\hbox{\lower4pt\hbox{$\sim$}}}\hbox{$>$}}}}

\newcommand{\mysim}{\mathord{\sim}}
\newcommand{\mylesssim}{\mathord{\lesssim}}

\newcommand{\myapprox}{\mathord{\approx}}

\title[The structure of detonation waves]{The structure of detonation waves in supernovae revisited}

\author[D. Kushnir]{
Doron Kushnir$^{1}$\thanks{E-mail: doron.kushnir@weizmann.ac.il}
\\
$^{1}$Dept. of Particle Phys. \& Astrophys., Weizmann Institute of
Science, Rehovot 76100, Israel
}

\date{Accepted XXX. Received YYY; in original form ZZZ}

\pubyear{2017}

\begin{document}
\label{firstpage}
\pagerange{\pageref{firstpage}--\pageref{lastpage}}
\maketitle

\begin{abstract}
The structure of a thermonuclear detonation wave can be solved accurately and, thus, may serve as a test bed for studying different approximations that are included in multidimensional hydrodynamical simulations of supernova. We present the structure of thermonuclear detonations for the equal mass fraction of $^{12}$C and $^{16}$O (CO) and for pure $^{4}$He (He) over a wide range of upstream plasma conditions. The lists of isotopes we constructed allow us to determine the detonation speeds, as well as the final states for these detonations, with an uncertainty of the percent level (obtained here for the first time). We provide our results with a numerical accuracy of $\mysim0.1\%$, which provides an efficient benchmark for future studies. We further show that CO detonations are pathological for all upstream density values, which differs from previous studies, which concluded that for low upstream densities CO detonations are of the Chapman--Jouget (CJ) type. We provide an approximate condition, independent of reaction rates, that allows to estimate whether arbitrary upstream values will support a detonation wave of the CJ type. Using this argument, we are able to show that CO detonations are pathological and to verify that He detonations are of the CJ type, as was previously claimed for He. Our analysis of the reactions that control the approach to nuclear statistical equilibrium, which determines the length-scale of this stage, reveals that at high densities, the reactions $^{11}$B$+p\leftrightarrow3^{4}$He plays a significant role, which was previously unknown. 

\end{abstract}




\section{Introduction}
\label{sec:Introduction}

Thermonuclear detonation waves are believed to play a key role in supernovae \citep[][]{Hoyle1960,Fowler1964}. The detonation wave structure is important for the energy release and for the nucleosynthesis during the explosion, and it is therefore a crucial ingredient for supernovae modelling \citep[see][for a recent review]{Seitenzahl2017}. However, resolving the detonation wave structure in a multidimensional hydrodynamical simulation of a supernova is currently impossible. This is because the fast thermonuclear burning dictates a burning length-scale that is much smaller than the size of the star, and because the number of isotopes participating in the thermonuclear burning is very large. These problems led to the introduction of various approximations that allow multidimensional hydrodynamical simulations of full stars. The error introduced by these approximations, however, is not well understood. Most notably, a small number ($10-20$) of isotopes is usually included in the multidimensional hydrodynamical simulations, and the method for choosing these isotopes has not yet been firmly established. 

\vspace{5mm}

A relevant, much simpler, problem to analyse is the structure of a steady-state, planar detonation wave, given by the ZND theory \citep{ZelDovich1940,vonNeumann47,Doring1943}, on which we concentrate in this work. This problem can be solved accurately for the case of a thermonuclear detonation wave, and thus can serve as a test bed for studying different approximations that are included in multidimensional hydrodynamical simulations. For example, we can calibrate lists of isotopes that allow the calculation of a thermonuclear detonation wave with some prescribed accuracy. We assume that the reader is familiar with the basic physics of thermonuclear detonation waves, as this topic has been heavily discussed over the past several decades. The theory of detonation waves in general is described in the text book of \citet{Fickett1979} and the fundamental physics of thermonuclear detonation waves is discussed by \citet[][]{Khokhlov89}. 

We consider two compositions for the upstream plasma that show dramatic differences in the structure of the detonation wave and are both relevant for supernova modelling. The first one is the equal mass fraction of $^{12}$C and $^{16}$O (CO) and the second is pure $^{4}$He (He). Other variants of the initial composition can be handled with the same tools described in this work. 

The structure of steady-state, planar, thermonuclear detonation waves has been studied by numerous authors. \citet{Imshennik1984} studied detonation waves in pure $^{12}$C, \citet{Khokhlov89} studied detonation waves in CO and He, and \citet{Townsley2016} presented solutions for detonation waves in CO (with a small initial mass fraction of $^{22}$Ne). Other studies employed a simplified reaction network (usually an $\alpha$-net composed of $13$ isotopes) to calculate steady-state, planar detonation waves in different mixtures \citep{Bruenn1975,Sharpe1999,Gamezo99,Dursi2006,Noel2007,Dominguez2011,Townsley2012,Dunkley2013}. Since the final state of thermonuclear detonation waves can be dominated by isotopes that are not $\alpha$-elements, the uncertainty with using $\alpha$-net can be significant. \citet{Sharpe1999} studied detonation waves in CO with a specific emphasis on a method to traverse the pathological point. 

One of our objective here is to calculate Chapman--Jouget (CJ) detonations with an uncertainty in the order of the percent level over a wide range of upstream plasma conditions that are relevant for supernovae. The parameters of CJ detonations have been already calculated for CO \citep{Bruenn71,Khokhlov88} and He \citep{Mazurek1973,Khokhlov88}. By comparing our results to those of previous works, we demonstrate that we are the first to reach an uncertainty level of $1$ percent. In fact, we show that the equation of state (EOS) used by \citet{Mazurek1973} is not accurate enough, and that the EOS used by \citet{Khokhlov88} is apparently inconsistent with other available EOSs. \citet{Timmes2000} calculated a few properties for CJ detonations in He, and they claim to agree with the results obtained by \citet{Mazurek1973} and \citet{Khokhlov88}. Although \citet{Timmes2000} do not provide the required information to reproduce their results, they probably did not use a tight criteria for agreement, as advocated here, to expose the apparent inconsistencies of \citet{Mazurek1973} and \citet{Khokhlov88}. 

We further calculate the structure of the detonation waves for both CO and He. Our determination of the pathological detonation speed for CO, as well as the final state of these detonations, is with a level of uncertainty of the percent level. We show that previous studies of the detonation wave structure with a detailed reaction network for both CO \citep{Khokhlov89,Townsley2016} and He \citep{Khokhlov89} are less accurate. Our results for the detonation wave speeds and for the final states are reported with a numerical accuracy of $\mysim0.1\%$, representing an efficient benchmark for future studies. We provide all the relevant information needed to fully reproduce our results. 

Besides providing accurate results and highlighting a few shortcomings of previous works, we present here a few new insights into the structure of thermonuclear detonation waves. We show that CO detonations are pathological for all upstream densities values, as far as our numerical accuracy allows us to test this. This is different from previous studies \citep{Imshennik1984,Khokhlov89,Sharpe1999,Gamezo99,Dunkley2013}, which concluded that for low upstream densities, CO detonations are of the CJ type. We explain why these claims were probably due to a loose definition for burning completion and/or low numerical accuracy. We provide an approximate condition, independent of reaction rates, that allows to estimate whether arbitrary upstream values (including composition) will support a detonation of the CJ type. Using this argument, we are able to show that CO detonations are pathological for all upstream densities and to verify that He detonations are of the CJ type, as was previously claimed for He \citep{Khokhlov89}. We show conclusively for the first time that in the case of CO detonations, the sonic point changes position in a discontinuous manner from $x\sim100\,\textrm{cm}$ to $x\sim10^{4}\,\textrm{cm}$ around the upstream density of $\myapprox2.7\times10^{7}\,\textrm{g}\,\textrm{cm}^{-3}$. 

The calculations in this work were performed with a modified version of the {\sc MESA} code\footnote{version r7624; https://sourceforge.net/projects/mesa/files/releases/} \citep{Paxton2011,Paxton2013,Paxton2015}. 

The definition of the problem to be solved is described in Section~\ref{sec:definition}. The required input physics for an accurate calculation of the detonation wave structure is described in Section~\ref{sec:input physics}. We study CJ detonations in Section~\ref{sec:CJ detonations} and the full structure of the detonation waves in Section~\ref{sec:structure}. We discuss the approximate condition needed in order to estimate whether arbitrary upstream values will support a detonation of the CJ type in Section~\ref{sec:CJ condition} and the role of weak reactions in Section~\ref{sec:weak}. We summarize our results in Section~\ref{sec:discussion}. 


\section{Definition of the problem}
\label{sec:definition}

The structure of a detonation wave can be found by integration, where the initial conditions are the downstream values of the leading shock. We assume that the pressure, $P$, and the internal energy per unit mass, $\varepsilon$, are given as a function of the independent variables: density, $\rho$, temperature, $T$, and the mass fraction of the isotopes, $X_{i}$ ($\sum_{i}X_{i}=1$ and, unless stated otherwise, the sum goes over all isotopes). For planar, steady-state, non-relativistic hydrodynamics, the equations to integrate are \citep[see e.g.][]{Khokhlov89}:
\begin{eqnarray}\label{eq:ZND}
d\rho&=&\frac{\frac{\partial P}{\partial T}\left(\frac{\partial\varepsilon}{\partial T}\right)^{-1}\left(dq-\sum_{i}\frac{\partial \varepsilon}{\partial X_{i}}dX_{i}\right)+\sum_{i}\frac{\partial P}{\partial X_{i}}dX_{i}}{u^{2}-c_{s}^{2}},\nonumber\\
dT&=&\left(\frac{\partial P}{\partial T}\right)^{-1}\left[\left(u^{2}-\frac{\partial P}{\partial\rho}\right)d\rho-\sum_{i}\frac{\partial P}{\partial X_{i}}dX_{i}\right],
\end{eqnarray}
where $c_{s}$ is the frozen (constant composition), non-relativistic speed of sound, $u$ is the velocity in the shock rest frame
\begin{eqnarray}\label{eq:u}
u=\frac{\rho_{0}}{\rho}D,
\end{eqnarray}
$\rho_{0}$ is the upstream density, $D$ is the shock velocity in the lab frame (in which the upstream fuel is at rest), $q$ is the average binding energy:
\begin{eqnarray}\label{eq:q}
q=N_{A}\sum_{i}Q_{i}Y_{i},
\end{eqnarray}
$Q_{i}$ are the binding energies of the nuclei, $Y_{i}\approx X_{i}/A_{i}$ are the molar fractions of the nuclei (see discussion in Section~\ref{sec:accuracy}), $A_{i}$ are the nucleon numbers and $N_{A}$ is Avogadro's number. Upstream values will be denoted with subscript $0$, CJ values with subscript $\textrm{CJ}$ and pathological values with subscript $*$. We further define the equilibrium speed of sound, $c_{s}^{e}$. Unless stated otherwise, the partial derivatives are taken with the rest of the independent variables remaining constant. \citet{Sharpe1999} pointed out that since $\sum_{i}X_{i}=1$, not all $X_{i}$ are independent, and he consequently eliminated from the integration the mass fraction of one isotope and instead determined it from $\sum_{i}X_{i}=1$. In this paper, we choose to treat all $X_{i}$ as independent variables, while using $\sum_{i}X_{i}=1$ only for the initial conditions. This approach is valid, since the equations that determine $dX_{i}$ must satisfy $\sum_{i}dX_{i}=0$, leading to  $\sum_{i}X_{i}=1$ throughout the integration, up to a numerical error that can be controlled. Equations~\eqref{eq:ZND}-\eqref{eq:q} are accurate as long as there is no heat transfer nor particle exchange with the environment. Specifically, these equations assume the absence of weak reactions. 

The form of Equations~\eqref{eq:ZND} demonstrates that following some change in composition $dX_{i}$ (that determines some nuclear energy release $dq$) the changes in $d\rho$ and $dT$ are independent of the rate in which this change took place. It follows that if all reaction rates are slower by some factor, then the fluid reaches the exact same state but over a time longer by the same factor. The burning limiter for hydrodynamical simulation suggested by \citet{Kushnir2013} multiplies all reaction rates by some factor to prevent unstable numerical burning and, therefore, accurately describes detonation waves over scales larger than those in which the limiter operates. 

In order to calculate the structure of the detonation wave, a full derivative in time of Equations~\eqref{eq:ZND} is taken:
\begin{eqnarray}\label{eq:ZND t}
\frac{d\rho}{dt}&=&\frac{\frac{\partial P}{\partial T}\left(\frac{\partial\varepsilon}{\partial T}\right)^{-1}\left(\frac{dq}{dt}-\sum_{i}\frac{\partial \varepsilon}{\partial X_{i}}\frac{dX_{i}}{dt}\right)+\sum_{i}\frac{\partial P}{\partial X_{i}}\frac{dX_{i}}{dt}}{u^{2}-c_{s}^{2}}\nonumber\\
&\equiv&\frac{\phi}{u^{2}-c_{s}^{2}},\nonumber\\
\frac{dT}{dt}&=&\left(\frac{\partial P}{\partial T}\right)^{-1}\left[\left(u^{2}-\frac{\partial P}{\partial\rho}\right)\frac{d\rho}{dt}-\sum_{i}\frac{\partial P}{\partial X_{i}}\frac{dX_{i}}{dt}\right].
\end{eqnarray}
The integration of Equations~\eqref{eq:ZND t} yields the state of a fluid element as a function of the time since it was shocked, given the reaction rates
\begin{eqnarray}\label{eq:fi}
dX_{i}/dt=f_{i}(\rho,T,\{X_{j}\}).
\end{eqnarray}
Equation~\eqref{eq:fi} includes the complexity of the problem, as many isotopes have to be included in the integration with many reactions. We present our results as a function of the distance behind the shock wave, $x$, connected to the time through $u=dx/dt$.

We briefly mention here the possible solutions of Equations~\eqref{eq:ZND t} \citep{Wood1960}. In the final state of the detonation wave all isotopes are in equilibrium, i.e. $dX_{i}/dt=0$ (for the case of a thermonuclear detonation wave, this state is nuclear statistical equilibrium (NSE), see Section~\ref{sec:nse}). The equilibrium composition is a function of the thermodynamic variables only, so there exist an equilibrium Hugoniot adiabat that connects to the upstream values. For a given shock velocity, $D$, the Rayleigh line that passes through the upstream values either does not intersect the equilibrium Hugoniot, is tangent to it (one point of intersection), or intersects it twice. The shock velocity for which there is one intersection is called the CJ velocity, and it is independent of reaction rates. In this work we find $D_{\textrm{CJ}}$, as well as the corresponding equilibrium state, by numerically iterating over the value of $D$. If during the integration of Equations~\eqref{eq:ZND t} with $D=D_{\textrm{CJ}}$ the flow is always subsonic, then the minimal possible shock velocity is $D_{\textrm{CJ}}$. However, if during the integration the flow becomes sonic, then from Equations~\eqref{eq:ZND t}, we must require $\phi=0$ at the sonic point. The minimal shock velocity for which this condition is satisfied is called the pathological shock velocity, $D_{*}$, and it can only be found by integrating Equations~\eqref{eq:ZND t} (and so it depends on reaction rates). Overdriven detonations, which are solutions with higher shock velocities than the minimal shock velocity, either $D_{\textrm{CJ}}$ or $D_{*}$, exist as well, and they are subsonic throughout the integration. It can be shown that for pathological detonations $\phi$ changes sign while crossing the sonic point. While $\phi$ can change sign multiple times along the integration, for all known examples of thermonuclear detonations waves, for CJ detonations $\phi>0$ throughout the integration and for pathological detonations $\phi<0$ following the sonic point crossing. We provide in Section~\ref{sec:CO scan scale} an example of a pathological detonation in which $\phi$ changes sign twice before the sonic point crossing. Finally, note that the equilibrium state is only approached asymptotically at an infinite distance behind the shock wave. As we discuss in Section~\ref{sec:structure}, previous authors provide a finite distance behind the shock wave in which the equilibrium state is obtained, which could be due to a loose definition for burning completion.  

We use the following definitions for the average nucleon number and proton number:
\begin{eqnarray}
\bar{A}=\frac{1}{\sum_i X_{i}/A_{i}},\;\;\bar{Z}=\bar{A}\sum_i Z_{i}X_{i}/A_{i},
\end{eqnarray}
where $Z_{i}$ is the proton number of isotope $i$. We also define for the heavy isotopes:
\begin{eqnarray}\label{eq:Ytilde def}
\tilde{Y}=\sum_{i,i\ne n,p,\alpha}Y_{i},\;\;\tilde{A}=\frac{1}{\tilde{Y}}\sum_{i,i\ne n,p,\alpha}X_{i}.
\end{eqnarray}
It is convenient to normalize densities, $\rho_{7}=\rho[\textrm{g}\,\textrm{cm}^{-3}]/10^{7}$, and temperatures, $T_{9}=T[\textrm{K}]/10^{9}$.

\subsection{The level of accuracy}
\label{sec:accuracy}

We differentiate between the numerical accuracy (or convergence) of the results, which depends on the numerical scheme, and their uncertainty, which depends on the level of approximations that we introduce, as well as on the uncertainty of the input physics. Our aim, for a given set of input physics, is to reach a numerical accuracy of $\mysim10^{-3}$. This degree of numerical accuracy is appropriate for benchmarking and code checking. This numerical accuracy can be (and for many cases is) much higher than the uncertainty of the EOS and of the reaction rates that dominate the uncertainty budget.

The approximation of non-relativistic hydrodynamics is expected to introduce an error of $\textrm{MeV}/m_{p}c^{2}\sim10^{-3}$ for thermonuclear detonation waves. We further approximate the nuclear masses as $m_{i}\approx A_{i}m_{u}$, where $m_{u}$ is the atomic mass unit, unless stated otherwise. This approximation is always better than $1\%$ for each isotope, and the relevant isotopes with significant errors are: $n$ (error of $\myapprox8.6\times10^{-3}$ ), $p$ ($\myapprox7.8\times10^{-3}$), $^{2}$H ($\myapprox7.0\times10^{-3}$), $^{3}$H ($\myapprox5.3\times10^{-3}$), $^{3}$He ($\myapprox5.3\times10^{-3}$), $^{6}$Li ($\myapprox2.5\times10^{-3}$), $^{7}$Li ($\myapprox2.3\times10^{-3}$) and $^{7}$Be ($\myapprox2.4\times10^{-3}$). Since the total mass fraction of these isotopes is at most a few percent under the conditions relevant for thermonuclear detonation waves, the approximation of $m_{i}\approx A_{i}m_{u}$ introduces an error smaller than $\mysim10^{-3}$. The total mass fraction of other isotopes with a similar significant deviation from $m_{i}\approx A_{i}m_{u}$ is always small. The level of error introduced by the absence of weak reactions is discussed in Section~\ref{sec:weak}.


\section{Input physics}
\label{sec:input physics}

\subsection{Nuclear statistical equilibrium (NSE)}
\label{sec:nse}

NSE is the unique nuclear composition of a system when strong and electromagnetic interactions are in a state of detailed balance for a given set of thermodynamic state variables and electron fraction. Applying a detailed balance to the reaction that breaks up a nucleus with a nucleon number $A_{i}$ and a proton number $Z_{i}$ into free nucleons $(A_{i},Z_{i})\leftrightarrow Z_{i}p+N_{i}n$, where $N_{i}=A_{i}-Z_{i}$, yields a relation between the chemical potential of the nucleus $\mu_{i}$ and the chemical potential of free protons $\mu_{p}$ and neutrons $\mu_{n}$: $\mu_{i}=Z_{i}\mu_{p}+N_{i}\mu_{n}$ \citep{Clifford1965}. The last relation can be written as
\begin{equation}\label{eq:NSE1}
Z_{i}\mu_{p}+N_{i}\mu_{n}=m_{i}c^{2}+k_{B}T\ln\left[\frac{n_{i}}{w_{i}(T)}\left(\frac{h^{2}}{2\upi m_{i}k_{B}T}\right)^{3/2}\right]+\mu_{i}^{\textrm{coul}},
\end{equation}
where $k_{B}$ is Boltzmann's constant, $h$ is Planck's constant, $n_{i}$ is the number density and $\mu_{i}^{\textrm{coul}}$ is a Coulomb interaction term \citep{Calder2007,Seitenzahl2009}. The Coulomb term and the conditions under which Equation~\eqref{eq:NSE1} is valid are discussed in Section~\ref{sec:screening}. The mass fractions of all nuclei in an NSE can therefore be expressed in terms of the chemical potential of the protons and the neutrons and the nuclear binding energies $Q_{i}=(Z_{i}m_{p}+N_{i}m_{n}-m_{i})c^{2}$:
 \begin{eqnarray}\label{eq:NSE2}
X_{i}&=&\frac{m_{i}}{\rho}w_{i}(T)\left(\frac{2\upi m_{i}kT}{h^{2}}\right)^{3/2}\\ \nonumber
&\times&\exp\left[\frac{Z_{i}(\mu_{p}+\mu_{p}^{\textrm{coul}})+N_{i}\mu_{n}-\mu_{i}^{\textrm{coul}}+Q_{i}}{kT}\right],
\end{eqnarray}
where $w_{i}(T)$ are the nuclear partition functions and here we take the accurate nuclear masses for $m_{i}$. Since the mass fractions of all nuclei must sum to one, $\sum_{i}X_{i}=1$, and the nuclear composition has the prescribed electron fraction, $Y_{e}\approx\sum_{i}X_{i}Z_{i}/A_{i}$, for a given $\rho$, $T$, and $Y_{e}$, the mass fractions of all the isotopes can be found by solving for the neutron and proton chemical potentials that satisfy the two constraints. The NSE state is found in this work by using a modified version of the NSE routine of Frank Timmes\footnote{http://cococubed.asu.edu/}. Specifically, we include in Eq.~\eqref{eq:NSE2} the ion--ion Coulomb interaction terms of \citet[][see detailed discussion in Section~\ref{sec:screening}]{Chabrier1998}. 

The nuclear masses and partition functions were taken from the file \textsc{winvn\_v2.0.dat}, which is available through the JINA reaclib data base\footnote{http://jinaweb.org/reaclib/db/} \citep[JINA,][]{Cyburt2010}. For those isotopes whose $m_{i}$ values in \textsc{winvn\_v2.0.dat} differed from the most updated values given in the ENSDF database\footnote{https://www.nndc.bnl.gov/ensdf/}, $\tilde{m}_{i}$, we used the latter values instead\footnote{The differences are probably because new experimental values became available since the last time \textsc{winvn\_v2.0.dat} was updated. (Schatz, private communication).}. The list of isotopes for which $m_{i}$ and $\tilde{m}_{i}$ differ is given in Table~\ref{tbl:mass} of Appendix~\ref{sec:mass and spin}, together with their mass (excess) values. The file \textsc{winvn\_v2.0.dat} provides the values of $w_{i}(T)$ over some specified temperature grid in the $[10^{8},10^{10}]\,\textrm{K}$ range. For numerical stability, it is better to fit the $w_{i}(T)$ values to some function rather than interpolate. We use the functional form suggested by \citet{Woosley1978}:
\begin{eqnarray}\label{eq:wi}
w_{i}(T)&=&\left(2J_{i,0}+1\right)\left(1+\sum_{k}E_{i,k}\exp(-F_{i,k}/T_{9})\right)\times\nonumber\\
&&\exp\left(a_i/T_{9}+b_i+c_i T_{9}+d_i T_{9}^{2}\right),
\end{eqnarray}
where $(2J_{i,0}+1)$ is the statistical weight for the ground state of isotope $i$ and $a_i$ is negative. We initially used an extended list of $581$ isotopes (see Table~\ref{tbl:nets}) to find suitable sets of isotopes for the integration of Eqs.~\eqref{eq:ZND t} (see Section~\ref{sec:net}). We could usually fit the nuclear partition function for the extended list of isotopes with $E_{i,k}$ being equal to zero to better than $10\%$ over the relevant temperature range $[1.5\times10^{9},10^{10}]\,\textrm{K}$. In the case that such a fit was not possible, low-lying excited levels with $J_{i,k}$ and the excitation energy $\varepsilon_{i,k}\,[\textrm{MeV}]$ were added, where $E_{i,k}=(2J_{i,k}+1)/(2J_{i,0}+1)$ and $F_{i,k}=11.6045\varepsilon_{i,k}$. The addition of, at most, three low-lying excited levels typically sufficed to fit to better than $10\%$. For two isotopes, the fit was slightly worse: $^{78}$As ($\mysim12.7\%$) and $^{89}$Kr ($\mysim19.6\%$). The inaccuracies of the fit functions negligibly effect the results (see discussion in Section~\ref{sec:CJ detonations}). We make the fit parameters for all isotopes publicly available\footnote{The file \textsc{isotopes\_pfit.data} is included in the online-only supporting information and is also available through https://www.dropbox.com/sh/i6js2c0i96j8vgg/ AACrk93NR8i2LyDyYO91Eu4ma?dl=0}. We note that for some isotopes, the values of $J_{i,0}$ in \textsc{winvn\_v2.0.dat} differ from the most updated values given in the ENSDF data base\footnote{Once again, the differences are probably because new experimental values became available since the last time \textsc{winvn\_v2.0.dat} was updated. (Schatz, private communication).}. In these cases, we used the values of ENSDF, $\tilde{J}_{i,0}$, and normalized the $w_{i}(T)$ values from \textsc{winvn\_v2.0.dat} to $\tilde{w}_{i}(T)$ as follows\footnote{Suggested by Hendrik Schatz.}:
\begin{eqnarray}\label{eq:wi nor}
\tilde{w}_{i}(T)=1+\frac{2\tilde{J}_{i,0}+1}{2J_{i,0}+1}\left(w_{i}(T)-1\right).
\end{eqnarray}
The list of isotopes for which $J_{i,0}$ and $\tilde{J}_{i,0}$ differ is given in Table~\ref{tbl:spin} of Appendix~\ref{sec:mass and spin}, together with their spin values.

When nearing a state of NSE, the plasma may be in an intermediate state of nuclear-statistical-quasi-equilibrium \citep[NSQE;][]{Bodansky1968}, in which a group of heavy isotopes are in detailed balance. We assume that at NSQE there is an equilibrium of neutrons, protons, and $\alpha$-particles, $\mu_{\alpha}=2\mu_{p}+2\mu_{n}$, and that the rest of the isotopes are in a detailed balance, such that the chemical potentials of every two of them, $i$ and $j$, satisfy $\mu_{i}-\mu_{j}=(N_{i}-N_{j})\mu_{n}+(Z_{i}-Z_{j})\mu_{p}$. In particular, under this assumption the state of NSQE is uniquely determined by specifying $\rho$, $T$, $Y_{e}$ and $\tilde{Y}$ \citep[for a detailed discussion, see][]{Khokhlov89}.

\subsection{Nuclear reaction network}
\label{sec:net}

Previous studies of thermonuclear detonation waves employed lists of isotopes that were considered extensive enough. However, this assumption was not backed up by any quantitate calculation, so one cannot estimate the error introduced by these lists of isotopes. Moreover, inclusion of irrelevant isotopes can decrease the numerical accuracy. We, therefore, aim at finding a reasonably short list of isotopes that allows the calculation of a thermonuclear detonation wave with some prescribed degree of accuracy. 

We first define an extended list of $581$ isotopes (see Table~\ref{tbl:nets}), which includes all the available isotopes with $Z\le14$ from the file \textsc{winvn\_v2.0.dat} that satisfy the following two conditions:
\begin{enumerate}
\item JINA includes strong reactions that connect the isotope to the bulk of the isotopes (say to $^{56}$Ni). In other words, a subnet of a few isotopes is not allowed.
\item The isotope's decay time is longer than $1\,\textrm{ns}$ (which is roughly the carbon-burning time-scale in CO detonations).
 \end{enumerate}
We further add to our list of isotopes an extended pool of isotopes with $Z>14$ that is sufficient in terms of the conditions described below. Next, given some minimal abundance $Y_{\min}$, we include in the list every isotope that has an NSE number abundance that is $Y_{i}>Y_{\min}$ for some $\rho$, $T$ and $Y_{e}$ within the ranges $T\in[2\times10^{9},3\times10^{10}]\,\textrm{K}$, $\rho\in[100,10\times10^{10}]\,\textrm{g}/\textrm{cm}^{3}$ and $Y_{e}\in[0.495,0.5]$. We obtained lists for a few values of $Y_{\min}=10^{-y}$ ($y=4,5,6,7$). These lists have to be supplemented with other isotopes that, while not represented in the NSE state, are significant for the burning process. Specifically, the relaxation to an NSE state is controlled by slow reactions between low-$Z$ isotopes \citep[][who suggested that $^{12}$C$\leftrightarrow3\,^{4}$He is the most important one; see the discussion in Sections~\ref{sec:CO scan scale} and~\ref{sec:He scan scale}]{Khokhlov89}. We, therefore, add to the lists of isotopes obtained from the NSE condition more isotopes, in several stages, which are as follows.

We define an isotope list $\alpha$-ext that describes burning through $\alpha$-elements, which includes:
\begin{enumerate}
\item  $n$, $p$ and the $\alpha$-isotopes $^{4}$He, $^{12}$C, $^{16}$O, $^{20}$Ne, $^{24}$Mg, $^{28}$Si, $^{32}$S, $^{36}$Ar, $^{40}$Ca, $^{44}$Ti, $^{48}$Cr, $^{52}$Fe and $^{56}$Ni.
\item All isotopes that differ from $\alpha$-isotopes by $n$, $p$ or $\alpha$. 
\item $^{22}$Ne, since it has a significant mass fraction for some initial conditions.  
\item All isotopes of an element between the minimal and the maximal nucleon numbers determined from the previous steps.  
\item We exclude $^{5}$He and $^{9}$B from the list, see below.  
\end{enumerate}
The obtained $\alpha$-ext list includes 78 isotopes and is presented in Table~\ref{tbl:nets}. NSE$7$ is the combination of all species that meet the $Y_{\min}=10^{-7}$ threshold and all species from the $\alpha$-ext (actually, the only isotope from $\alpha$-ext that does not meet the $Y_{\min}=10^{-7}$ threshold is $^{19}$Ne). NSE$y$ $(y=4,5,6)$ is the combination of all species with $Z>14$ that meet the $Y_{\min}=10^{-y}$ NSE threshold, those with $Z\le14$ that meet the $Y_{\min}=10^{-7}$ NSE threshold, and all species from the $\alpha$-ext list. The inclusion of all isotopes with $Z\le14$ that meet the $Y_{\min}=10^{-7}$ NSE threshold in NSE$4-6$ only slightly increases the sizes of these nets and improves the calculation of the low-$Z$ isotopes. The obtained lists are presented in Table~\ref{tbl:nets}. One can verify that the extended list includes for each element at least one additional isotope with a smaller (larger) nucleon number compared to our most detailed NSE$7$ list, or that there are no more isotopes with smaller or larger nucleon numbers (bold numbers in Table~\ref{tbl:nets}). Furthermore, the extended list contains isotopes of Br and Kr, none of which survive in NSE$7$. Unless otherwise stated, the NSE$7$ net is the one used from this point on in the text.  

Finally, in order to verify that we are not missing any important low-$Z$ isotopes, we add to the NSE$7$ list all the isotopes with a $Z\le14$ from the extended list that have a measured mass and ground-state spin (not calculated)\footnote{$^{10}$C is excluded for a reason that is discussed in Section~\ref{sec:structure}.}. We call this list NSE$7$Si and it is presented in Table~\ref{tbl:nets}.
 
\begin{table*}
\begin{minipage}{170mm}
\caption{The lists of isotopes used in this work. Bold numbers mark the minimal or maximal nucleon numbers available in \textsc{winvn\_v2.0.dat}.}
\begin{tabular}{|c||c||c||c||c||c||c||c|}
\hline
$\textrm{Element}$  & $\textrm{Extended}$ & $\textrm{NSE}7\textrm{Si}$  & $\textrm{NSE}7$ & $\textrm{NSE}6$ & $\textrm{NSE}5$ & $\textrm{NSE}4$ & $\alpha-\textrm{ext}$\\ 
 & $581\,\textrm{isotopes}$ & $344\,\textrm{isotopes}$  & $260\,\textrm{isotopes}$ & $218\,\textrm{isotopes}$ & $179\,\textrm{isotopes}$ & $137\,\textrm{isotopes}$ & $78\,\textrm{isotopes}$ \\ \hline
$\textrm{n}$ & \textbf{1} & \textbf{1} & \textbf{1} & \textbf{1} & \textbf{1} & \textbf{1} & \textbf{1} \\ 
$\textrm{H}$ & \textbf{1}--\textbf{3} & \textbf{1}--\textbf{3} & \textbf{1}--\textbf{3} &\textbf{1}--\textbf{3} & \textbf{1}--\textbf{3} & \textbf{1}--\textbf{3} &  \textbf{1}--\textbf{3}\\ 
$\textrm{He}$ & \textbf{3}--4, \textbf{6}      & \textbf{3}--4, \textbf{6} & \textbf{3}--4, \textbf{6} & \textbf{3}--4, \textbf{6} & \textbf{3}--4, \textbf{6} &  \textbf{3}--4, \textbf{6} &  \textbf{3}--4 \\  
$\textrm{Li}$ & \textbf{6}--\textbf{9}            & \textbf{6}--\textbf{9} &\textbf{6}--7 &\textbf{6}--7 &\textbf{6}--7 & \textbf{6}--7 & --\\  
$\textrm{Be}$ & \textbf{7}, 9--\textbf{13} &\textbf{7}, 9--\textbf{13} & \textbf{7}, 9--10 & \textbf{7}, 9--10 & \textbf{7}, 9--10 & \textbf{7}, 9--10 & --\\ 
$\textrm{B}$ & \textbf{8}, 10--14\footnote{$^{17-18}$B form a subnet.} & \textbf{8}, 10--14 & 10--11 & 10--11 & 10--11 & 10--11 & 11 \\ 
$\textrm{C}$ & \textbf{9}--17\footnote{$^{20-21}$C form a subnet.} & \textbf{9}, 11--16\footnote{$^{10}$C is excluded for a reason that is discussed in Section~\ref{sec:structure}.} & 11--14 &11--14 & 11--14 & 11--14 & 11--13 \\ 
$\textrm{N}$ & \textbf{12}--\textbf{20} & \textbf{12}--19 & 13--15 & 13--15 & 13--15 & 13--15 & 13--15 \\ 
$\textrm{O}$ & \textbf{13}--24\footnote{$^{25-28}$O are too short lived.} & \textbf{13}--24 & 15--18 & 15--18 & 15--18 & 15--18 & 15--17\\
$\textrm{F}$ & \textbf{14}--27\footnote{$^{28}$F is too short lived.}  & \textbf{14}--27 & 17--19 & 17--19 & 17--19 & 17--19 & 17--19\\  
$\textrm{Ne}$ & \textbf{17}--34\footnote{$^{35-38}$Ne are too short lived.} & \textbf{17}--31 & 19--23 & 19--23 & 19--23 & 19--23 & 19--22\\ 
$\textrm{Na}$ & 19--37\footnote{$^{18,38-42}$Na are too short lived.} & 19--33 & 21--25 & 21--25 & 21--25 & 21--25 & 21--23\\ 
$\textrm{Mg}$ & \textbf{20}--40\footnote{$^{41-45}$Mg are too short lived.} & \textbf{20}--36 & 23--28 & 23--28 & 23--28 & 23--28 & 23--25\\  
$\textrm{Al}$ & \textbf{22}--43\footnote{$^{44-48}$Al are too short lived.} & 23--35 & 25--30 & 25--30 & 25--30 & 25--30 & 25--27\\ 
$\textrm{Si}$ & \textbf{23}--44\footnote{$^{45-51}$Si are too short lived.} & 24--40 & 27--33 & 27--33 & 27--33 & 27--33 & 27--29\\ 
$\textrm{P}$ & 26--40 & 29--35 & 29--35 & 29--34 & 29--33 & 29--31 & 29--31\\ 
$\textrm{S}$ & 28--45 & 30--37 & 30--37 & 31--37 & 31--36 & 31--33 & 31--33\\ 
$\textrm{Cl}$ & 31--46 & 32--39 & 32--39 & 33--39 & 33--37 & 33--35 & 33--35\\  
$\textrm{Ar}$ & 32--49 & 34--42 & 34--42 & 35--41 & 35--39 & 35--37 & 35--37\\ 
$\textrm{K}$ & 35--51 & 37--45 & 37--45 & 37--44 & 37--41 & 37--39 & 37--39\\ 
$\textrm{Ca}$ & 36--54 & 38--48 & 38--48 & 39--47 & 39--45 & 39--41 & 39--41\\  
$\textrm{Sc}$ & 40--56 & 41--51 & 41--51 & 41--50 & 41--48 & 41--43 & 41--43\\ 
$\textrm{Ti}$ & 40--58 & 43--53 & 43--53 & 43--52 & 43--51 & 43--50 & 43--45\\  
$\textrm{V}$ & 42--58 & 45--55 & 45--55 & 45--54 & 43--53 & 45--51 & 45--47\\ 
$\textrm{Cr}$ & 44--59 & 47--57 & 47--57 & 47--56 & 47--55 & 47--54 & 47--49\\ 
$\textrm{Mn}$ & 46--60 & 49--59 & 49--59 & 49--58 & 49--57 & 49--56 & 49--51\\ 
$\textrm{Fe}$ & 48--64 & 50--62 & 50--62 & 51--61 & 51--59 & 51--58 & 51--53\\ 
$\textrm{Co}$ & 50--65 & 52--64 & 52--64 & 53--63 & 53--61 & 53--60 & 53--55\\ 
$\textrm{Ni}$ & 52--71 & 54--66 & 54--66 & 55--65 & 55--64 & 55--61 & 55--57\\ 
$\textrm{Cu}$ & 54--72 & 56--68 & 56--68 & 57--67 & 57--65 & 57--61 & 57\\ 
$\textrm{Zn}$ & 56--77 & 58--70 & 58--70 & 59--69 & 60--67 & -- & --\\ 
$\textrm{Ga}$ & 58--78 & 61--72 & 61--72 & 62--70 & -- & -- & --\\ 
$\textrm{Ge}$ & 60--82 & 64--74 & 64--74 & 69--71 & -- & -- & --\\ 
$\textrm{As}$ & 62--83 & 69--75 & 69--75 & -- & -- & -- & --\\
$\textrm{Se}$ & 64--86 & 75 & 75 & -- & -- & --  & --\\  
$\textrm{Br}$ & 70--86 & -- & -- & -- & -- & -- & --\\ 
$\textrm{Kr}$ & 71--91 & -- & -- & -- & -- & -- & -- \\ 
\end{tabular}
\centering
\label{tbl:nets}
\end{minipage}
\end{table*}

The forward reaction rates are taken from JINA (the default library of 2017 October 20). All strong reactions that connect between isotopes from the list are included (this requires some modification of the relevant subroutines of {\sc MESA}). To allow the plasma to reach an NSE, inverse reaction rates were determined according to a detailed balance. We modified the relevant subroutine of {\sc MESA} so as to be exactly compatible with Equation~\eqref{eq:NSE2}. Enhancement of the reaction rates due to screening corrections and their compatibility with Equation~\eqref{eq:NSE2} are described in Section~\ref{sec:screening}.

\begin{figure}
\includegraphics[width=0.48\textwidth]{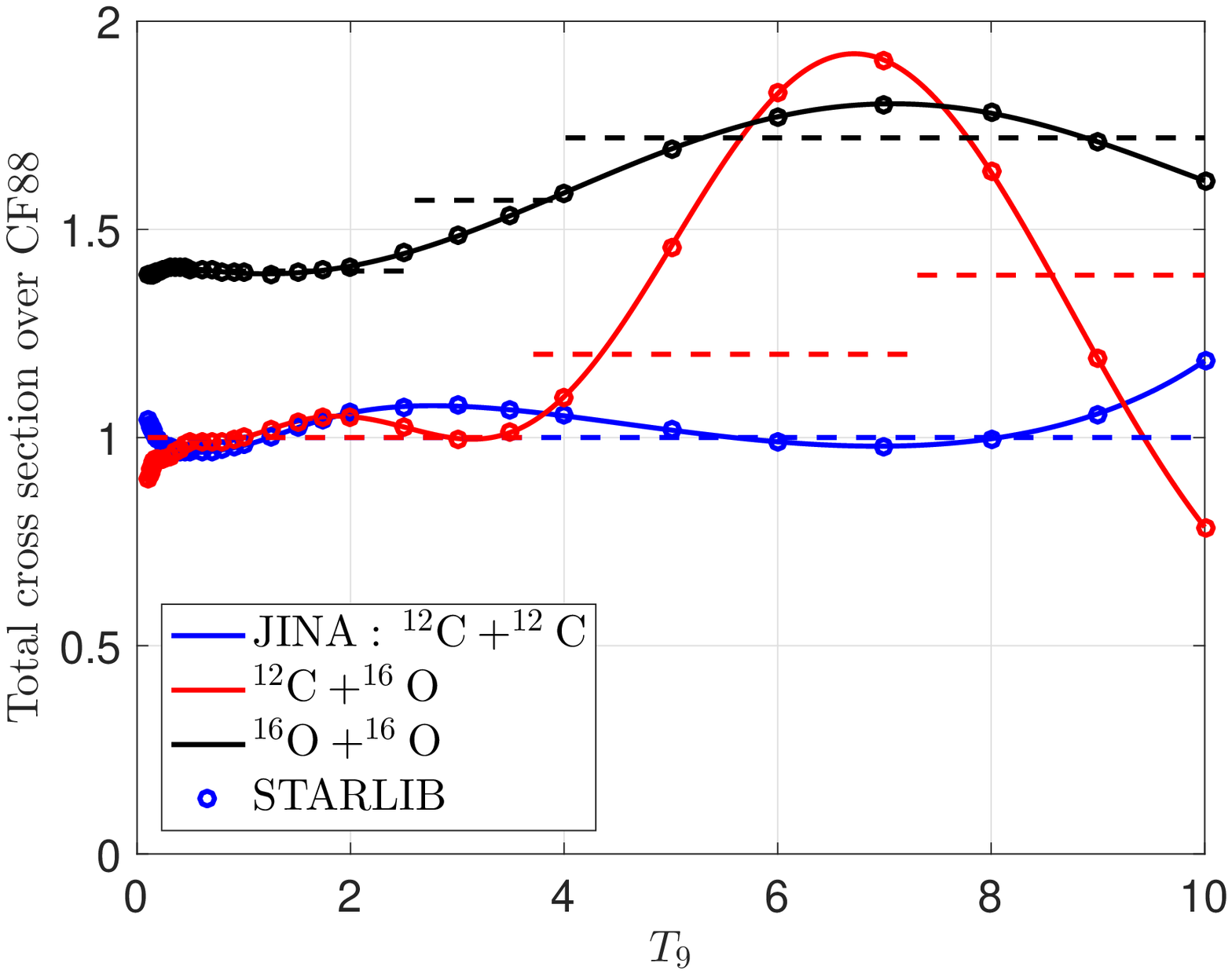}
\caption{Total cross-sections of $^{12}$C+$^{12}$C (blue), $^{12}$C+$^{16}$O (red) and $^{16}$O+$^{16}$O (black) from the JINA reaclib database (solid lines) and from STARLIB (points) divided by the total cross-sections of \citet{CF88}. The ratios for $^{12}$C+$^{16}$O and for $^{16}$O+$^{16}$O are larger by factors that roughly equal the $n$, $p$, and $\alpha$ yields of the reactions \citep[dashed lines, as given by][]{CF88}, which suggest that a choice was made to conserve the yields of $n$, $p$, and $\alpha$ instead of conserving the total cross-section. 
\label{fig:CF88}}
\end{figure}

A note is in place regarding the total cross-sections for the reactions $^{12}$C+$^{16}$O and $^{16}$O+$^{16}$O given by JINA. According to JINA, these rates are taken from \citet[][CF88]{CF88}. \citet{CF88} provide the total cross-section for these reactions, as well as the yields of $n$, $p$, and $\alpha$ for these reactions. They note that the sum of these yields can exceed unity because of reactions such as $^{16}$O$(^{16}$O$,np)^{30}$P and $^{16}$O$(^{16}$O$,2p)^{30}$Si. This should not be confused with branching ratios for different channels that always sums up to unity. Since the branching ratios are not given by \citet{CF88} for the $^{12}$C+$^{16}$O and $^{16}$O+$^{16}$O reactions, it is not clear how the branching ratios were determined for the $n$, $p$, and $\alpha$ channels provided by JINA for these reactions (other channels, such as $np$ and $2p$, are not provided). Moreover, as shown in Figure~\ref{fig:CF88}, the total cross-sections for these reactions (sum over all channels)\footnote{the $^{12}$C$(^{16}$O$,n)^{27}$Si rate is calculated as the reverse rate of $^{27}$Si$(n,^{12}$C$)^{16}$O.} provided by JINA are significantly larger from the ones given by \citet{CF88}. The JINA total cross-sections are larger by factors that roughly equal the total yields (dashed lines in Figure~\ref{fig:CF88}), which suggests that a choice was made to conserve the yields of $n$, $p$, and $\alpha$ instead of conserving the total cross-section. For comparison, we also present in Figure~\ref{fig:CF88} the reaction  $^{12}$C+$^{12}$C, where the total yields sum up to unity. We also present in Figure~\ref{fig:CF88}, the total cross-sections provided by \textsc{v65a\_090817} of STARLIB\footnote{https://starlib.github.io/Rate-Library/} \citep{Sallaska2013}, which are similar to the total cross-sections provided by JINA. For the purposes of burning in supernovae, it seems more obvious to favour the correct  total cross-sections rather than the correct yields of $n$, $p$, and $\alpha$, so in this work, we normalized all the channels of the $^{12}$C+$^{16}$O and $^{16}$O+$^{16}$O reactions such that the total cross-sections are identical to the ones provided by \citet{CF88} while keeping the branching ratios provided by JINA. 
 
\subsection{Equation of state}
\label{sec:eos}

The EOS is composed of contributions from electron--positron plasma, radiation, ideal gas for the nuclei, Coulomb corrections and nuclear level excitations:
\begin{eqnarray}\label{eq:EOS}
\varepsilon&=&\varepsilon_{ep}+\varepsilon_{\textrm{rad}}+\varepsilon_{\textrm{ion}}+\varepsilon_{\textrm{cou}}+\varepsilon_{ex},\nonumber\\
p&=&p_{ep}+p_{\textrm{rad}}+p_{\textrm{ion}}+p_{\textrm{cou}},\nonumber\\
S&=&S_{ep}+S_{\textrm{rad}}+S_{\textrm{ion}}+S_{\textrm{cou}}+S_{ex}.
\end{eqnarray}
We use the Timmes EOS\footnote{http://cococubed.asu.edu/} \citep{Timmes1999} for the electron--positron plasma and the EOS provided by {\sc MESA} for the ideal gas part of the nuclei, for the radiation and for the Coulomb corrections (but based on \citet{Chabrier1998} and not on \citet{Yakovlev1989}, see detailed discussion in Section~\ref{sec:screening}). We further include the nuclear level excitation energy of the ions and a more accurate expression for the entropy of the ions. As demonstrated in Section~\ref{sec:CJ detonations}, the nuclear level excitations can be the most important correction term for an ideal EOS for the relevant thermodynamic states. Although this term was included in \citet{Khokhlov88} and probably also in \citet[][see discussion in Section~\ref{sec:Khokhlov89}]{Khokhlov89}, it is not part of the EOS routines provided by {\sc FLASH} \citep{Fryxell2000} and {\sc MESA}. In fact, this term is not even mentioned in \citet{Fryxell2000} as a relevant correction for an ideal EOS, who argued that the most important correction is the ion--ion Coulomb interaction term. We show below that nuclear level excitations can be a more important correction to the energy than the Coulomb correction (but since nuclear level excitations do not contribute to the pressure, the Coulomb correction is the most important correction to the pressure). We make our EOS publicly available\footnote{The files are available through https://www.dropbox.com/sh/ oiwalp3f4qoy8lo/AABz7LJC-4fUjnb9OoWG3UvPa?dl=0}.

An alternative for using the Timmes EOS is using the more efficient \textit{Helmholtz} EOS \citep{Timmes00}, which is a table interpolation of the Helmholtz free energy as calculated by the Timmes EOS over a density-temperature grid. Although the accuracy of the interpolation is better than $\mysim10^{-7}$ for the relevant density--temperature region with dense enough grid, an internal inconsistency of the \textit{Helmholtz} EOS precludes obtaining a numerical accuracy of $\mysim10^{-3}$ for our results within some regions of the relevant parameter space. This issue may also be relevant for other applications, and it is discussed in Appendix~\ref{sec:helm}.

\subsubsection{Nuclear level excitations}
\label{sec:nuclear level excitations}

The nuclear level excitation energy is given by \citep{Landau80}:
\begin{equation}\label{eq:eex}
\varepsilon_{ex}=N_{A}k_{B}T\sum_{i}Y_{i}\frac{\partial \ln w_{i}(T)}{\partial \ln T}.
\end{equation}
The nuclear level excitations do not contribute to the pressure, but they do contribute to the entropy:
\begin{equation}\label{eq:sex}
S_{ex}=\varepsilon_{ex}/T.
\end{equation}

The input parameters for the EOS routines in {\sc MESA} are $\rho$, $T$, $\bar{A}$ and $\bar{Z}$. In order to calculate $\varepsilon_{ex}$, the routines must be modified to include $X_{i}$ as input parameters. The routines were further modified to supply partial derivatives with respect to $X_{i}$, in order to integrate Equations~\eqref{eq:ZND t}.

\subsubsection{A more accurate expression for the entropy of the ions}
\label{sec:new entropy}

The entropy of the ions (not including the nuclear level excitations) is given by \citep[see e.g. ][]{Shapiro1983}:
\begin{eqnarray}\label{eq:ion entropy}
S_{\textrm{ion}}\approx k_{B}N_{A}\sum_{i}\frac{X_{i}}{A_{i}}\ln\left[\frac{e^{5/2}}{h^{3}}\frac{(2\upi k_{B}T)^{3/2}}{\rho X_{i}}\left(\frac{A_{i}}{N_{A}}\right)^{5/2}w_{i}(T)\right].
\end{eqnarray}
This expression can be compared with the one used by {\sc MESA}:
\begin{eqnarray}\label{eq:ion entropy helm}
S_{\textrm{ion}}=\frac{k_{B}N_{A}}{\bar{A}}\ln\left[\frac{e^{5/2}}{h^{3}}\frac{(2\upi k_{B}T)^{3/2}}{\rho}\left(\frac{\bar{A}}{N_{A}}\right)^{5/2}\right],
\end{eqnarray}
which assumes $w_{i}(T)=1$ and averages in some sense over the mass fractions. This is a reasonable choice in the case that $X_{i}$ are not given, but since $X_{i}$ are required in order to calculate the nuclear level excitations, we use the more accurate expression for the entropy, Equation~\eqref{eq:ion entropy}. 

\subsection{Coulomb corrections}
\label{sec:screening}

For the plasma conditions relevant to thermonuclear supernovae, the ion--electron interaction, $\bar{Z}e^{2}\left(4\upi n_{e}/3\right)^{1/3}$, where $e$ is the electron charge and $n_{e}$ is the electron number density, is weak compared to the kinetic energy of the electrons ($\mylesssim10\%$ at most). Assuming commutatively of the kinetic and potential operators and the separation of the traces of the electronic and ionic parts of the Hamiltonian, the non-ideal corrections to the plasma due to the Coulomb interaction can be divided into exchange correlation of the electron fluid (electron--electron), ion--electron (polarisation) interaction and ion--ion interaction \citep[see e.g.][]{Chabrier1998}. The relevant conditions for thermonuclear supernovae include both the relativity parameter, $p_{F}/m_{e}c$, where $p_{F}$ is the zero-temperature Fermi momentum of electrons, and the degeneracy parameter, $T/T_{F}$, where $T_{F}$ is the Fermi temperature, larger or smaller than unity.  

An analytical parameterization of the electron--electron term (exchange and correlation) was given for non-relativistic electrons by \citet[][]{Ichimaru1987} and by \citet{Stolzmann2000}. For relativistic electrons, the exchange part was given for high degeneracy by \citet{Stolzmann2000}\footnote{Note that their equation~(82) is wrong by a minus sign, and their equation~(85) should be $u_{ee}^{x}=f_{ee}^{x}(1+V_{\lambda}^{b}/V^{b}+W_{\lambda}^{b}/W^{b})$.} and the full term (exchange and correlation) was given by \citet{Jancovici1962} for zero temperature. As far as we know, there is no available parameterization of the correlation part for relativistic electrons at finite temperatures, nor for the exchange part for relativistic electrons at slight degeneracy, as they are expected to be small. Since these regimes are relevant for thermonuclear supernovae, we inspected the available exchange and correlation terms near these regimes and found them to be a correction smaller than $0.1\%$. However, we cannot verify that they are on the sub-percent level throughout these regimes. For regimes where a parameterization of the electron--electron term is available, the correction is larger than $1$ percent only for low densities $\rho_{7}\simlt0.03$ and low temperatures $T_{9}\simlt0.2$. We will hereunder avoid these regions (unless stated otherwise), and, therefore, neglect the electron--electron term, which introduces a sub-percent order of uncertainty. We also neglect the ion--electron term, given for arbitrary degeneracy and relativity of the electrons by \citet{Potekhin2000}, as it introduces a correction smaller than $3\times10^{-3}$ for the relevant conditions of thermonuclear supernovae.

\begin{figure}
\includegraphics[width=0.46\textwidth]{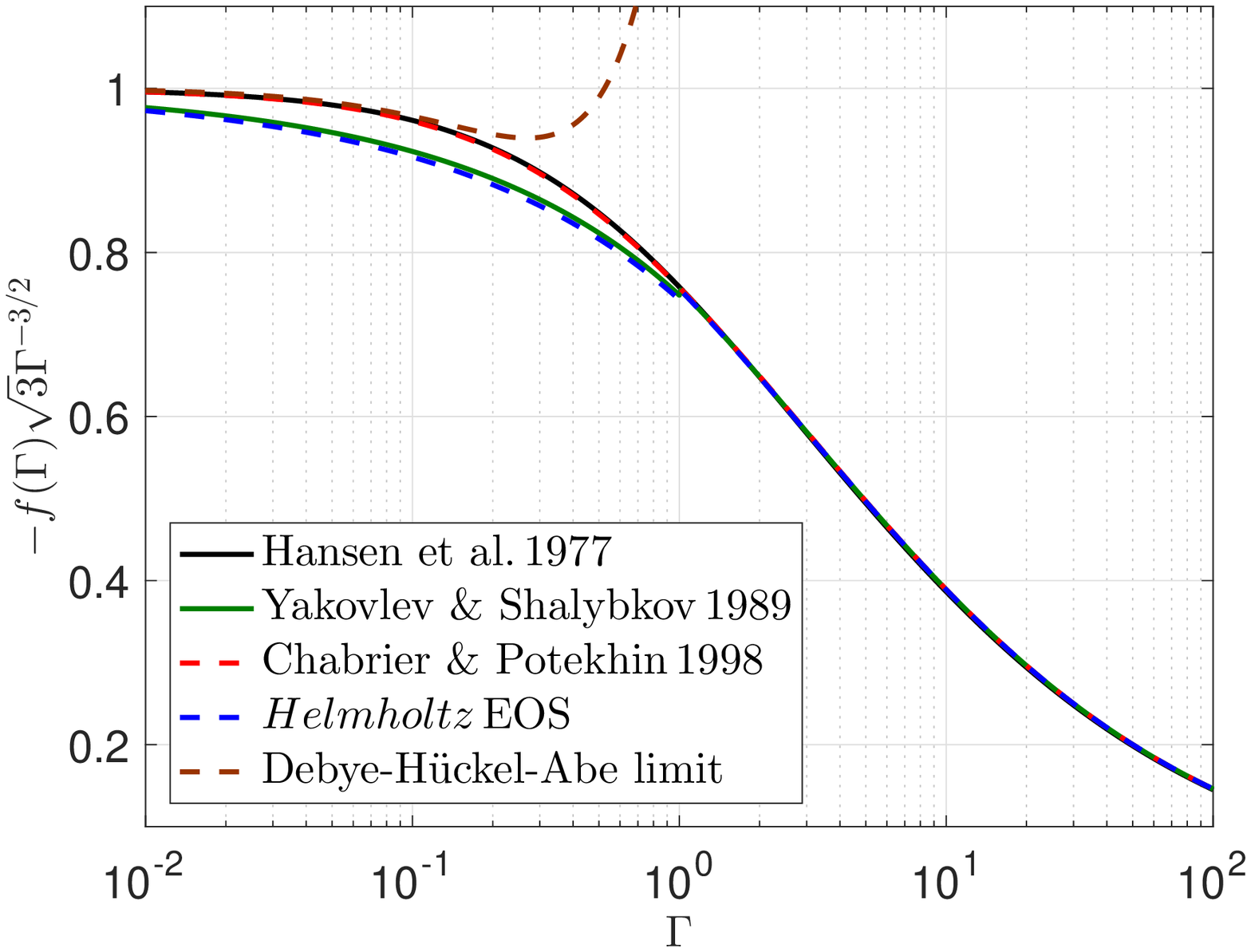}
\caption{Different fits for $f(\Gamma)$: \citet[][black]{Hansen1977}, \citet[][green]{Yakovlev1989}, \citet[][red]{Chabrier1998}, the fit implemented in \textit{Helmholtz} EOS (blue) and the Debye--H{\"u}ckel--Abe \citep{Abe1959} limit for $\Gamma\ll1$.
\label{fig:cou}}
\end{figure}

The ion--ion interaction term for a plasma with only one type of $N_{i}$ ions is given as the dimensionless Helmholtz free energy $F_{i}/N_{i}k_{B}T\equiv f_{i}=f(\Gamma_{i})$, with an ion coupling parameter $\Gamma_{i}=Z_{i}^{5/3}\Gamma_{e}$ and an electron coupling parameter $\Gamma_{e}\approx(4\upi\rho N_{A} Y_{e}/3)^{1/3}e^{2}/k_{B}T$. It is useful to note that $\Gamma_{i}\approx1.1(T/2\times10^{8}\,\textrm{K})^{-1}(Y_{e}\rho/10^{9}\,\textrm{g}\,\textrm{cm}^{-3})^{1/3}Z_{i}^{5/3}$. A useful four-parameter fit for $f(\Gamma)$ was given by \citet{Hansen1977}, which is shown in Figure~\ref{fig:cou}. The fit interpolates between the Debye--H{\"u}ckel--Abe \citep{Abe1959} result in the weak coupling limit ($\Gamma\ll1$) and the strong coupling limit ($\Gamma\gg1$) that can be simulated. The fit is not valid above the melting point ($\Gamma\approx175$). Later on, \citet{Yakovlev1989} provided a fit for $f(\Gamma)$ with a different functional form. Their results do not deviate by more than $4\%$ from the fit of \citet{Hansen1977}, but their fit is not continuous at $\Gamma=1$; see Figure~\ref{fig:cou}. This is because they required continuity only for $\Gamma df/d\Gamma$, but this leads, for example, to a discontinuity in the entropy. The \textit{Helmholtz} EOS uses the same functional form of \citet{Yakovlev1989} with somewhat different numerical values, and suffers from the same problem. \citet{Chabrier1998} used the fit of \citet{Hansen1977} with three parameters, and their results do not deviate by more than $1$ percent from the fit of \citet{Hansen1977}. Finally, \citet{Potekhin2000} introduced a seven-parameter fit\footnote{Note that the term $-B_{2}\ln(1+\Gamma/B_{1})$ in their equation~(16) should be replaced with $-B_{2}\ln(1+\Gamma/B_{2})$.} that deviates from the three-parameter fit of \citet{Chabrier1998} by less than a percent. We hereunder use the fit for $f(\Gamma)$ of \citet{Chabrier1998}, since it is the simplest one and it is accurate to better than a percent. 

When the plasma comprises a mixture of different ions, there are situations where the linear mixing rule (LMR), which states that the correction is a number weighted linear sum of one component plasma, is a good approximation \citep{Hansen1977}. If the LMR applies, then the Coulomb correction to the chemical potential of each ion is given by $\mu_{i}^{\textrm{coul}}=k_{B}Tf_{i}$ and is independent of the other ions. Nevertheless, at the weak coupling regime the LMR fails, as the Debye--H{\"u}ckel limit is non-linear. \citet{Potekhin2009a} and \citet{Potekhin2009b} studied the transition to the Debye--H{\"u}ckel limit and showed that the LMR is accurate to better than $10$ percent for $\langle\Gamma\rangle=\langle Z_{i}^{5/3}\rangle \Gamma_{e}>1$, where $\langle Z_{i}^{5/3}\rangle$ is a number weighted sum. The relevant NSE state of the detonation waves are in the regime $0.1\lesssim\langle\Gamma\rangle\lesssim1$, where the LMR can introduce deviations of up to $\mysim30\%$. Even larger deviations can be obtained for $0.01\lesssim\langle\Gamma\rangle\lesssim0.1$, which is typical of the post-shock conditions of helium detonations (although the plasma includes mainly helium ions there). \citet{Potekhin2009b} suggested a modification of $f_{i}$ to accurately describe the transition to the Debye--H{\"u}ckel limit. This modification makes $\mu_{i}^{\textrm{coul}}$ dependent on other ions in the plasma, which significantly complicates the calculation of the NSE state \citep{Nadyozhin2005}. We show later that the Coulomb correction changes the NSE state by a few percent, which means that the modification of the LMR is usually a sub-percent correction (but could be higher). We, therefore, choose in this work to adopt the LMR.

Once the ion--ion terms are determined, the correction of the EOS, the correction of the NSE relation, Equation~\eqref{eq:NSE2}, and the screening of the thermonuclear reaction can be calculated self-consistently. Usually, however, this is not the case. Sometimes only the corrections to the EOS are considered \citep[e.g., as in][]{Khokhlov88,Khokhlov89}, and sometimes all corrections are considered but not is a consistent way (see below). Here we consider all corrections in a consistent way. Following \citet[][]{Khokhlov88}, we approximate the LMR correction to the EOS by $f(\Gamma)$ for a `mean' nucleus $\Gamma=\bar{Z}^{5/3}\Gamma_{e}$. This introduces an error of only a few percent compared with LMR (i.e. summing over all ions) and significantly simplifies the calculation of these corrections. For the NSE relation, we use $\mu_{i}^{\textrm{coul}}=k_{B}Tf_{i}$, and this determines, from detailed balance, the screening factors of all thermonuclear reactions \citep{KushnirScreen}. In brief, consider the screening of a reaction with reactants $i=1,..,N$ with charges $Z_{i}$. The screening factor for this reaction is identical to the screening factor of a reaction in which all reactants form a single isotope $j$ with a charge $Z_{j}=\sum_{i=1}^{N}Z_{i}$ and a photon. The inverse reaction, photodisintegration, is not screened, and, therefore, from the detailed balance condition we get for the screening factor:
\begin{eqnarray}\label{eq:NSE screening}
\exp\left(\frac{\sum_{i=1}^{N}\mu_{i}^{C}-\mu_{j}^{C}}{k_{B}T}\right)
\end{eqnarray}
\citep[same as equation~(15) of][for the case of $N=2$]{Dewitt1973}. The screening routines available in {\sc MESA} are not compatible with our choice of $\mu_{i}$, and they also include `quantum' corrections \citep{Alastuey1978}. Although these screening factors can still be enforced to satisfy a detailed balance \citep{Calder2007}, we choose to use Equation~\eqref{eq:NSE screening} as it is consistent with our NSE relation and as the `quantum' corrections have a negligible effect on thermonuclear detonation waves. We hereunder refer to both the inclusion of the Coulomb correction terms for the NSE and the screening of thermonuclear reaction as the `Coulomb correction term for the NSE state'.  


\section{CJ Detonations}
\label{sec:CJ detonations}

In this section, we calculate several properties of the CJ detonations. This is useful because CJ detonations are independent of reaction rates, which allows an efficient benchmarking for the EOS and the NSE routines. Furthermore, even for initial conditions where the unsupported detonation is pathological, the final CJ NSE conditions provide a good approximation for the pathological NSE conditions. We numerically determined the CJ detonation speed, $D_{\textrm{CJ}}$, to an accuracy of $\mysim10^{-6}$, which allows benchmarking to the accuracy level we aimed for, $10^{-3}$. In Section~\ref{sec:CO CJ}, we consider the initial composition of CO. We further compare our results to \citet[][Section~\ref{sec:Bruenn72}]{Bruenn72}, \citet[][Section~\ref{sec:Khokhlov88}]{Khokhlov88} and \citet[][Section~\ref{sec:Gamezo99}]{Gamezo99}. In Section~\ref{sec:He CJ}, we consider the initial composition of pure helium, and compare our results to \citet[][Section~\ref{sec:Mazurek}]{Mazurek1973} and \citet[][Section~\ref{sec:Khokhlov88_He}]{Khokhlov88}. We exploit the comparisons to previous works to highlight the sensitivity of the results to various assumptions. 

\subsection{CJ detonations of carbon-oxygen mixtures}
\label{sec:CO CJ}

\begin{figure}
\includegraphics[width=0.48\textwidth]{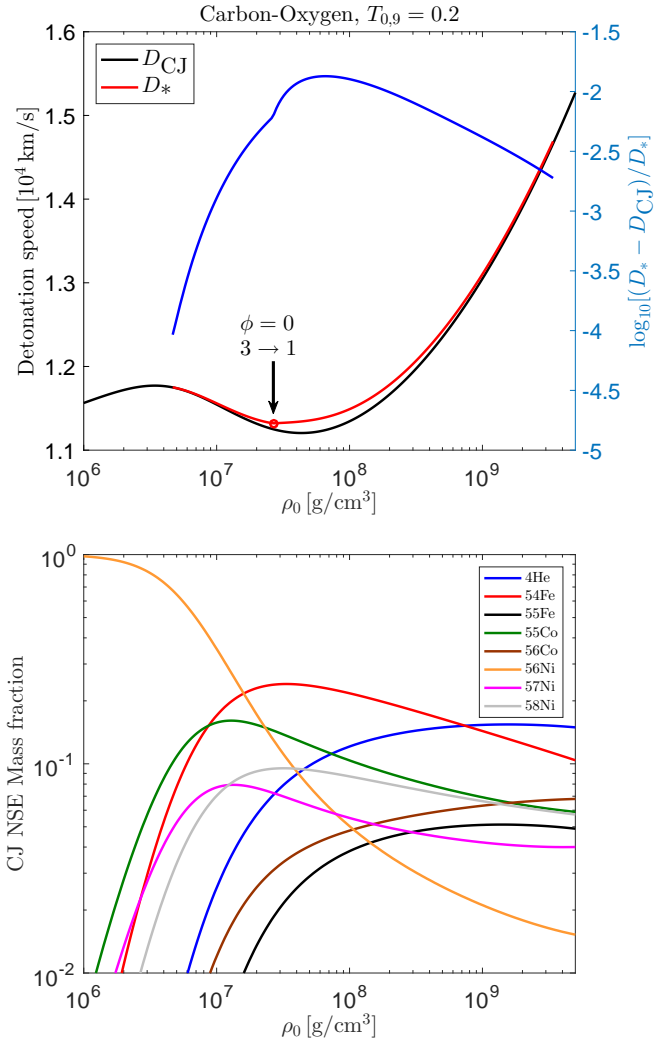}
\caption{Upper panel: $D_{\textrm{CJ}}$ (black) and $D_{\textrm{*}}$ (red) for CO and upstream temperature of $T_{0,9}=0.2$ as a function of the upstream density. The deviation between $D_{\textrm{CJ}}$ and $D_{*}$ (blue) is always smaller than $\myapprox1.4\%$. We are unable to determine $D_{\textrm{*}}$ with a high enough degree of accuracy for densities above $\rho_{0,7}=340$ and below $\rho_{0,7}=0.47$. Nevertheless, at high densities, the decrease in the deviation as a function of the upstream density is smaller than exponential, which suggests that even at larger upstream densities the detonation remains pathological. At low densities, the deviation decreases exponentially with $1/\rho_{0}$ (see Figure~\ref{fig:CJProof}), which suggests that the detonation remains pathological even at lower upstream densities. The minimum of $D_{*}$ corresponds to a discontinuous behaviour of the sonic point location (see Section~\ref{sec:CO scan scale}). Bottom panel: mass fractions of key isotopes at the CJ NSE state for the same upstream conditions. We only present the mass fraction of isotopes that have a mass fraction larger than $5\times10^{-2}$ at some $\rho_{0}$ within the inspected range. 
\label{fig:DetonationSpeed_CJAbu}}
\end{figure}

The calculated $D_{\textrm{CJ}}$ for CO is presented in the upper panel of Figure~\ref{fig:DetonationSpeed_CJAbu} for an upstream temperature of $T_{0,9}=0.2$ and an upstream density in the relevant range for supernovae, $[10^{6},5\times10^{9}]\,\textrm{g}/\textrm{cm}^{3}$. Similarly to \citet{Gamezo99} and \citet{Dunkley2013}, we find that $D_{\textrm{CJ}}$ is not a monotonic function of $\rho_{0}$ and that it has a maximum at $\rho_{0,7}\approx0.35$ and a minimum at  $\rho_{0,7}\approx4.3$ (the minimum can also be extracted from table IV of \citet{Khokhlov88}). Key isotopes at the CJ NSE state are presented in the bottom panel of Figure~\ref{fig:DetonationSpeed_CJAbu} for the same upstream values. We only present the mass fraction of isotopes that have a mass fraction larger than $5\times10^{-2}$ at some $\rho_{0}$ within the inspected range. At low densities, the NSE state is dominated by $^{56}$Ni (with $\bar{A}\approx55$ and $\tilde{A}\approx56$ at $\rho_{0,7}=0.1$), while at higher densities the NSE state is mainly a mixture of $^{4}$He, $^{54}$Fe, $^{55}$Co and $^{58}$Ni (with $\bar{A}\approx12$ and $\tilde{A}\approx52$ at $\rho_{0,7}=500$). A few key parameters of these CJ detonations are given in Table~\ref{tbl:CO CJ}. The temperature at the CJ NSE state increases monotonically with $\rho_{0}$, which decreases both the $\bar{A}$ at these states and the released thermonuclear energy compared with the initial states, $q_{01,\textrm{CJ}}$. It is also demonstrated that the nuclear excitation energy contribution to the energy at the NSE state is slightly greater than a percent for the high densities, and is slightly larger in magnitude than the Coulomb correction.   

\begin{table*}
\begin{minipage}{180mm}
\caption{Key parameters of CJ (upper rows for each upstream density) and pathological (lower rows for each upstream density, if available) detonations for CO and upstream temperature of $T_{0,9}=0.2$}
\begin{tabular}{|c||c||c||c||c||c||c||c||c||c||c||c||c|}
\hline
$\rho_{0}$  & $P_{0}/\rho_{0}$ & $ \gamma_{0} $\footnote{$\gamma_0=c_{s,0}^{2}\rho_{0}/P_{0}$}  & $D$ & $P/\rho_{0}$   & $\rho/\rho_{0}$ & $T$ & $\gamma^{e}$\footnote{$\gamma^{e}=\left(c_{s}^{e}\right)^{2}\rho/P$} &$q_{01}$ & $\bar{A}$ & $\tilde{A}$ & $f_{\textrm{coul}}$ \footnote{$f_{\textrm{coul}}=\log_{10}\left(-\frac{\varepsilon_{\textrm{coul}}}{\varepsilon}\right)$} &  $f_{\textrm{ex}}$ \footnote{$f_{\textrm{ex}}=\log_{10}\left(\frac{\varepsilon_{\textrm{ex}}}{\varepsilon}\right)$}\\ 
$[\textrm{g}/\textrm{cm}^{3}]$  & $[\textrm{MeV}/m_{p}]$ &  & $[10^{4}\,\textrm{km}/\textrm{s}]$ & $[\textrm{MeV}/m_{p}]$   &  & $[10^{9}\,\textrm{K}]$ & & $[\textrm{MeV}/m_{p}]$ & & & &  \\ \hline
$1\times10^{6}$	&	$0.02958$	&	$1.5666$	&	$1.1564$	&	$0.6103$	&	$1.7122$	&	$3.140$	&	$1.3389$	&	$0.8186$	&	$55.22$	&	$55.96$	&	$-2.3$	&	$-4.5$	\\
$3\times10^{6}$	&	$0.05065$	&	$1.5146$	&	$1.1767$	&	$0.6522$	&	$1.7127$	&	$4.010$	&	$1.2984$	&	$0.7954$	&	$46.67$	&	$55.67$	&	$-2.2$	&	$-3.4$	\\
$1\times10^{7}$	&	$0.08847$	&	$1.4533$	&	$1.1545$	&	$0.6784$	&	$1.7360$	&	$5.058$	&	$1.1844$	&	$0.6763$	&	$25.67$	&	$54.56$	&	$-2.2$	&	$-2.5$	\\
	&		&		&	$1.1560$	&	$0.7124$	&	$1.8090$	&	$5.107$	&	$1.1802$	&	$0.6667$	&	$24.76$	&	$54.49$	&	$-2.3$	&	$-2.5$	\\
$3\times10^{7}$	&	$0.1402$	&	$1.4072$	&	$1.1231$	&	$0.6706$	&	$1.6743$	&	$5.866$	&	$1.1747$	&	$0.5425$	&	$17.07$	&	$53.75$	&	$-2.3$	&	$-2.1$	\\
	&		&		&	$1.1325$	&	$0.7585$	&	$1.8581$	&	$5.979$	&	$1.1747$	&	$0.5162$	&	$16.01$	&	$53.65$	&	$-2.3$	&	$-2.1$	\\
$1\times10^{8}$	&	$0.2224$	&	$1.3730$	&	$1.1345$	&	$0.7049$	&	$1.5602$	&	$6.637$	&	$1.2230$	&	$0.4480$	&	$13.87$	&	$53.28$	&	$-2.2$	&	$-2.0$	\\
	&		&		&	$1.1490$	&	$0.8169$	&	$1.7585$	&	$6.763$	&	$1.2263$	&	$0.4231$	&	$13.21$	&	$53.22$	&	$-2.2$	&	$-2.0$	\\
$3\times10^{8}$	&	$0.3302$	&	$1.3546$	&	$1.1933$	&	$0.8038$	&	$1.4675$	&	$7.362$	&	$1.2621$	&	$0.4002$	&	$12.71$	&	$52.97$	&	$-2.2$	&	$-1.9$	\\
	&		&		&	$1.2030$	&	$0.8963$	&	$1.5991$	&	$7.456$	&	$1.2639$	&	$0.3858$	&	$12.38$	&	$52.94$	&	$-2.2$	&	$-1.9$	\\
$1\times10^{9}$	&	$0.5010$	&	$1.3437$	&	$1.3049$	&	$0.9954$	&	$1.3853$	&	$8.265$	&	$1.2898$	&	$0.3736$	&	$12.19$	&	$52.70$	&	$-2.1$	&	$-1.9$	\\
	&		&		&	$1.3103$	&	$1.0688$	&	$1.4637$	&	$8.333$	&	$1.2905$	&	$0.3660$	&	$12.03$	&	$52.68$	&	$-2.1$	&	$-1.9$	\\
$3\times10^{9}$	&	$0.7274$	&	$1.3386$	&	$1.4475$	&	$1.2651$	&	$1.3260$	&	$9.247$	&	$1.3055$	&	$0.3660$	&	$12.11$	&	$52.51$	&	$-2.1$	&	$-1.9$	\\
	&		&		&	$1.4505$	&	$1.3240$	&	$1.3729$	&	$9.297$	&	$1.3058$	&	$0.3620$	&	$12.03$	&	$52.50$	&	$-2.1$	&	$-1.9$	\\
$5\times10^{9}$	&	$0.8639$	&	$1.3372$	&	$1.5273$	&	$1.4288$	&	$1.3021$	&	$9.769$	&	$1.3107$	&	$0.3673$	&	$12.18$	&	$52.46$	&	$-2.1$	&	$-1.9$	\\
 \hline
\end{tabular}
\centering
\label{tbl:CO CJ}
\end{minipage}
\end{table*}

The results do not depend much on the initial upstream temperature. The $D_{\textrm{CJ}}$ values for $T_{0,9}=0.04$ (the reason for choosing this temperature is explained in Section~\ref{sec:CO structure}) deviate from the results for $T_{0,9}=0.2$ by less than $10^{-3}$, and the key parameters of Table~\ref{tbl:CO CJ} deviate by less than $0.6\%$, where the largest deviation is obtained for $q_{01,\textrm{CJ}}$ at $\rho_{0,7}=500$.

The results calculated with the NSE$4$ (NSE$5$, NSE$7$Si) isotope list deviate from the results presented above by less than $10^{-3}$ ($3\times10^{-4}$, $7\times10^{-4}$), which suggests that our isotope list is converged to better than $10^{-3}$. The most uncertain input physics in this calculation is the Coulomb corrections. The contribution of the Coulomb corrections to the initial state is of the order of a few percent (highest contribution in the lowest densities). A slightly smaller contribution is obtained at the NSE state (see Table~\ref{tbl:CO CJ}). The Coulomb interaction terms also change the NSE state by a few percent (see Sections~\ref{sec:Bruenn72} and~\ref{sec:Khokhlov88}). We, therefore, estimate the uncertainty of the results to be on the sub-percent level (see Section~\ref{sec:screening}).

\subsubsection{Comparing CO CJ detonations to \citet{Bruenn72}}
\label{sec:Bruenn72}

\citet{Bruenn72} calculated CJ detonations for an initial composition of $X(^{12}\textrm{C})=X(^{16}\textrm{O})=0.49$, $X(^{22}\textrm{Ne})=0.02$, an upstream temperature of $T_{0,9}=0.3$ and a few values of the upstream density in the range of $[5\times10^{6},3\times10^{10}]\,\textrm{g}/\textrm{cm}^{3}$. We calculated the CJ NSE states for the same initial conditions by following the input physics of \citet{Bruenn72} as closely as possible. The EOS that was used for the CJ NSE values did not include nuclear-level excitation terms and probably did not include Coulomb terms as well. The list of isotopes included $341$ isotopes\footnote{In figure 1 of \citep{Bruenn72}, only 337 isotopes are shown; together with $n$, $p$ and $^{4}$He, one isotope is missing. We assume that $^{44}$S is missing from figure 1 since both $^{43}$S and $^{45}$S are included, so we add it to the list of isotopes.}. When possible, the binding energies are taken from \citet{Mattauch1965}\footnote{Some of the values in \citet[][p. 13]{Mattauch1965} are not clearly visible in the online scanned version. In these cases, we used the modern values, since the values in this paper are almost identical to the modern ones.}, and for the remainder, the exponential mass formula of \citet{Cameron1965} was being used. Actually, the mass formula of \citet{Cameron1965} seems to contain possible errors, so we apply a few corrections to it (see Appendix~\ref{sec:CE bugs}). We assume that these correction were applied by \citet{Bruenn72} as well. Finally, the nuclear partition functions of \citet{Clifford1965} were used. 

The results of our calculations with the same input physics of \citet{Bruenn72} are compared to the results of \citet{Bruenn72} in Table~\ref{tbl:B72} for a few representing upstream densities (compare rows `B$72$ setup' to rows `B$72$'). The obtained $P_{\textrm{CJ}}$ and $\rho_{\textrm{CJ}}$ from our calculations are systematically larger than the results of \citet{Bruenn72} (by $9-22\%$ and $7-20\%$, respectively), while $q_{01,\textrm{CJ}}$ is systematically lower (by $4-6\%$). We show below that the reason for this discrepancy is the NSE calculation and not the EOS. But first, let us compare the results obtained with the input physics of \citet{Bruenn72} to the calculation of the same initial conditions with our default input physics (the row `Default'), which highlights the sensitivity of the results to various assumptions. We concentrate on the $q_{01,\textrm{CJ}}$ values for $\rho_{0,7}=500$ that shows the largest sensitivity. The value for the input physics of \citet{Bruenn72} deviates from the default input physics value by $\myapprox19\%$. The Coulomb term of the NSE reduces the deviation to $\myapprox13\%$ and the Coulomb term of the EOS reduces the deviation even further, to $\myapprox9\%$. This demonstrates that the sensitivity of the result to the Coulomb corrections can reach as high as $10$ percent. Including the nuclear level excitations terms in the EOS (with the modern values of the partition functions) reduces the deviation to $\myapprox1.6\%$, demonstrating the importance of these terms. The remaining discrepancy is reduced to $\myapprox0.2\%$ by using the modern values for the partition functions instead of the nuclear partition functions of \citet{Clifford1965} for the calculation of the NSE.

\begin{table*}
\begin{minipage}{150mm}
\caption{Parameters of CJ detonations for an initial composition of $X(^{12}\textrm{C})=X(^{16}\textrm{O})=0.49$, $X(^{22}\textrm{Ne})=0.02$ and an upstream temperature of $T_{0,9}=0.3$ for a few representing upstream densities. For each upstream density, we present the results of \citet[][B$72$]{Bruenn72}, the results of our calculations with the same input physics of \citet[][B$72$ setup]{Bruenn72}, B$72$ setup with the addition of the Coulomb correction term for the NSE (B$72$ setup + Coul. NSE), the additional inclusion of the Coulomb correction terms for the EOS (B$72$ setup + Coul. NSE + Coul. EOS), the additional inclusion of the nuclear level excitations terms in the EOS (using the modern values of the partition functions, B$72$ setup + Coul. NSE + Coul. EOS + $\varepsilon_{ex}$), and by further using the modern values for the partition functions instead of the nuclear partition functions of \citet[][]{Clifford1965} for the calculation of the NSE (B$72$ setup + Coul. NSE + Coul. EOS + $\varepsilon_{ex}$ + part.). The upper rows for each upstream density are the results with our default input physics.}
\begin{tabular}{|c||c||c||c||c||c|}
\hline
$\rho_{0}$  & \textrm{Case} &$P_{\textrm{CJ}}/P_{0}$ & $ \rho_{\textrm{CJ}}/\rho_{0} $  & $T_{\textrm{CJ}}$ &$q_{01,\textrm{CJ}}$ \\ 
$[\textrm{g}/\textrm{cm}^{3}]$  &  & &  & $[10^{9}\,\textrm{K}]$ & $[10^{17}\,\textrm{erg}/\textrm{g}]$  \\ \hline
$5\times10^{6}$	&	$\textrm{Default}$	&	$10.23$	&	$1.727$	&	$4.472$	&	$7.290$	\\
	&	$\textrm{B}72\,\textrm{setup}+\textrm{Coul. NSE}+\textrm{Coul. EOS}+\varepsilon_{ex}+\textrm{part.}$	&	$10.23$	&	$1.726$	&	$4.473$	&	$7.299$	\\
	&	$\textrm{B}72\,\textrm{setup}+\textrm{Coul. NSE}+\textrm{Coul. EOS}+\varepsilon_{ex}$	&	$10.22$	&	$1.725$	&	$4.471$	&	$7.293$	\\
	&	$\textrm{B}72\,\textrm{setup}+\textrm{Coul. NSE}+\textrm{Coul. EOS}$	&	$10.22$	&	$1.724$	&	$4.472$	&	$7.292$	\\
	&	$\textrm{B}72\,\textrm{setup}+\textrm{Coul. NSE}$	&	$10.08$	&	$1.723$	&	$4.468$	&	$7.297$	\\
	&	$\textrm{B}72\,\textrm{setup}$	&	$10.08$	&	$1.726$	&	$4.467$	&	$7.271$	\\
	&	$\textrm{B}72$	&	$8.058$	&	$1.411$	&	$4.265$	&	$7.720$	\\
\hline$2\times10^{8}$	&	$\textrm{Default}$	&	$2.654$	&	$1.500$	&	$7.107$	&	$3.975$	\\
	&	$\textrm{B}72\,\textrm{setup}+\textrm{Coul. NSE}+\textrm{Coul. EOS}+\varepsilon_{ex}+\textrm{part.}$	&	$2.653$	&	$1.500$	&	$7.106$	&	$3.974$	\\
	&	$\textrm{B}72\,\textrm{setup}+\textrm{Coul. NSE}+\textrm{Coul. EOS}+\varepsilon_{ex}$	&	$2.646$	&	$1.499$	&	$7.080$	&	$3.937$	\\
	&	$\textrm{B}72\,\textrm{setup}+\textrm{Coul. NSE}+\textrm{Coul. EOS}$	&	$2.660$	&	$1.500$	&	$7.104$	&	$3.844$	\\
	&	$\textrm{B}72\,\textrm{setup}+\textrm{Coul. NSE}$	&	$2.633$	&	$1.498$	&	$7.113$	&	$3.800$	\\
	&	$\textrm{B}72\,\textrm{setup}$	&	$2.611$	&	$1.496$	&	$7.025$	&	$3.712$	\\
	&	$\textrm{B}72$	&	$2.280$	&	$1.338$	&	$6.970$	&	$3.890$	\\
\hline$5\times10^{9}$	&	$\textrm{Default}$	&	$1.653$	&	$1.302$	&	$9.801$	&	$3.519$	\\
	&	$\textrm{B}72\,\textrm{setup}+\textrm{Coul. NSE}+\textrm{Coul. EOS}+\varepsilon_{ex}+\textrm{part.}$	&	$1.653$	&	$1.302$	&	$9.799$	&	$3.518$	\\
	&	$\textrm{B}72\,\textrm{setup}+\textrm{Coul. NSE}+\textrm{Coul. EOS}+\varepsilon_{ex}$	&	$1.650$	&	$1.301$	&	$9.725$	&	$3.462$	\\
	&	$\textrm{B}72\,\textrm{setup}+\textrm{Coul. NSE}+\textrm{Coul. EOS}$	&	$1.666$	&	$1.305$	&	$9.823$	&	$3.213$	\\
	&	$\textrm{B}72\,\textrm{setup}+\textrm{Coul. NSE}$	&	$1.647$	&	$1.300$	&	$9.868$	&	$3.080$	\\
	&	$\textrm{B}72\,\textrm{setup}$	&	$1.628$	&	$1.295$	&	$9.497$	&	$2.902$	\\
	&	$\textrm{B}72$	&	$1.492$	&	$1.210$	&	$9.414$	&	$3.017$	\\
\hline
\end{tabular}
\centering
\label{tbl:B72}
\end{minipage}
\end{table*}

We turn now to analyse the reason for the differences between the `B$72$ setup' and `B$72$' results. A somewhat simpler case to study is the NSE state at some $\rho$, $T$, and $Y_{e}$ with the same input physics of \citet{Bruenn72}, given in \citet{Bruenn71}. We concentrate on the results with a neutron--proton ratio of $1$ ($Y_{e}\approx0.5$) from table 1 of \citet{Bruenn71}. The results of our calculations with the same input physics of \citet{Bruenn71} are compared to the results of \citet{Bruenn71} in Table~\ref{tbl:B71} (compare rows `B$71$ setup' to rows `B$71$'). Although our pressure calculations agree with those of \citet{Bruenn71} to better than $0.5\%$, the obtained $\bar{A}$ deviates at high temperatures by $5-10\%$. If we recalculate the pressure with the $\bar{A}$ values of \citet[][this only changes the small contributions of the ions]{Bruenn71}, then the pressures agree to better than $1.5\times10^{-3}$. This result suggests that our EOS is consistent with the EOS used by \citet{Bruenn71}. However, the different values of $\bar{A}$ demonstrate that the NSE states are different, which lead to different CJ NSE states. As the code that was used to calculate the results of \citet{Bruenn71} was lost\footnote{Bruenn (private communication).}, we were unable to identify the cause of this discrepancy. 

\begin{table*}
\begin{minipage}{100mm}
\caption{The NSE state for a few values of $\rho$, $T$, and $Y_{e}=0.5$. For each case, we present the results of \citet[][B$71$]{Bruenn71}, the results of our calculations with the same input physics of \citet[][B$71$ setup]{Bruenn71}, and the results of recalculating the pressure with the $\bar{A}$ values of \citet[][B$71$ setup + B$71$ $\bar{A}$]{Bruenn71}.}
\begin{tabular}{|c||c||c||c||c|}
\hline
$\rho$  & $T$ & \textrm{Case} & $P$ & $\bar{A}$\\ 
$[\textrm{g}/\textrm{cm}^{3}]$  & $[10^{9}\,\textrm{K}]$ & & $[\textrm{erg}/\textrm{cm}^{3}]$ &   \\ \hline

$1\times10^{7}$	&	$6$	&	$\textrm{B}71\,\textrm{setup}$	&	$9.423\times10^{24}$	&	$6.963$	\\
	&		&	$\textrm{B}71\,\textrm{setup} + \textrm{B}71\,\bar{A}$	&	$9.459\times10^{24}$	&	$6.634$	\\
	&		&	$\textrm{B}71$	&	$9.461\times10^{24}$	&	$6.634$	\\
\hline$1\times10^{8}$	&	$3$	&	$\textrm{B}71\,\textrm{setup}$	&	$2.507\times10^{25}$	&	$55.93$	\\
	&		&	$\textrm{B}71\,\textrm{setup} + \textrm{B}71\,\bar{A}$	&	$2.507\times10^{25}$	&	$55.98$	\\
	&		&	$\textrm{B}71$	&	$2.511\times10^{25}$	&	$55.98$	\\
\hline$1\times10^{8}$	&	$7$	&	$\textrm{B}71\,\textrm{setup}$	&	$4.972\times10^{25}$	&	$7.609$	\\
	&		&	$\textrm{B}71\,\textrm{setup} + \textrm{B}71\,\bar{A}$	&	$4.980\times10^{25}$	&	$7.527$	\\
	&		&	$\textrm{B}71$	&	$4.983\times10^{25}$	&	$7.527$	\\
\hline$2\times10^{8}$	&	$3$	&	$\textrm{B}71\,\textrm{setup}$	&	$1.262\times10^{27}$	&	$55.97$	\\
	&		&	$\textrm{B}71\,\textrm{setup} + \textrm{B}71\,\bar{A}$	&	$1.262\times10^{27}$	&	$55.64$	\\
	&		&	$\textrm{B}71$	&	$1.262\times10^{27}$	&	$55.64$	\\
\hline$2\times10^{9}$	&	$8$	&	$\textrm{B}71\,\textrm{setup}$	&	$1.476\times10^{27}$	&	$14.41$	\\
	&		&	$\textrm{B}71\,\textrm{setup} + \textrm{B}71\,\bar{A}$	&	$1.468\times10^{27}$	&	$15.69$	\\
	&		&	$\textrm{B}71$	&	$1.470\times10^{27}$	&	$15.69$	\\
\hline
\end{tabular}
\centering
\label{tbl:B71}
\end{minipage}
\end{table*}

\subsubsection{Comparing CO CJ detonations to \citet{Khokhlov88}}
\label{sec:Khokhlov88}

\citet{Khokhlov88} calculated CJ detonations for CO, an upstream temperature of $T_{0,9}=0.2$ and a few values of the upstream density in the range of $[10^{7},5\times10^{9}]\,\textrm{g}/\textrm{cm}^{3}$. We calculated the CJ NSE states for the same initial conditions by following the input physics of \citet{Khokhlov88}. The difference between our Coulomb terms and those used by \citet{Khokhlov88} is smaller than a percent, and since the Coulomb corrections are a few percent at most, this difference can lead to deviations that are smaller than $10^{-3}$. The list of isotopes included $83$ isotopes, and we used the modern values of the binding energies and partition functions.

Our comparison of the results of our calculations with the same input physics of \citet{Khokhlov88} to those of \citet{Khokhlov88} in Table~\ref{tbl:K88} (i.e., comparison of rows `K$88$ setup' to rows `K$88$') reveals large deviations at low densities (up to $13\,\%$ in $q_{01,\textrm{CJ}}$, for example). We suggest below that the reason for the discrepancy is a possible error in the EOS used by \citet{Khokhlov88}. Before we do so, we compare the results obtained with the input physics of \citet{Khokhlov88} to the calculation of the same initial conditions with our default input physics (the row `Default'). The $q_{01,\textrm{CJ}}$ values for the input physics of \citet{Khokhlov88} deviate from the default input physics value by $1-6\,\%$. The Coulomb term for the NSE reduces the deviation to below $10^{-3}$. This once again demonstrates that the sensitivity of the result to the Coulomb corrections is on the order of a few percent. 

\begin{table*}
\begin{minipage}{150mm}
\caption{Parameters of CJ detonations for CO and an upstream temperature of $T_{0,9}=0.2$ for a few upstream densities. For each upstream density, we present the results of \citet[][K$88$]{Khokhlov88}, the results of our calculations with the same input physics as that used by \citet[][K$88$ setup]{Khokhlov88} and by adding the Coulomb correction term for the NSE (K$88$ setup + Coul. NSE). The upper rows for each upstream density are the results obtained with our default input physics.}
\begin{tabular}{|c||c||c||c||c||c||c||c|}
\hline
$\rho_{0}$  & \textrm{Case} &$P_{\textrm{CJ}}/P_{0}$ & $ \rho_{0}/\rho_{\textrm{CJ}} $  & $T_{\textrm{CJ}}$ &$q_{01,\textrm{CJ}}$ & $\gamma^{e}_{\textrm{CJ}}$ & $D_{\textrm{CJ}}$\\ 
$[\textrm{g}/\textrm{cm}^{3}]$  &  & &  & $[10^{9}\,\textrm{K}]$ & $[10^{17}\,\textrm{erg}/\textrm{g}]$ &  & $[10^{4}\,\textrm{km}/\textrm{s}]$  \\ \hline
$1\times10^{7}$	&	$\textrm{Default}$	&	$7.668$	&	$0.5760$	&	$5.058$	&	$6.478$	&	$1.1844$	&	$1.1545$	\\
	&	$\textrm{K}88\,\textrm{setup}+\textrm{Coul. NSE}$	&	$7.675$	&	$0.5756$	&	$5.059$	&	$6.477$	&	$1.1843$	&	$1.1546$	\\
	&	$\textrm{K}88\,\textrm{setup}$	&	$7.656$	&	$0.5749$	&	$5.049$	&	$6.430$	&	$1.1810$	&	$1.1518$	\\
	&	$\textrm{K}88$	&	$7.95$	&	$0.59$	&	$4.73$	&	$7.10$	&	$1.23$	&	$1.19$	\\
\hline$3\times10^{7}$	&	$\textrm{Default}$	&	$4.782$	&	$0.5973$	&	$5.866$	&	$5.197$	&	$1.1747$	&	$1.1231$	\\
	&	$\textrm{K}88\,\textrm{setup}+\textrm{Coul. NSE}$	&	$4.782$	&	$0.5973$	&	$5.866$	&	$5.197$	&	$1.1747$	&	$1.1232$	\\
	&	$\textrm{K}88\,\textrm{setup}$	&	$4.756$	&	$0.5970$	&	$5.838$	&	$5.131$	&	$1.1731$	&	$1.1190$	\\
	&	$\textrm{K}88$	&	$5.15$	&	$0.59$	&	$5.60$	&	$5.84$	&	$1.18$	&	$1.16$	\\
\hline$1\times10^{8}$	&	$\textrm{Default}$	&	$3.169$	&	$0.6409$	&	$6.637$	&	$4.291$	&	$1.2230$	&	$1.1345$	\\
	&	$\textrm{K}88\,\textrm{setup}+\textrm{Coul. NSE}$	&	$3.168$	&	$0.6412$	&	$6.637$	&	$4.292$	&	$1.2230$	&	$1.1345$	\\
	&	$\textrm{K}88\,\textrm{setup}$	&	$3.146$	&	$0.6415$	&	$6.578$	&	$4.214$	&	$1.2224$	&	$1.1294$	\\
	&	$\textrm{K}88$	&	$3.33$	&	$0.63$	&	$6.47$	&	$4.67$	&	$1.21$	&	$1.16$	\\
\hline$3\times10^{8}$	&	$\textrm{Default}$	&	$2.434$	&	$0.6814$	&	$7.362$	&	$3.833$	&	$1.2621$	&	$1.1933$	\\
	&	$\textrm{K}88\,\textrm{setup}+\textrm{Coul. NSE}$	&	$2.435$	&	$0.6813$	&	$7.362$	&	$3.833$	&	$1.2621$	&	$1.1934$	\\
	&	$\textrm{K}88\,\textrm{setup}$	&	$2.415$	&	$0.6825$	&	$7.261$	&	$3.740$	&	$1.2617$	&	$1.1873$	\\
	&	$\textrm{K}88$	&	$2.49$	&	$0.68$	&	$7.18$	&	$4.02$	&	$1.25$	&	$1.21$	\\
\hline$1\times10^{9}$	&	$\textrm{Default}$	&	$1.987$	&	$0.7219$	&	$8.265$	&	$3.578$	&	$1.2898$	&	$1.3049$	\\
	&	$\textrm{K}88\,\textrm{setup}+\textrm{Coul. NSE}$	&	$1.987$	&	$0.7218$	&	$8.264$	&	$3.578$	&	$1.2899$	&	$1.3051$	\\
	&	$\textrm{K}88\,\textrm{setup}$	&	$1.970$	&	$0.7233$	&	$8.093$	&	$3.454$	&	$1.2895$	&	$1.2974$	\\
	&	$\textrm{K}88$	&	$2.01$	&	$0.72$	&	$8.05$	&	$3.59$	&	$1.28$	&	$1.31$	\\
\hline$3\times10^{9}$	&	$\textrm{Default}$	&	$1.739$	&	$0.7542$	&	$9.247$	&	$3.505$	&	$1.3055$	&	$1.4475$	\\
	&	$\textrm{K}88\,\textrm{setup}+\textrm{Coul. NSE}$	&	$1.740$	&	$0.7541$	&	$9.246$	&	$3.505$	&	$1.3056$	&	$1.4479$	\\
	&	$\textrm{K}88\,\textrm{setup}$	&	$1.723$	&	$0.7566$	&	$8.972$	&	$3.338$	&	$1.3053$	&	$1.4384$	\\
	&	$\textrm{K}88$	&	$1.75$	&	$0.75$	&	$8.95$	&	$3.59$	&	$1.30$	&	$1.45$	\\
\hline$5\times10^{9}$	&	$\textrm{Default}$	&	$1.654$	&	$0.7680$	&	$9.769$	&	$3.518$	&	$1.3107$	&	$1.5273$	\\
	&	$\textrm{K}88\,\textrm{setup}+\textrm{Coul. NSE}$	&	$1.654$	&	$0.7681$	&	$9.768$	&	$3.518$	&	$1.3108$	&	$1.5279$	\\
	&	$\textrm{K}88\,\textrm{setup}$	&	$1.639$	&	$0.7704$	&	$9.430$	&	$3.322$	&	$1.3106$	&	$1.5175$	\\
	&	$\textrm{K}88$	&	$1.66$	&	$0.77$	&	$9.45$	&	$3.35$	&	$1.30$	&	$1.53$	\\
\hline
\end{tabular}
\centering
\label{tbl:K88}
\end{minipage}
\end{table*}

In order to analyse the reason for the differences between the `K$88$ setup' and the `K$88$' results, we calculate the pressure and $q_{01,\textrm{CJ}}$ at the NSE state for the values of $\rho_{\textrm{CJ}}$ and $T_{\textrm{CJ}}$ as given by \citet{Khokhlov88}. The results of our calculations with the same input physics of \citet{Khokhlov88} are compared to the results of \citet{Khokhlov88} in Table~\ref{tbl:K88EOS} (compare rows `K$88$ setup' to rows `K$88$'). The values of $q_{01,\textrm{CJ}}$ usually deviate by less than $\myapprox2\%$ (only for $\rho_{0,7}=300$ a deviation of $\myapprox5\%$ is obtained), which suggests that the compositions of the NSE states are similar. However, the deviation in the pressures are large for low densities and reach $\myapprox21\%$ for $\rho_{0,7}=1$. Since the agreement between the \citet{Nadyozhin74} electron--positron EOS used by \citet{Khokhlov88} and the EOS used by us is better than $0.1\%$ \citep{Timmes1999}, the difference is possibly because of some numerical bug. In fact, the difference between the pressures is almost exactly the radiation pressure (compare rows `K$88$ setup + twice $p_{\textrm{rad}}$' to rows `K$88$'). We conclude that the reason for the discrepancy is an apparent bug in the EOS used by \citet{Khokhlov88}\footnote{\citet{Khokhlov88} claims that the discrepancy between his results and the results of \citet{Bruenn72} at low densities is because of the approximate EOS used by \citet{Bruenn72}, while, in fact, we find that the EOS used by \citet{Bruenn72} is accurate and the one used by \citet{Khokhlov88} may contains an error.}.

\begin{table*}
\begin{minipage}{100mm}
\caption{The pressure and $q_{01,\textrm{CJ}}$ at the NSE state for the values of $\rho_{\textrm{CJ}}$ and $T_{\textrm{CJ}}$ as given by \citet{Khokhlov88} for CO. For each case, we present the results of \citet[][K$88$]{Khokhlov88}, the results of our calculations with the same input physics of \citet[][K$88$ setup]{Khokhlov88}, and the results of recalculating the pressure with twice the radiation pressure (K$88$ setup + twice $p_{\textrm{rad}}$).}
\begin{tabular}{|c||c||c||c|}
\hline
$\rho_{0}$  & \textrm{Case} &$P_{\textrm{CJ}}/P_{0}^{\textrm{K}88}$ & $q_{01,\textrm{CJ}}$\\ 
$[\textrm{g}/\textrm{cm}^{3}]$  &  & & $[10^{17}\,\textrm{erg}/\textrm{g}]$  \\ \hline
$1\times10^{7}$	&	$\textrm{K}88\,\textrm{setup}$	&	$6.46$	&	$7.08$	\\
	&	$\textrm{K}88\,\textrm{setup}+\textrm{twice}\,p_{\textrm{rad}}$	&	$7.95$	&		\\
	&	$\textrm{K}88$	&	$7.95$	&	$7.10$	\\
\hline$3\times10^{7}$	&	$\textrm{K}88\,\textrm{setup}$	&	$4.46$	&	$5.90$	\\
	&	$\textrm{K}88\,\textrm{setup}+\textrm{twice}\,p_{\textrm{rad}}$	&	$5.08$	&		\\
	&	$\textrm{K}88$	&	$5.15$	&	$5.84$	\\
\hline$1\times10^{8}$	&	$\textrm{K}88\,\textrm{setup}$	&	$3.14$	&	$4.64$	\\
	&	$\textrm{K}88\,\textrm{setup}+\textrm{twice}\,p_{\textrm{rad}}$	&	$3.34$	&		\\
	&	$\textrm{K}88$	&	$3.33$	&	$4.67$	\\
\hline$3\times10^{8}$	&	$\textrm{K}88\,\textrm{setup}$	&	$2.40$	&	$4.03$	\\
	&	$\textrm{K}88\,\textrm{setup}+\textrm{twice}\,p_{\textrm{rad}}$	&	$2.47$	&		\\
	&	$\textrm{K}88$	&	$2.49$	&	$4.02$	\\
\hline$1\times10^{9}$	&	$\textrm{K}88\,\textrm{setup}$	&	$1.97$	&	$3.60$	\\
	&	$\textrm{K}88\,\textrm{setup}+\textrm{twice}\,p_{\textrm{rad}}$	&	$1.99$	&		\\
	&	$\textrm{K}88$	&	$2.01$	&	$3.59$	\\
\hline$3\times10^{9}$	&	$\textrm{K}88\,\textrm{setup}$	&	$1.74$	&	$3.41$	\\
	&	$\textrm{K}88\,\textrm{setup}+\textrm{twice}\,p_{\textrm{rad}}$	&	$1.75$	&		\\
	&	$\textrm{K}88$	&	$1.75$	&	$3.59$	\\
\hline$5\times10^{9}$	&	$\textrm{K}88\,\textrm{setup}$	&	$1.64$	&	$3.28$	\\
	&	$\textrm{K}88\,\textrm{setup}+\textrm{twice}\,p_{\textrm{rad}}$	&	$1.65$	&		\\
	&	$\textrm{K}88$	&	$1.66$	&	$3.35$	\\
\hline
\end{tabular}
\centering
\label{tbl:K88EOS}
\end{minipage}
\end{table*}

\subsubsection{Comparing CO CJ detonations to \citet{Gamezo99}}
\label{sec:Gamezo99}

\citet{Gamezo99} calculated CJ detonations for CO, an upstream temperature of $T_{0,9}=0.2$ and a few values of the upstream density in the range of $[3\times10^{5},3\times10^{9}]\,\textrm{g}/\textrm{cm}^{3}$. The list of isotopes included $13$ $\alpha$-nuclei, and Coulomb corrections were probably not included. We calculate the CJ NSE states for the same initial conditions by following the input physics of \citet{Gamezo99}. We use the modern values of the binding energies and partition functions.

\begin{figure}
\includegraphics[width=0.48\textwidth]{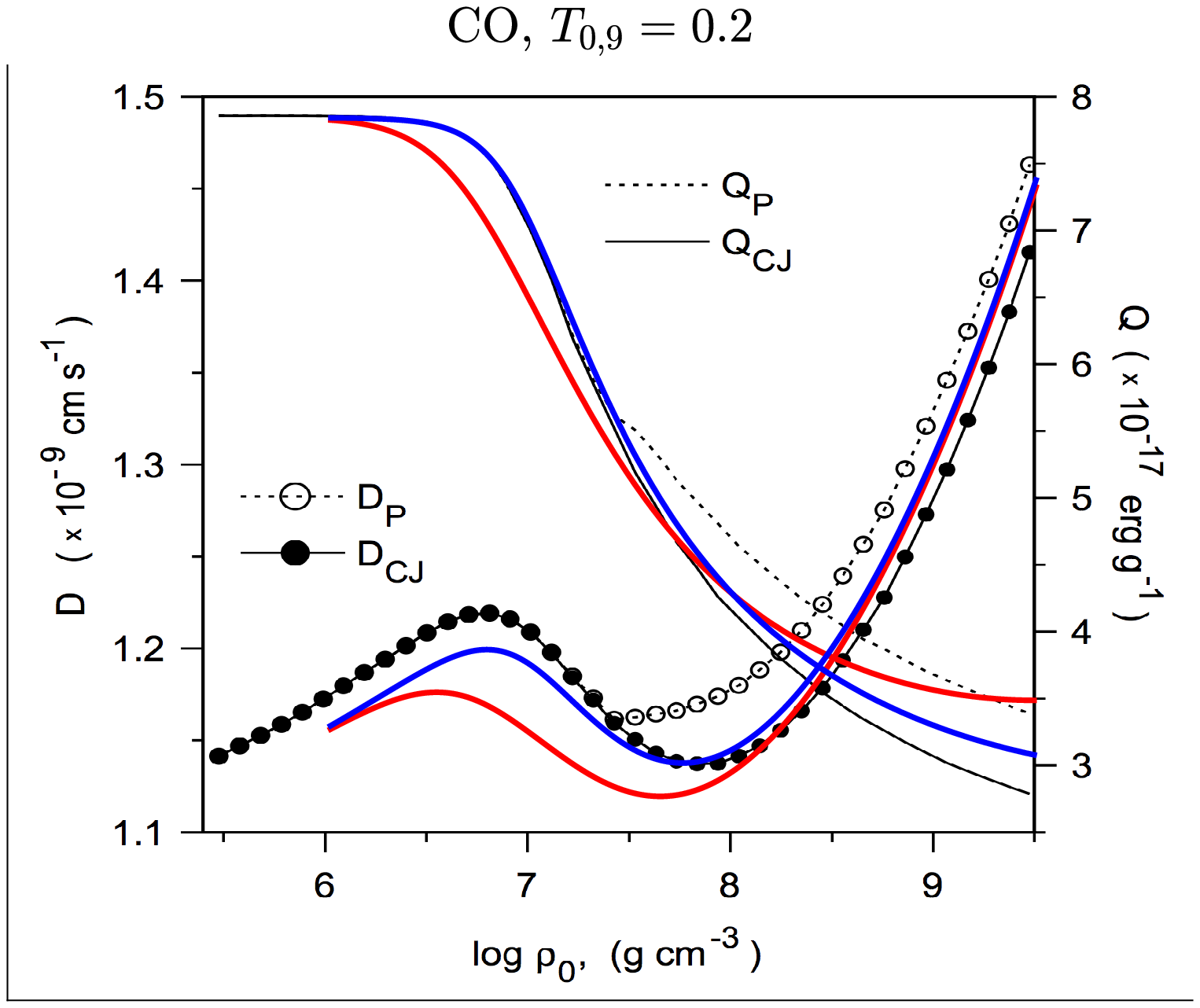}
\caption{Figure 2 from \citet{Gamezo99} ($\copyright$ AAS. Reproduced with permission). $D_{\textrm{CJ}}$ and $q_{01,\textrm{CJ}}$ for CO and $T_{0,9}=0.2$ as a function of the upstream density. Black lines represent the results of \citet{Gamezo99}, blue lines are our results with the input physics of \citet{Gamezo99}, and red lines reflect results with our default input physics.
\label{fig:Gamezo99}}
\end{figure}

The results of our calculations with the same input physics of \citet{Gamezo99} are compared to the results of \citet{Gamezo99} in Figure~\ref{fig:Gamezo99} (compare the black lines to the blue lines) and in Table~\ref{tbl:G99} (compare rows `G$99$ setup' to rows `G$99$'). The general behaviour of both $D_{\textrm{CJ}}$ and $q_{01,\textrm{CJ}}$ is similar. Deviations of up to $\myapprox2\%$ are obtained in $D_{\textrm{CJ}}$ and large deviations are obtained at high densities in $q_{01,\textrm{CJ}}$ ($\myapprox7\,\%$ for $\rho_{0,7}=100$). Below, we try to analyse the reason for the discrepancy. 

\begin{table*}
\begin{minipage}{150mm}
\caption{Parameters of CJ detonations for CO and upstream temperature of $T_{0,9}=0.2$ for a few upstream densities. For each upstream density, we present the results of \citet[][G$99$]{Gamezo99}, the results of our calculations with the same input physics of \citet[][G$99$ setup]{Gamezo99}, the G$99$ setup with the addition of the Coulomb correction term for the NSE (G$99$ setup + Coul. NSE), and the additional inclusion of the Coulomb correction term for the EOS (G$99$ setup + Coul. NSE + Coul. EOS), and also the addition of the nuclear level excitations (G$99$ setup + Coul. NSE + Coul. EOS + $\varepsilon_{ex}$). The upper rows for each upstream density are the results with our default input physics.}
\begin{tabular}{|c||c||c||c|}
\hline
$\rho_{0}$  & \textrm{Case} & $q_{01,\textrm{CJ}}$ & $D_{\textrm{CJ}}$\\ 
$[\textrm{g}/\textrm{cm}^{3}]$  &  & $[10^{17}\,\textrm{erg}/\textrm{g}]$ &  $[10^{4}\,\textrm{km}/\textrm{s}]$  \\ \hline
$1\times10^{7}$	&	$\textrm{Default}$	&	$6.478$	&	$1.155$	\\
     	&	$\textrm{G}99\,\textrm{setup}+\textrm{Coul. NSE}+\textrm{Coul. EOS}+\varepsilon_{ex}$	&	$7.118$	&	$1.192$	\\
     	&	$\textrm{G}99\,\textrm{setup}+\textrm{Coul. NSE}+\textrm{Coul. EOS}$	&	$7.113$	&	$1.193$	\\
     	&	$\textrm{G}99\,\textrm{setup}+\textrm{Coul. NSE}$	&	$7.120$	&	$1.195$	\\
     	&	$\textrm{G}99\,\textrm{setup}$	&	$7.059$	&	$1.192$	\\
     	&	$\textrm{G}99$	&	$7.03$	&	$1.21$	\\
     \hline$1\times10^{8}$	&	$\textrm{Default}$	&	$4.291$	&	$1.134$	\\
     	&	$\textrm{G}99\,\textrm{setup}+\textrm{Coul. NSE}+\textrm{Coul. EOS}+\varepsilon_{ex}$	&	$4.467$	&	$1.151$	\\
     	&	$\textrm{G}99\,\textrm{setup}+\textrm{Coul. NSE}+\textrm{Coul. EOS}$	&	$4.427$	&	$1.153$	\\
     	&	$\textrm{G}99\,\textrm{setup}+\textrm{Coul. NSE}$	&	$4.396$	&	$1.154$	\\
     	&	$\textrm{G}99\,\textrm{setup}$	&	$4.290$	&	$1.146$	\\
     	&	$\textrm{G}99$	&	$4.17$	&	$1.14$	\\
     \hline$1\times10^{9}$	&	$\textrm{Default}$	&	$3.578$	&	$1.305$	\\
     	&	$\textrm{G}99\,\textrm{setup}+\textrm{Coul. NSE}+\textrm{Coul. EOS}+\varepsilon_{ex}$	&	$3.650$	&	$1.315$	\\
     	&	$\textrm{G}99\,\textrm{setup}+\textrm{Coul. NSE}+\textrm{Coul. EOS}$	&	$3.542$	&	$1.319$	\\
     	&	$\textrm{G}99\,\textrm{setup}+\textrm{Coul. NSE}$	&	$3.459$	&	$1.320$	\\
     	&	$\textrm{G}99\,\textrm{setup}$	&	$3.310$	&	$1.309$	\\
     	&	$\textrm{G}99$	&	$3.08$	&	$1.28$	\\
     \hline
\end{tabular}
\centering
\label{tbl:G99}
\end{minipage}
\end{table*}

First, it is not clear how the CJ values were actually calculated by \citet{Gamezo99}, since they claim to integrate the reaction equations to obtain the CJ values. Besides the fact that this is not required, as the CJ values are independent of reaction rates, it is also not possible for pathological detonations, as the integration hits a sonic point for $D<D_{*}$. Let us now concentrate on the $\rho_{0,7}=1$ case, where we obtain a similar $q_{01,\textrm{CJ}}$ but a lower $D_{\textrm{CJ}}$. We find from the upper panel of figure 3 of \citet{Gamezo99} that $u_{\textrm{CJ}}\approx0.68\times10^{4}\,\textrm{km}/\textrm{s}$ and that $c_{s,\textrm{CJ}}\approx0.75\times10^{4}\,\textrm{km}/\textrm{s}$. With these $u_{\textrm{CJ}}$ and $D_{\textrm{CJ}}$ figures, we get from Equation~\eqref{eq:u} that $\rho_{\textrm{CJ},7}\approx1.79$. We can now use our EOS (without the Coulomb correction) to find $T_{\textrm{CJ}}$ in two ways. For the value of $c_{s,\textrm{CJ}}$, we find that $T_{\textrm{CJ},9}\approx5.31$, and for the value of $c_{s,\textrm{CJ}}^{e}=u_{\textrm{CJ}}$, we find that $T_{\textrm{CJ},9}\approx5.10$. This discrepancy demonstrates that the calculation of \citet{Gamezo99} is inconsistent (to the level of a few percent). 

Because of these unresolved discrepancies, we did not try to reproduce the results of \citet{Gamezo99} for the pathological case. We will just mention here that the $q_{01,*}$ values presented in figure 2 of \citet{Gamezo99} seem to be inconsistent. Our calculations always yield a $q_{01,*}<q_{01,\textrm{CJ}}$. This is because at higher detonation speeds, the temperature of the NSE state is higher and, therefore, more $^{4}$He nuclei are present, which decreases $q_{01}$. \citet{Gamezo99} obtained $q_{01,*}>q_{01,\textrm{CJ}}$, which seems to be inconsistent. Moreover, from figure 2 of \citet{Gamezo99}, we can extract $q_{01,*}\approx4.70\times10^{17}\,\textrm{erg}/\textrm{g}$ for $\rho_{0,7}=10$, while from the bottom panel of figure 3 of \citet{Gamezo99} we find that $q_{01,*}\approx3.57\times10^{17}\,\textrm{erg}/\textrm{g}$ for the same $\rho_{0}$.

Let us go back now to Table~\ref{tbl:G99} and compare the results obtained with the input physics of \citet{Gamezo99} to the calculation with our default input physics (the row `Default'; compare also the blue and the red lines in Figure~\ref{fig:Gamezo99}). The $q_{01,\textrm{CJ}}$ values for the input physics of \citet{Gamezo99} deviate from the default input physics values by up to $\myapprox9\,\%$. The Coulomb terms and the nuclear level excitations terms change the values of $q_{01,\textrm{CJ}}$ by up to a few percent each. Finally, extending the isotope list to our default list changes the values of $q_{01,\textrm{CJ}}$ by $\myapprox2-9\%$. The reason for this alteration is that $\alpha$-nuclei cannot correctly represent the NSE state, as a significant fraction of the mass can be stored in different isotopes (see bottom panel of Figure~\ref{fig:DetonationSpeed_CJAbu}). This inability is compensated for by artificially increasing the mass fractions of all the elements with a $Z_{i}\ge14$, especially $^{56}$Ni and $^{52}$Fe. For this reason, calculations with $\alpha$-nuclei are inadequate for the accurate analysis that we aim for in this work.


\subsection{CJ detonations of pure helium}
\label{sec:He CJ}

\begin{figure}
\includegraphics[width=0.48\textwidth]{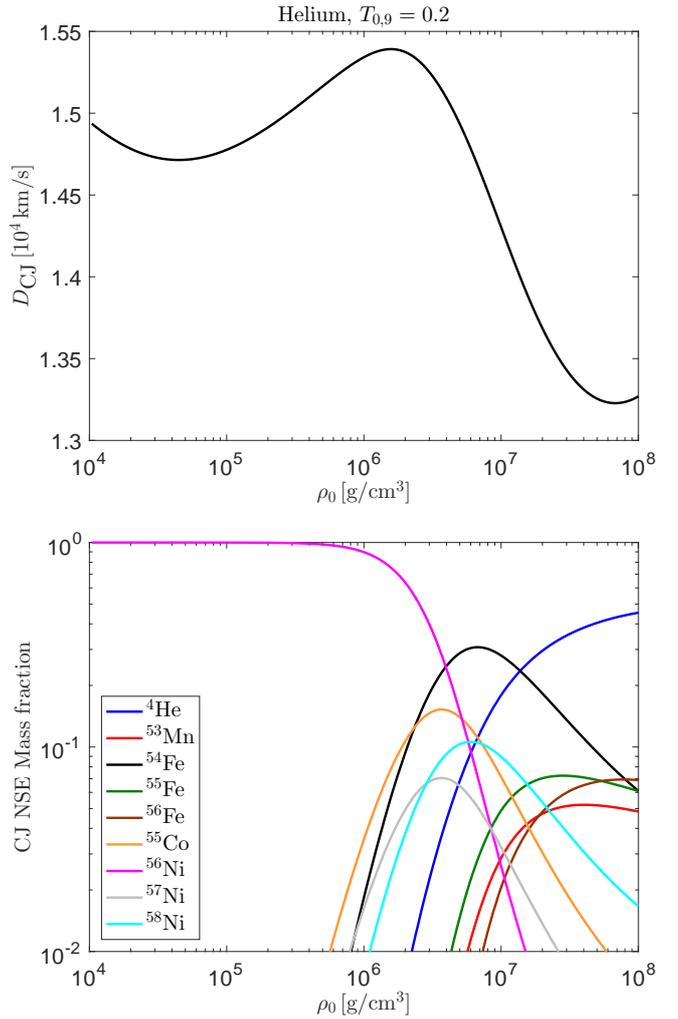}
\caption{Upper panel: $D_{\textrm{CJ}}$ for He and upstream temperature of $T_{0,9}=0.2$ as a function of the upstream density. Bottom panel: mass fractions of key isotopes at the CJ NSE state for the same upstream conditions. We only present the mass fraction of isotopes that have a mass fraction larger than $5\times10^{-2}$ at some $\rho_{0}$ within the inspected range. 
\label{fig:DetonationSpeed_CJAbu_Helium}}
\end{figure}

The calculated $D_{\textrm{CJ}}$ for He is presented in the upper panel of Figure~\ref{fig:DetonationSpeed_CJAbu_Helium} for an upstream temperature of $T_{0,9}=0.2$ and an upstream density in the relevant range for supernovae of $[10^{4},10^{8}]\,\textrm{g}/\textrm{cm}^{3}$. Similarly to \citet{Dunkley2013}, we find that $D_{\textrm{CJ}}$ is not a monotonic function of $\rho_{0}$ and that it has a minimum at $\rho_{0,7}\approx4.5\times10^{-3}$ and a maximum at $\rho_{0,7}\approx0.16$. There is another minimum at $\rho_{0,7}\approx7$, which can also be extracted from table IV of \citet{Khokhlov88}. Key isotopes at the CJ NSE state are presented in the bottom panel of Figure~\ref{fig:DetonationSpeed_CJAbu_Helium} for the same upstream values. We only present the mass fraction of isotopes that have a mass fraction larger than $5\times10^{-2}$ at some $\rho_{0}$ within the inspected range. At low densities, the NSE state is dominated by $^{56}$Ni (with $\bar{A}\approx\tilde{A}\approx56$ at $\rho_{0,7}=10^{-3}$), while at higher densities the NSE state is mainly a mixture of $^{4}$He, $^{54}$Fe, $^{55}$Fe and $^{56}$Fe (with $\bar{A}\approx6.5$ and $\tilde{A}\approx53$ at $\rho_{0,7}=10$). A few key parameters of these CJ detonations are given in Table~\ref{tbl:He CJ}. The temperature at the CJ NSE state increases monotonically with $\rho_{0}$, which decreases both the $\bar{A}$ at these states and the released thermonuclear energy compared with the initial states, $q_{01,\textrm{CJ}}$. It is also demonstrated that the nuclear excitation energy contribution to the energy at the NSE state can reach $1$ percent for the high densities, and becomes much larger in magnitude than the Coulomb correction.   

\begin{table*}
\begin{minipage}{180mm}
\caption{Key parameters of CJ detonations for He and upstream temperature of $T_{0,9}=0.2$.}
\begin{tabular}{|c||c||c||c||c||c||c||c||c||c||c||c||c|}
\hline
$\rho_{0}$  & $P_{0}/\rho_{0}$ & $ \gamma_{0}$\footnote{$\gamma_0=c_{s,0}^{2}\rho_0/P_0$}  & $D_{\textrm{CJ}}$ & $P_{\textrm{CJ}}/\rho_{0}$   & $\rho_{\textrm{CJ}}/\rho_{0}$ & $T_{\textrm{CJ}}$ & $\gamma^{e}_{\textrm{CJ}}$\footnote{$\gamma^{e}_{\textrm{CJ}}=\left(c_{s,\textrm{CJ}}^{e}\right)^{2}\rho_{\textrm{CJ}}/P_{\textrm{CJ}}$} &$q_{01,\textrm{CJ}}$ & $\bar{A_{\textrm{CJ}}}$ & $\tilde{A_{\textrm{CJ}}}$ & $f_{\textrm{coul}}$ \footnote{$f_{\textrm{coul}}=\log_{10}\left(-\frac{\varepsilon_{\textrm{coul}}}{\varepsilon}\right)$ at the CJ state} &  $f_{\textrm{ex}}$ \footnote{$f_{\textrm{ex}}=\log_{10}\left(\frac{\varepsilon_{\textrm{ex}}}{\varepsilon}\right)$ at the CJ state}\\ 
$[\textrm{g}/\textrm{cm}^{3}]$  & $[\textrm{MeV}/m_{p}]$ &  & $[10^{4}\,\textrm{km}/\textrm{s}]$ & $[\textrm{MeV}/m_{p}]$   &  & $[10^{9}\,\textrm{K}]$ & & $[\textrm{MeV}/m_{p}]$ & & & &  \\ \hline
$1\times10^{4}$	&	$0.01373$	&	$1.5967$	&	$1.4948$	&	$1.0459$	&	$1.7937$	&	$1.341$	&	$1.2448$	&	$1.5803$	&	$56.00$	&	$56.00$	&	$-3.2$	&	$-10.5$	\\
$3\times10^{4}$	&	$0.01404$	&	$1.6178$	&	$1.4731$	&	$1.0070$	&	$1.7804$	&	$1.689$	&	$1.2659$	&	$1.5803$	&	$56.00$	&	$56.00$	&	$-3.1$	&	$-8.4$	\\
$1\times10^{5}$	&	$0.01590$	&	$1.6211$	&	$1.4776$	&	$0.9977$	&	$1.7566$	&	$2.180$	&	$1.3006$	&	$1.5803$	&	$55.99$	&	$56.00$	&	$-2.9$	&	$-6.6$	\\
$3\times10^{5}$	&	$0.02075$	&	$1.6082$	&	$1.5018$	&	$1.0222$	&	$1.7400$	&	$2.776$	&	$1.3255$	&	$1.5797$	&	$55.77$	&	$55.98$	&	$-2.7$	&	$-5.3$	\\
$1\times10^{6}$	&	$0.03312$	&	$1.5718$	&	$1.5342$	&	$1.0737$	&	$1.7346$	&	$3.653$	&	$1.3213$	&	$1.5656$	&	$50.12$	&	$55.81$	&	$-2.6$	&	$-4.0$	\\
$3\times10^{6}$	&	$0.05443$	&	$1.5201$	&	$1.5251$	&	$1.1194$	&	$1.7812$	&	$4.684$	&	$1.2203$	&	$1.4462$	&	$26.54$	&	$54.80$	&	$-2.7$	&	$-2.9$	\\
$1\times10^{7}$	&	$0.09261$	&	$1.4585$	&	$1.4304$	&	$1.0650$	&	$1.8357$	&	$5.860$	&	$1.0939$	&	$1.1083$	&	$11.44$	&	$53.75$	&	$-2.9$	&	$-2.3$	\\
$3\times10^{7}$	&	$0.1449$	&	$1.4116$	&	$1.3430$	&	$0.9588$	&	$1.7614$	&	$6.753$	&	$1.1151$	&	$0.8232$	&	$7.78$	&	$53.58$	&	$-2.9$	&	$-2.1$	\\
$1\times10^{8}$	&	$0.2279$	&	$1.3764$	&	$1.3269$	&	$0.9421$	&	$1.6356$	&	$7.666$	&	$1.1932$	&	$0.6564$	&	$6.57$	&	$53.35$	&	$-2.8$	&	$-2.0$	\\ \hline
\end{tabular}
\centering
\label{tbl:He CJ}
\end{minipage}
\end{table*}

The results do not depend much on the initial upstream temperature. The $D_{\textrm{CJ}}$ values for $T_{0,9}=0.01$\footnote{The electron--electron term is neglected here, and the correction is in the range of a few percent for $\rho_{0,7}\lesssim0.027$ and $T_{0,9}=0.01$.} deviate from the results for $T_{0,9}=0.2$ by less than $2\times10^{-3}$, and the key parameters of Table~\ref{tbl:CO CJ} deviate by less than $8.5\times10^{-3}$, where the largest deviation is obtained for $q_{01,\textrm{CJ}}$ at $\rho_{0,7}=10$.

The results calculated with the NSE$4$ (NSE$5$, NSE$7$Si) isotope list deviate from the results presented above by less than $2.5\times10^{-3}$ ($2\times10^{-4}$, $3\times10^{-8}$), which suggests that our isotope list is converged to $\mysim10^{-3}$. The most uncertain input physics in this calculation is the Coulomb corrections. The contribution of the Coulomb corrections to the initial state is of the order of a few percent (highest contribution for the lowest densities), with a smaller contribution obtained at the NSE state (see Table~\ref{tbl:He CJ}). The Coulomb interaction terms change the NSE composition in the order of a few percent as well (see Section~\ref{sec:Khokhlov88_He}). We, therefore, estimate the uncertainty of the results to be on the sub-percent level (see Section~\ref{sec:screening}).

\subsubsection{Comparing He CJ detonations to \citet{Mazurek1973}}
\label{sec:Mazurek}

\citet{Mazurek1973} calculated CJ detonations for He, an upstream temperature of $T_{0}=0.05$ and a few values of the upstream density in the range of $[10^{6},5\times10^{9}]\,\textrm{g}/\textrm{cm}^{3}$. The details of the input physics used by \citet{Mazurek1973} are given in \citet{MazurekPhDT} and includes a list of $155$ isotopes without Coulomb correction terms nor the nuclear-level excitation. The source for the values of the binding energies and partition functions is not given, so we use the modern values. We calculate the CJ NSE states for the same initial conditions of \citet{Mazurek1973} by following the input physics described above (M$73$ setup hereafter). The results of our calculations are compared to the results of \citet{Mazurek1973} in Table~\ref{tbl:M73_Helium} (compare rows `M$73$ setup' to rows `M$73$'). Large deviations are obtained at low densities (up to $11\,\%$ in $\rho_{\textrm{CJ}}/\rho_0$, for example). In these cases ($0.1\le\rho_{0,7}\le1$), our calculated $D_{\textrm{CJ}}$ is significantly lower (by up to $10\%$) than the values of \citet{Mazurek1973}. 

In order to analyse the reason for the discrepancy, we calculate the pressure, $q_{01,\textrm{CJ}}$, the electron--positron pressure and the electron--positron energy at the NSE state for the values of $\rho_{\textrm{CJ}}$ and $T_{\textrm{CJ}}$ as given by \citet{Mazurek1973}. The results of our calculations with the M$73$ setup are compared to the results of \citet{Mazurek1973} in Table~\ref{tbl:M73EOS_Helium} (compare rows `M$73$ setup' to rows `M$73$'). In order to calculate the electron--positron terms for M$73$, we assume that the CJ conditions hold and we use the analytical terms for the radiation and the ions (with the M$73$ setup values for $\bar{A}$). The values of $q_{01,\textrm{CJ}}$ deviate by less than $\myapprox5\%$, which suggests that the compositions of the NSE states are similar, and the difference between the pressure levels is below $\myapprox1\%$, which suggests that our pressure calculation is consistent with the one used by \citet{Mazurek1973}. Indeed, when we directly compare the electron--positron pressures, the deviation is smaller than $1.5\%$, which also suggests that the deviation in $\bar{A}$ is small. However, the electron--positron energies deviate by up to $33\%$, with the largest deviation obtained for $\rho_{0,7}=0.1$. We believe that this is because of inaccuracies in the EOS used by \citet{Mazurek1973} for the high positron-to-proton ratio, $n_{+}/n_{p}$. \citet{MazurekPhDT} admits that his EOS becomes less accurate in higher $n_{+}/n_{p}$, although the error is estimated to be $\mysim10^{-4}$ for $n_{+}/n_{p}\approx10$, where even for $\rho_{0,7}=0.1$ we only have $n_{+}/n_{p}\approx0.68$ (see Table~\ref{tbl:M73EOS_Helium}). \citet{MazurekPhDT} estimated the level of accuracy of his EOS by comparing it to Table A.4.1 in Appendix A.4 of \citet{Cox68}, and he claimed that his results match exactly the results there, except for regions with $n_{+}/n_{p}>50$ (there are really only three entries with $n_{+}/n_{p}>50$ in the tables of \citet{Cox68}). We can verify almost directly in the case $\rho_{0,7}=1$ that the results of \citet{Mazurek1973} are not accurate. This is done by using the following values; $\rho_{\textrm{CJ},7}/\mu_{e}=0.82$ ($\mu_{e}=2$) and $T_{\textrm{CJ},9}=5.81$ as given by \citet{Mazurek1973} with similar values to the entries $\eta=0$, $\beta=0$ ($T_{9}\approx5.93$) and $\rho_{m}/\mu_{e}=9.243\times10^{6}\,\textrm{g}/\textrm{cm}^{3}$ in the tables of \citet{Cox68}. There we find $p_{ep}/\varepsilon_{ep}\rho=0.3787$, which does not seem to change too much for $\mysim10\%$ changes in $T$ and $\rho$. Comparing this to the M$73$ setup value ($\myapprox0.38$) and the M$73$ value ($\myapprox0.46$) suggests that the electron--positron energy terms are not accurately calculated by \citet{Mazurek1973}.

We also compare the results obtained with the M$73$ setup to the calculation of the same initial conditions but with our default input physics (the row `Default' in Table~\ref{tbl:M73_Helium}). The $q_{01,\textrm{CJ}}$ values for the M$73$ setup deviate from the default input physics value at high densities by up to $\myapprox7\%$. The Coulomb correction term for the NSE reduces the deviation to less than $4\%$, and the Coulomb correction term for the EOS reduces it further to below $3\%$. 

\begin{table*}
\begin{minipage}{150mm}
\caption{Parameters of CJ detonations for He and upstream temperature of $T_{0,9}=0.05$ for a few upstream densities. For each upstream density, we present the results of \citet[][M$73$]{Mazurek1973}, the results of our calculations with the M$73$ setup (M$73$ setup), M$73$ setup with the addition of the Coulomb correction term to the NSE (M$73$ setup + Coul. NSE) and with the further addition of the Coulomb correction term to the EOS (M$73$ setup + Coul. NSE + Coul. EOS). The upper rows for each upstream density are the results obtained with our default input physics.}
\begin{tabular}{|c||c||c||c||c||c||c|}
\hline
$\rho_{0}$  & \textrm{Case} &$P_{\textrm{CJ}}/P_{0}$ & $ \rho_{\textrm{CJ}}/\rho_{0} $  & $T_{\textrm{CJ}}$ &$q_{01,\textrm{CJ}}$ & $D_{\textrm{CJ}}$\\ 
$[\textrm{g}/\textrm{cm}^{3}]$  &  & &  & $[10^{9}\,\textrm{K}]$ & $[10^{17}\,\textrm{erg}/\textrm{g}]$ & $[10^{4}\,\textrm{km}/\textrm{s}]$  \\ \hline
$1\times10^{6}$	&	$\textrm{Default}$	&	$37.786$	&	$1.7385$	&	$3.651$	&	$14.998$	&	$1.5339$	\\
	&	$\textrm{M}73\,\textrm{setup}+\textrm{Coul. NSE}+\textrm{Coul. EOS}$	&	$37.780$	&	$1.7381$	&	$3.651$	&	$14.998$	&	$1.5340$	\\
	&	$\textrm{M}73\,\textrm{setup}+\textrm{Coul. NSE}$	&	$37.476$	&	$1.7379$	&	$3.649$	&	$14.999$	&	$1.5344$	\\
	&	$\textrm{M}73\,\textrm{setup}$	&	$37.459$	&	$1.7376$	&	$3.649$	&	$14.994$	&	$1.5342$	\\
	&	$\textrm{M}73$	&	$39.00$	&	$1.56$	&	$3.69$	&	$15.01$	&	$1.70$	\\
\hline$5\times10^{6}$	&	$\textrm{Default}$	&	$17.191$	&	$1.8145$	&	$5.193$	&	$12.671$	&	$1.4936$	\\
	&	$\textrm{M}73\,\textrm{setup}+\textrm{Coul. NSE}+\textrm{Coul. EOS}$	&	$17.200$	&	$1.8120$	&	$5.194$	&	$12.664$	&	$1.4953$	\\
	&	$\textrm{M}73\,\textrm{setup}+\textrm{Coul. NSE}$	&	$17.090$	&	$1.8115$	&	$5.193$	&	$12.667$	&	$1.4958$	\\
	&	$\textrm{M}73\,\textrm{setup}$	&	$17.074$	&	$1.8156$	&	$5.190$	&	$12.596$	&	$1.4931$	\\
	&	$\textrm{M}73$	&	$16.90$	&	$1.64$	&	$5.22$	&	$12.95$	&	$1.59$	\\
\hline$1\times10^{7}$	&	$\textrm{Default}$	&	$11.979$	&	$1.8387$	&	$5.857$	&	$10.639$	&	$1.4310$	\\
	&	$\textrm{M}73\,\textrm{setup}+\textrm{Coul. NSE}+\textrm{Coul. EOS}$	&	$11.987$	&	$1.8346$	&	$5.860$	&	$10.616$	&	$1.4335$	\\
	&	$\textrm{M}73\,\textrm{setup}+\textrm{Coul. NSE}$	&	$11.935$	&	$1.8367$	&	$5.861$	&	$10.609$	&	$1.4340$	\\
	&	$\textrm{M}73\,\textrm{setup}$	&	$11.898$	&	$1.8415$	&	$5.849$	&	$10.499$	&	$1.4293$	\\
	&	$\textrm{M}73$	&	$11.10$	&	$1.64$	&	$5.81$	&	$10.84$	&	$1.49$	\\
\hline$5\times10^{7}$	&	$\textrm{Default}$	&	$5.423$	&	$1.7101$	&	$7.135$	&	$7.095$	&	$1.3270$	\\
	&	$\textrm{M}73\,\textrm{setup}+\textrm{Coul. NSE}+\textrm{Coul. EOS}$	&	$5.439$	&	$1.7095$	&	$7.145$	&	$7.024$	&	$1.3298$	\\
	&	$\textrm{M}73\,\textrm{setup}+\textrm{Coul. NSE}$	&	$5.414$	&	$1.7097$	&	$7.146$	&	$7.011$	&	$1.3303$	\\
	&	$\textrm{M}73\,\textrm{setup}$	&	$5.365$	&	$1.7084$	&	$7.096$	&	$6.893$	&	$1.3235$	\\
	&	$\textrm{M}73$	&	$5.16$	&	$1.64$	&	$7.07$	&	$6.95$	&	$1.33$	\\
\hline$1\times10^{8}$	&	$\textrm{Default}$	&	$4.206$	&	$1.6394$	&	$7.661$	&	$6.327$	&	$1.3287$	\\
	&	$\textrm{M}73\,\textrm{setup}+\textrm{Coul. NSE}+\textrm{Coul. EOS}$	&	$4.221$	&	$1.6394$	&	$7.675$	&	$6.241$	&	$1.3317$	\\
	&	$\textrm{M}73\,\textrm{setup}+\textrm{Coul. NSE}$	&	$4.198$	&	$1.6381$	&	$7.677$	&	$6.226$	&	$1.3321$	\\
	&	$\textrm{M}73\,\textrm{setup}$	&	$4.162$	&	$1.6374$	&	$7.604$	&	$6.104$	&	$1.3250$	\\
	&	$\textrm{M}73$	&	$4.16$	&	$1.63$	&	$7.60$	&	$6.07$	&	$1.33$	\\
\hline$5\times10^{8}$	&	$\textrm{Default}$	&	$2.788$	&	$1.5095$	&	$9.017$	&	$5.478$	&	$1.4210$	\\
	&	$\textrm{M}73\,\textrm{setup}+\textrm{Coul. NSE}+\textrm{Coul. EOS}$	&	$2.802$	&	$1.5111$	&	$9.044$	&	$5.344$	&	$1.4252$	\\
	&	$\textrm{M}73\,\textrm{setup}+\textrm{Coul. NSE}$	&	$2.788$	&	$1.5096$	&	$9.048$	&	$5.314$	&	$1.4257$	\\
	&	$\textrm{M}73\,\textrm{setup}$	&	$2.761$	&	$1.5069$	&	$8.899$	&	$5.177$	&	$1.4171$	\\
	&	$\textrm{M}73$	&	$2.75$	&	$1.50$	&	$8.87$	&	$5.16$	&	$1.42$	\\
\hline$1\times10^{9}$	&	$\textrm{Default}$	&	$2.459$	&	$1.4652$	&	$9.698$	&	$5.342$	&	$1.4913$	\\
	&	$\textrm{M}73\,\textrm{setup}+\textrm{Coul. NSE}+\textrm{Coul. EOS}$	&	$2.471$	&	$1.4663$	&	$9.732$	&	$5.179$	&	$1.4964$	\\
	&	$\textrm{M}73\,\textrm{setup}+\textrm{Coul. NSE}$	&	$2.460$	&	$1.4652$	&	$9.739$	&	$5.140$	&	$1.4969$	\\
	&	$\textrm{M}73\,\textrm{setup}$	&	$2.435$	&	$1.4621$	&	$9.542$	&	$4.988$	&	$1.4875$	\\
	&	$\textrm{M}73$	&	$2.42$	&	$1.46$	&	$9.50$	&	$4.98$	&	$1.49$	\\
\hline
\end{tabular}
\centering
\label{tbl:M73_Helium}
\end{minipage}
\end{table*}

\begin{table*}
\begin{minipage}{120mm}
\caption{The pressure, $q_{01,\textrm{CJ}}$, the electron--positron pressure and the electron--positron energy at the NSE state for the values of $\rho_{\textrm{CJ}}$ and $T_{\textrm{CJ}}$ as given by \citet{Mazurek1973}. In order to calculate the electron--positron terms for M$73$, we assume that the CJ conditions hold and we use the analytical terms for the radiation and the ions (with M$73$ setup values for $\bar{A}$). For each case, we present the results of \citet[][M$73$]{Mazurek1973}, and the results of our calculations with the M$73$ setup. We also present the positron-to-proton ratio, $n_{+}/n_{p}$, as calculated for the M$73$ setup.}
\begin{tabular}{|c||c||c||c||c||c||c|}
\hline
$\rho_{0}$  & \textrm{Case} &$P_{\textrm{CJ}}/P_{0}^{\textrm{M}73}$ & $q_{01,\textrm{CJ}}$ & $p_{ep,\textrm{CJ}}$ & $\varepsilon_{ep,\textrm{CJ}}$ & $\left(n_{+}/n_{p}\right)_{\textrm{CJ}}$\\ 
$[\textrm{g}/\textrm{cm}^{3}]$  &  & & $[10^{17}\,\textrm{erg}/\textrm{g}]$ & $[\textrm{MeV}/m_{p}]$  & $[\textrm{MeV}/m_{p}]$ & \\ \hline
$1\times10^{6}$	&	$\textrm{M}73\,\textrm{setup}$	&	$38.67$	&	$14.96$	&	$0.39$	&	$1.21$	&	$6.82\times10^{-1}$	\\
	&	$\textrm{M}73$	&	$39.00$	&	$15.01$	&	$0.40$	&	$0.87$	\\
$5\times10^{6}$	&	$\textrm{M}73\,\textrm{setup}$	&	$17.00$	&	$12.33$	&	$0.41$	&	$1.13$	&	$3.56\times10^{-1}$	\\
	&	$\textrm{M}73$	&	$16.90$	&	$12.95$	&	$0.40$	&	$0.94$	\\
$1\times10^{7}$	&	$\textrm{M}73\,\textrm{setup}$	&	$11.13$	&	$10.45$	&	$0.38$	&	$1.00$	&	$2.02\times10^{-1}$	\\
	&	$\textrm{M}73$	&	$11.10$	&	$10.84$	&	$0.38$	&	$0.90$	\\
$5\times10^{7}$	&	$\textrm{M}73\,\textrm{setup}$	&	$5.16$	&	$6.88$	&	$0.38$	&	$0.95$	&	$2.90\times10^{-2}$	\\
	&	$\textrm{M}73$	&	$5.16$	&	$6.95$	&	$0.38$	&	$0.93$	\\
$1\times10^{8}$	&	$\textrm{M}73\,\textrm{setup}$	&	$4.15$	&	$6.08$	&	$0.42$	&	$1.06$	&	$1.01\times10^{-2}$	\\
	&	$\textrm{M}73$	&	$4.16$	&	$6.07$	&	$0.42$	&	$1.05$	\\
$5\times10^{8}$	&	$\textrm{M}73\,\textrm{setup}$	&	$2.74$	&	$5.26$	&	$0.58$	&	$1.53$	&	$6.81\times10^{-4}$	\\
	&	$\textrm{M}73$	&	$2.75$	&	$5.16$	&	$0.58$	&	$1.53$	\\
$1\times10^{9}$	&	$\textrm{M}73\,\textrm{setup}$	&	$2.42$	&	$5.12$	&	$0.69$	&	$1.85$	&	$1.79\times10^{-4}$	\\
	&	$\textrm{M}73$	&	$2.42$	&	$4.98$	&	$0.69$	&	$1.86$	\\
\hline
\end{tabular}
\centering
\label{tbl:M73EOS_Helium}
\end{minipage}
\end{table*}

\subsubsection{Comparing He CJ detonations to \citet{Khokhlov88}}
\label{sec:Khokhlov88_He}

\citet{Khokhlov88} calculated CJ detonations for He, an upstream temperature of $T_{0}=0.1$ and a few values of the upstream density in the range of $[10^{6},10^{9}]\,\textrm{g}/\textrm{cm}^{3}$. We calculated the CJ NSE states for the same initial conditions by following the input physics of \citet{Khokhlov88}. The results of our calculations with the same input physics of \citet{Khokhlov88} are compared to the results of \citet{Khokhlov88} in Table~\ref{tbl:K88_Helium} (compare rows `K$88$ setup' to rows `K$88$'). Large deviations are obtained (up to $15\,\%$ in $q_{01,\textrm{CJ}}$, for example). We showed in Section~\ref{sec:Khokhlov88} that the EOS used by \citet{Khokhlov88} apparently contains a numerical bugs, to which we attribute the differences between the results. We verified this again by calculating the pressure and $q_{01,\textrm{CJ}}$ at the NSE state for the values of $\rho_{\textrm{CJ}}$ and $T_{\textrm{CJ}}$ as given by \citet{Khokhlov88}. The results of our calculations with the same input physics of \citet{Khokhlov88} are compared to the results of \citet{Khokhlov88} in Table~\ref{tbl:K88EOS_Helium} (compare rows `K$88$ setup' to rows `K$88$'). The values of $q_{01,\textrm{CJ}}$ deviate by less than $\myapprox1\%$, which suggests that the compositions of the NSE states are similar. However, the deviation in the pressure levels are large for low densities and reach $\myapprox37\%$ for $\rho_{0,7}=0.1$. Once again, the difference between the pressure levels is almost exactly the radiation pressure (compare rows `K$88$ setup + twice $p_{\textrm{rad}}$' to rows `K$88$'). 

We also compare the results obtained with the input physics of \citet{Khokhlov88} to the calculation of the same initial conditions but with our default input physics (the row `Default' in Table~\ref{tbl:K88_Helium}). The $q_{01,\textrm{CJ}}$ values for the input physics of \citet{Khokhlov88} deviate from the default input physics value by up to $\myapprox3\%$. The Coulomb correction term for the NSE reduces the deviation to below $2.5\times10^{-3}$. 

\begin{table*}
\begin{minipage}{150mm}
\caption{Parameters of CJ detonations for He and an upstream temperature of $T_{0,9}=0.1$ for a few upstream densities. For each upstream density, we present the results of \citet[][K$88$]{Khokhlov88}, the results of our calculations with the same input physics of \citet[][K$88$ setup]{Khokhlov88}, and K$88$ setup with the addition of the Coulomb correction term to the NSE (K$88$ setup + Coul. NSE). The upper rows for each upstream density are the results obtained with our default input physics.}
\begin{tabular}{|c||c||c||c||c||c||c||c|}
\hline
$\rho_{0}$  & \textrm{Case} &$P_{\textrm{CJ}}/P_{0}$ & $ \rho_{0}/\rho_{\textrm{CJ}} $  & $T_{\textrm{CJ}}$ &$q_{01,\textrm{CJ}}$ & $\gamma^{e}_{\textrm{CJ}}$ & $D_{\textrm{CJ}}$\\ 
$[\textrm{g}/\textrm{cm}^{3}]$  &  & &  & $[10^{9}\,\textrm{K}]$ & $[10^{17}\,\textrm{erg}/\textrm{g}]$ &  & $[10^{4}\,\textrm{km}/\textrm{s}]$  \\ \hline
$1\times10^{6}$	&	$\textrm{Default}$	&	$35.989$	&	$0.5759$	&	$3.651$	&	$14.997$	&	$1.3214$	&	$1.5340$	\\
     	&	$\textrm{K}88\,\textrm{setup}+\textrm{Coul. NSE}$	&	$35.996$	&	$0.5758$	&	$3.651$	&	$14.997$	&	$1.3214$	&	$1.5340$	\\
     	&	$\textrm{K}88\,\textrm{setup}$	&	$36.005$	&	$0.5756$	&	$3.652$	&	$14.993$	&	$1.3208$	&	$1.5338$	\\
     	&	$\textrm{K}88$	&	$36.80$	&	$0.58$	&	$3.33$	&	$15.10$	&	$1.31$	&	$1.56$	\\
     \hline$3\times10^{6}$	&	$\textrm{Default}$	&	$21.671$	&	$0.5606$	&	$4.682$	&	$13.858$	&	$1.2204$	&	$1.5251$	\\
     	&	$\textrm{K}88\,\textrm{setup}+\textrm{Coul. NSE}$	&	$21.659$	&	$0.5609$	&	$4.682$	&	$13.860$	&	$1.2204$	&	$1.5251$	\\
     	&	$\textrm{K}88\,\textrm{setup}$	&	$21.658$	&	$0.5600$	&	$4.681$	&	$13.818$	&	$1.2175$	&	$1.5235$	\\
     	&	$\textrm{K}88$	&	$22.60$	&	$0.57$	&	$4.34$	&	$14.50$	&	$1.26$	&	$1.57$	\\
     \hline$1\times10^{7}$	&	$\textrm{Default}$	&	$11.815$	&	$0.5443$	&	$5.857$	&	$10.633$	&	$1.0942$	&	$1.4308$	\\
     	&	$\textrm{K}88\,\textrm{setup}+\textrm{Coul. NSE}$	&	$11.819$	&	$0.5441$	&	$5.858$	&	$10.632$	&	$1.0944$	&	$1.4308$	\\
     	&	$\textrm{K}88\,\textrm{setup}$	&	$11.769$	&	$0.5433$	&	$5.844$	&	$10.527$	&	$1.0906$	&	$1.4262$	\\
     	&	$\textrm{K}88$	&	$12.80$	&	$0.55$	&	$5.56$	&	$12.00$	&	$1.14$	&	$1.50$	\\
     \hline$3\times10^{7}$	&	$\textrm{Default}$	&	$6.735$	&	$0.5668$	&	$6.751$	&	$7.906$	&	$1.1154$	&	$1.3438$	\\
     	&	$\textrm{K}88\,\textrm{setup}+\textrm{Coul. NSE}$	&	$6.733$	&	$0.5671$	&	$6.752$	&	$7.913$	&	$1.1162$	&	$1.3441$	\\
     	&	$\textrm{K}88\,\textrm{setup}$	&	$6.682$	&	$0.5670$	&	$6.716$	&	$7.791$	&	$1.1148$	&	$1.3378$	\\
     	&	$\textrm{K}88$	&	$7.34$	&	$0.57$	&	$6.53$	&	$9.09$	&	$1.11$	&	$1.41$	\\
     \hline$1\times10^{8}$	&	$\textrm{Default}$	&	$4.182$	&	$0.6105$	&	$7.663$	&	$6.314$	&	$1.1933$	&	$1.3281$	\\
     	&	$\textrm{K}88\,\textrm{setup}+\textrm{Coul. NSE}$	&	$4.184$	&	$0.6106$	&	$7.668$	&	$6.325$	&	$1.1943$	&	$1.3287$	\\
     	&	$\textrm{K}88\,\textrm{setup}$	&	$4.146$	&	$0.6112$	&	$7.596$	&	$6.200$	&	$1.1942$	&	$1.3217$	\\
     	&	$\textrm{K}88$	&	$4.47$	&	$0.60$	&	$7.50$	&	$6.95$	&	$1.17$	&	$1.37$	\\
     \hline$3\times10^{8}$	&	$\textrm{Default}$	&	$3.097$	&	$0.6473$	&	$8.560$	&	$5.635$	&	$1.2433$	&	$1.3795$	\\
     	&	$\textrm{K}88\,\textrm{setup}+\textrm{Coul. NSE}$	&	$3.099$	&	$0.6475$	&	$8.571$	&	$5.649$	&	$1.2438$	&	$1.3805$	\\
     	&	$\textrm{K}88\,\textrm{setup}$	&	$3.071$	&	$0.6482$	&	$8.451$	&	$5.510$	&	$1.2442$	&	$1.3727$	\\
     	&	$\textrm{K}88$	&	$3.15$	&	$0.65$	&	$8.37$	&	$5.93$	&	$1.23$	&	$1.40$	\\
     \hline$1\times10^{9}$	&	$\textrm{Default}$	&	$2.451$	&	$0.6831$	&	$9.700$	&	$5.326$	&	$1.2761$	&	$1.4905$	\\
     	&	$\textrm{K}88\,\textrm{setup}+\textrm{Coul. NSE}$	&	$2.454$	&	$0.6828$	&	$9.717$	&	$5.340$	&	$1.2764$	&	$1.4917$	\\
     	&	$\textrm{K}88\,\textrm{setup}$	&	$2.431$	&	$0.6842$	&	$9.522$	&	$5.171$	&	$1.2771$	&	$1.4827$	\\
     	&	$\textrm{K}88$	&	$2.48$	&	$0.68$	&	$9.50$	&	$5.28$	&	$1.27$	&	$1.50$	\\
     \hline
\end{tabular}
\centering
\label{tbl:K88_Helium}
\end{minipage}
\end{table*}

\begin{table*}
\begin{minipage}{100mm}
\caption{The pressure and $q_{01,\textrm{CJ}}$ at the NSE state for the values of $\rho_{\textrm{CJ}}$ and $T_{\textrm{CJ}}$ as given by \citet{Khokhlov88} for He. For each case, we present the results of \citet[][K$88$]{Khokhlov88}, the results of our calculations with the same input physics of \citet[][K$88$ setup]{Khokhlov88}, and the results of recalculating the pressure with twice the radiation pressure (K$88$ setup + twice $p_{\textrm{rad}}$).}
\begin{tabular}{|c||c||c||c|}
\hline
$\rho_{0}$  & \textrm{Case} &$P_{\textrm{CJ}}/P_{0}^{\textrm{K}88}$ & $q_{01,\textrm{CJ}}$\\ 
$[\textrm{g}/\textrm{cm}^{3}]$  &  & & $[10^{17}\,\textrm{erg}/\textrm{g}]$  \\ \hline
$1\times10^{6}$	&	$\textrm{K}88\,\textrm{setup}$	&	$25.31$	&	$15.09$	\\
	&	$\textrm{K}88\,\textrm{setup}+\textrm{twice}\,p_{\textrm{rad}}$	&	$36.19$	&		\\
	&	$\textrm{K}88$	&	$36.80$	&	$15.10$	\\
\hline$3\times10^{6}$	&	$\textrm{K}88\,\textrm{setup}$	&	$16.51$	&	$14.53$	\\
	&	$\textrm{K}88\,\textrm{setup}+\textrm{twice}\,p_{\textrm{rad}}$	&	$22.56$	&		\\
	&	$\textrm{K}88$	&	$22.60$	&	$14.50$	\\
\hline$1\times10^{7}$	&	$\textrm{K}88\,\textrm{setup}$	&	$10.02$	&	$12.00$	\\
	&	$\textrm{K}88\,\textrm{setup}+\textrm{twice}\,p_{\textrm{rad}}$	&	$12.81$	&		\\
	&	$\textrm{K}88$	&	$12.80$	&	$12.00$	\\
\hline$3\times10^{7}$	&	$\textrm{K}88\,\textrm{setup}$	&	$6.20$	&	$9.05$	\\
	&	$\textrm{K}88\,\textrm{setup}+\textrm{twice}\,p_{\textrm{rad}}$	&	$7.32$	&		\\
	&	$\textrm{K}88$	&	$7.34$	&	$9.09$	\\
\hline$1\times10^{8}$	&	$\textrm{K}88\,\textrm{setup}$	&	$4.11$	&	$6.88$	\\
	&	$\textrm{K}88\,\textrm{setup}+\textrm{twice}\,p_{\textrm{rad}}$	&	$4.48$	&		\\
	&	$\textrm{K}88$	&	$4.47$	&	$6.95$	\\
\hline$3\times10^{8}$	&	$\textrm{K}88\,\textrm{setup}$	&	$3.02$	&	$5.94$	\\
	&	$\textrm{K}88\,\textrm{setup}+\textrm{twice}\,p_{\textrm{rad}}$	&	$3.14$	&		\\
	&	$\textrm{K}88$	&	$3.15$	&	$5.93$	\\
\hline$1\times10^{9}$	&	$\textrm{K}88\,\textrm{setup}$	&	$2.44$	&	$5.30$	\\
	&	$\textrm{K}88\,\textrm{setup}+\textrm{twice}\,p_{\textrm{rad}}$	&	$2.48$	&		\\
	&	$\textrm{K}88$	&	$2.48$	&	$5.28$	\\
\hline
\end{tabular}
\centering
\label{tbl:K88EOS_Helium}
\end{minipage}
\end{table*}

It is interesting to note that \citet{Townsley2012} calculated the $D_{\textrm{CJ}}$ for He, $\rho_{0,7}=0.5$, $T_{0,9}=0.2$, by using the \textit{Helmholtz} EOS and the $13$ $\alpha$-element network. They claim that their value, $1.54\times10^{4}\,\textrm{km}/\textrm{s}$, is consistent with the results of \citet{Khokhlov88}, as they interpolate between the entries $\rho_{0,7}=0.3$ and $\rho_{0,7}=1$ of Table~\ref{tbl:K88_Helium}. We verified that this is in fact a coincidence, because the apparent numerical bug in the EOS of \citet{Khokhlov88} compensates for the difference between the input physics of \citet{Townsley2012} and \citet{Khokhlov88}. 


\section{The structure of the detonation wave}
\label{sec:structure}

In this section, we present our calculation of the structure of the (possibly pathological) detonation waves. For a given detonation speed, in which the final state is NSE (and the solution does not cross the sonic point), the end state is known in advance, and is independent of the reaction rates. We use this fact to monitor the numerical accuracy of the integration. Another useful method is to monitor the energy conservation during the integration, which allows us to estimate that our numerical accuracy is better than $10^{-3}$. The numerical integration is performed with a fourth-order implicit Rosenbrock method (option \textsc{rodas4\_solver} of {\sc MESA}) with the parameters $\textrm{rtol}=10^{-7}$ (relative error tolerance) and $\textrm{atol}=10^{-8}$ (absolute error tolerance). In Section~\ref{sec:CO structure}, we consider the initial composition of CO, and in Section~\ref{sec:He structure}, we consider the initial composition of He.  

\subsection{The structure of the detonation wave in CO}
\label{sec:CO structure}

In this section, we present the structure of the detonation wave in CO. In Section~\ref{sec:CO example}, we present an example of the structure of a detonation wave for some specific initial conditions. In Sections~\ref{sec:CO scan D} and~\ref{sec:CO scan scale}, we calculate the pathological detonation speed, $D_{*}$, and the structure of the detonation wave, respectively, as a function of the upstream density. We comment on the uncertainty of the results in Section~\ref{sec:CO uncertain}. Finally, we compare out results to \citet{Khokhlov89} and \citet{Townsley2016} in Sections~\ref{sec:Khokhlov89} and~\ref{sec:Townsley2016}, respectively. 

\subsubsection{An example for CO: $\rho_{0,7}=1$ and $T_{0,9}=0.2$}
\label{sec:CO example}

\begin{figure}
\includegraphics[width=0.45\textwidth]{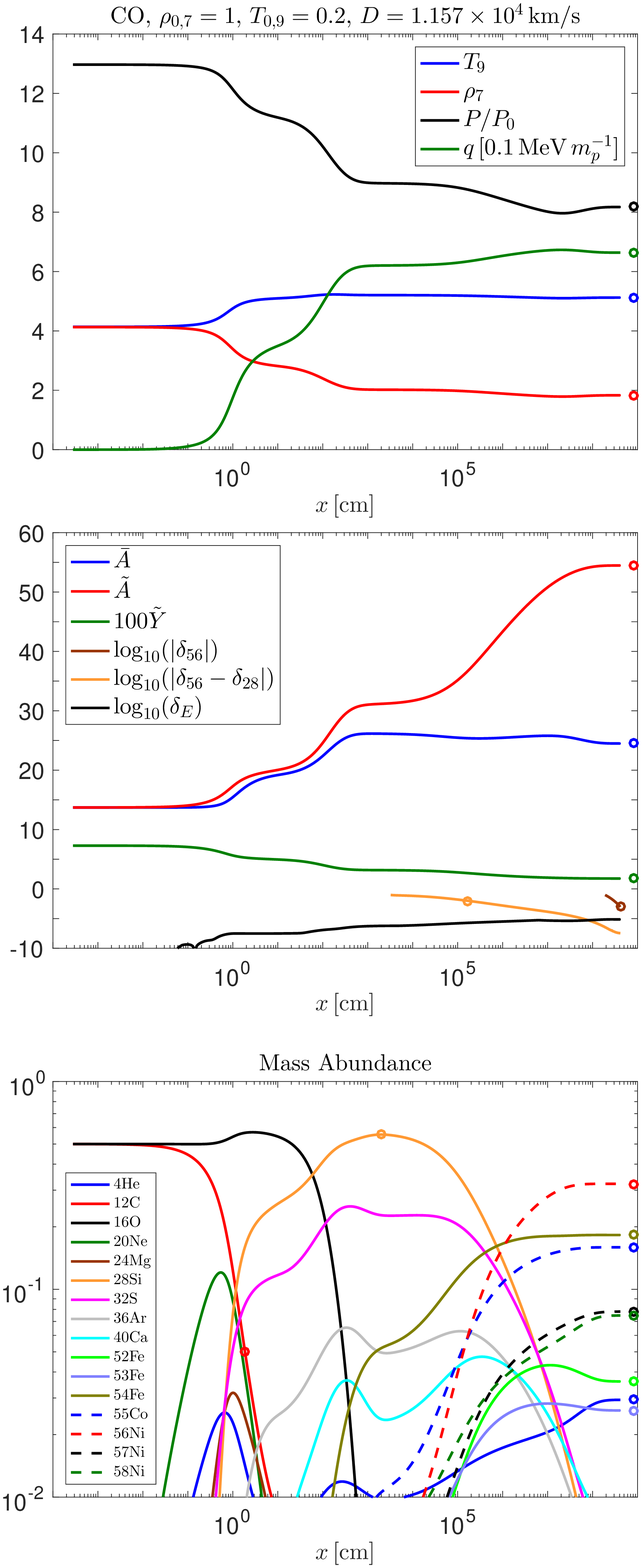}
\caption{The structure of an overdriven detonation wave as a function of the distance behind the shock. Upper panel: temperature (blue), density (red), pressure (black) and thermonuclear energy release (green). Middle panel: $\bar{A}$ (blue), $\tilde{A}$ (red), $\tilde{Y}$ (green), $\delta_{56}$ (see the text, brown), $\delta_{56}(x)-\delta_{28}(x)$ (which monitors the NSQE state, orange) and $\delta_{E}$ (which monitors energy conservation, black). The orange point marks the location where $|\delta_{56}-\delta_{28}|=10^{-2}$, and the brown point marks the location where $|\delta_{56}|=10^{-3}$. Bottom panel: mass fractions of a few key isotopes. The red (orange) point marks the location where the mass fraction of $^{12}$C ($^{28}$Si) reaches $0.05$ (is maximal). The points at the right edges of the panels represent the NSE values. At a distance of $x\approx2.0\times10^{7}\,\textrm{cm}$, the heat release becomes endothermic. This is connected with the minimum of the density at $x\approx2.2\times10^{7}\,\textrm{cm}$.
\label{fig:CO_1e7_2e8}}
\end{figure}

We first present in Figure~\ref{fig:CO_1e7_2e8}, as an example, the structure of a detonation wave as a function of the distance behind the shock, $x$, for CO, $\rho_{0,7}=1$, $T_{0,9}=0.2$ and a detonation speed of $D=1.157\times10^{4}\,\textrm{km}/\textrm{s}$ ($>D_{*}\approx1.1560\times10^{4}\,\textrm{km}/\textrm{s}$  for these upstream conditions, see below). Following some induction time, the $^{12}$C is consumed and its mass fraction reaches $0.05$ at $x\approx1.9\,\textrm{cm}$ (red point in the lower panel), where $\myapprox0.26\,\textrm{MeV}/m_{p}$ are released. This is followed by $^{16}$O burning, which synthesizes heavier elements, most notably $^{28}$Si. It is convenient to mark the end of this process as the state in which the mass fraction of $^{28}$Si is maximal ($x\approx2.1\times10^{3}\,\textrm{cm}$, orange point in the lower panel). This burning releases additional $\myapprox0.36\,\textrm{MeV}/m_{p}$. As the carbon and oxygen continue to burn, the number of heavy nuclei decreases ($\tilde{Y}$ decreases), while the average mass number $\bar{A}$ increases. During this stage only a minute amount of $^{4}$He is synthesized, such that $\bar{A}\lesssim\tilde{A}\approx30$, as $^{28}$Si is maximal. 

At this stage, the material is in a state of NSQE. Following the approach of \citet{Khokhlov89}, we monitor this by calculating $\delta_{56}(x)-\delta_{28}(x)$, where\footnote{Note that there is probably a typo in the definition of $\delta_{i}$ in \citet{Khokhlov89}.}
\begin{eqnarray}\label{eq:delta def}
\delta_{i}(x)=\ln\left(X_{i}(x)/X_{i}^{*}(x)\right),
\end{eqnarray}
$X_{i}^{*}(x)$ is calculated according to Eq.~\eqref{eq:NSE2} with $\rho(x)$, $T(x)$, $X_{n}(x)$, $X_{p}(x)$; to simplify the notation, we used $i=28,56$ for $^{28}$Si, $^{56}$Ni, respectively. The middle panel shows that $|\delta_{56}(x)-\delta_{28}(x)|=0.1$ slightly after the point in time when the mass fraction of $^{28}$Si is at a maximum, and it decreases as the solution approaches NSE ($|\delta_{56}(x)-\delta_{28}(x)|=0.01$ at $x\approx1.7\times10^{5}\,\textrm{cm}$, orange point). The middle panel shows that $\tilde{Y}$ slowly decreases towards the NSE value, and we verified that the decrease is controlled by the inverse triple-$\alpha$ reaction, $^{12}$C$\rightarrow3^{4}$He. During this slow burning, not much energy is released, with the heavy elements approaching $\tilde{A}\approx55$, while a significant amount of $^{4}$He is synthesized, leaving $\bar{A}\approx25$. 

The approach to NSE is monitored with $\delta_{56}(x)$. The middle panel shows that $|\delta_{56}|=0.1$ at $x\approx2\times10^{8}\,\textrm{cm}$. From that position, $|\delta_{56}|$ decreases exponentially with an $e$-folding distance of $l_{\textrm{56}}\approx5.5\times10^{7}\,\textrm{cm}$. The brown point marks the location where $|\delta_{56}|=10^{-3}$. We stop the integration when $\delta_{\max}=10^{-3}$, where
\begin{eqnarray}\label{eq:delta max}
\delta_{\max}=\max_{i}\left(|\delta_{i}|\right),
\end{eqnarray}
and we do not go over isotopes with either an $X_{i}<10^{-20}$ or an $X_{i}^{*}<10^{-20}$. It should be realized that the NSE state is only approached asymptotically at infinity, and there is no finite position in which the NSE state is obtained. The deviation of the solution parameters at the end of the integration from the NSE values (points at the right edges of the panels), which are calculated only from the conservation laws, is smaller than $10^{-4}$. This demonstrates the high accuracy of our integration. 

We mentioned in Section~\ref{sec:net} that $^{10}$C is not included in the isotope list NSE$7$Si. This isotope approaches its NSE value through the slow reaction $^{10}$C$(\alpha,n)^{13}$O. While this has a negligible effect on the solution, we would have to integrate it over long time periods in order to make sure that $\delta_{\max}=10^{-3}$. We, therefore, exclude this isotope. This example demonstrates that the distance needed to reach some prescribed deviation from the NSE state is sensitive to the list of isotopes. This is the reason why we monitor the approach to NSE with $\delta_{56}$, which is much less sensitive to the isotope list. 

Energy conservation during the integration is monitored by the parameter $\delta_{E}$, which is the deviation of the conserved quantity $\varepsilon-q+P/\rho+u^{2}/2$ (Bernoulli's law) from its initial value\footnote{We thank Dean Townsley for pointing us to this method.}. The middle panel shows that the value of $\delta_{E}$  increases towards the NSE and is smaller than $10^{-5}$ at the end of the integration. The loss of accuracy is caused by the detailed balance of fast reactions. The time derivative of the mass fraction of each isotope is a sum over all the reactions that involve that isotope. This sum is actually of the difference of forward and backward reactions, which should be equal at a detailed balance state. Consider such a difference between two fast reactions as the solution approaches a detailed balance. The accuracy in which this difference is calculated decreases, since it is the difference between two large numbers with many identical significant digits. For most cases, we are able to maintain a high enough numerical accuracy ($\delta_{E}<10^{-3}$) up to the time when $\delta_{\max}=10^{-3}$. This is enough to fully describe the approach to NSE, since at this stage all the solution parameters are approaching their NSE values exponentially, at an $e$-folding distance of $l_{\textrm{NSE}}$. However, for a few cases we were unable to maintain the high accuracy up to the time when $\delta_{\max}=10^{-3}$. It may be possible to find a specialized algorithm to calculate accurately the approach to NSE, but this is outside the scope of this paper \citep[performing the sums in extended precision is a possibility, see][]{Paxton2015}. 

At a distance of $x\approx2.0\times10^{7}\,\textrm{cm}$, the heat release becomes endothermic. This is connected with the minimum of the density ($\phi=0$ in Equations~\eqref{eq:ZND t}) at $x\approx2.2\times10^{7}\,\textrm{cm}$ and with the fact that the detonation speed of this solution is slightly above $D_{*}$. For a detonation speed that equals $D_{*}$, the position of the point where $\phi=0$ coincides with the sonic point ($u=c_{s}$). We numerically determine $D_{*}$ as the detonation speed for which integration with $D<D_{*}$ hits the sonic point, $|u^{2}-c_{s}^{2}|/u^{2}<10^{-3}$, while integration with $D>D_{*}$ reaches $\delta_{\max}=10^{-3}$. We estimate that these choices limit the numerical accuracy in the determination of $D_{*}$ to $\mysim10^{-3}$, since the order of magnitude of error in all terms of Equations~\eqref{eq:ZND t} should be similar. The sonic point location was determined as the sonic point of the integration with the highest detonation speed that is still smaller than $D_{*}$. However, because of the rapid change of the sonic point location as $D$ approaches $D_{*}$ \citep{Sharpe1999}, the numerical accuracy of the sonic point location is of the order of a few tens of percent. Other properties of the pathological detonation, which are far from the sonic point, are determined to a numerical accuracy that is similar to the numerical accuracy of  $D_{*}$ determination, i.e. $\mysim10^{-3}$.

\subsubsection{The dependence of $D_{*}$ on the upstream density}
\label{sec:CO scan D}

\begin{figure}
\includegraphics[width=0.48\textwidth]{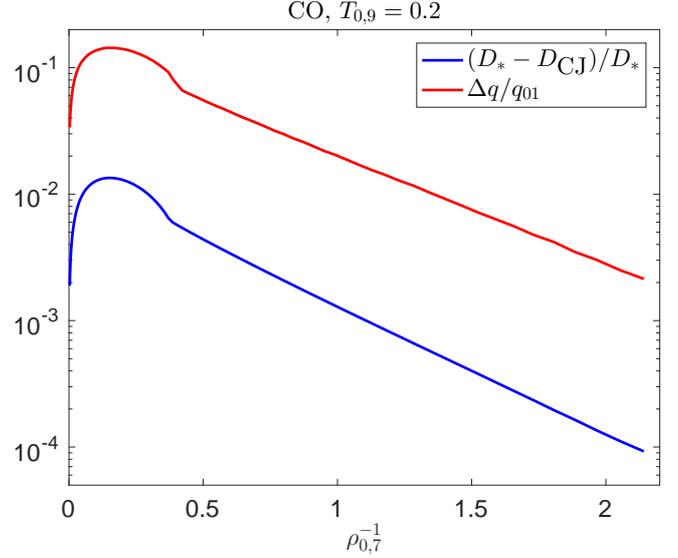}
\caption{The deviation between $D_{\textrm{CJ}}$ and $D_{*}$ (blue line) and the difference between the maximal value of $q$ along the integration and $q_{01}$ at the end of the integration, $\Delta q$ (red line), as a function of $1/\rho_{0}$. $(D_{*}-D_{\textrm{CJ}})/D_{*}$ decreases exponentially with $1/\rho_{0}$, which suggests that the detonation remains pathological even at lower upstream densities. $\Delta q/q_{01}$ decreases exponentially with $1/\rho_{0}$, which shows that higher and higher numerical accuracy is required for lower and lower upstream densities in order to determine whether a detonation is CJ based on $q(x)$ inspection. 
\label{fig:CJProof}}
\end{figure}

The calculated $D_{*}$ for CO is presented in the upper panel of Figure~\ref{fig:DetonationSpeed_CJAbu} for an upstream temperature of $T_{0,9}=0.2$. The deviation between $D_{\textrm{CJ}}$ and $D_{*}$ is always smaller than $\myapprox1.4\%$ (blue line). We are unable to integrate for densities above $\rho_{0,7}=340$ with a high enough accuracy, i.e., $\delta_{E}<10^{-3}$. Furthermore, at these high densities, the deviation between $D_{\textrm{CJ}}$ and $D_{*}$ approaches our numerical accuracy for $D_{*}$. Nevertheless, the decrease in the deviation as a function of the upstream density is smaller than exponential, which suggests that even at larger upstream densities the detonation remains pathological. At low densities, the deviation between $D_{\textrm{CJ}}$ and $D_{*}$ approaches $10^{-3}$ at $\rho_{0,7}\approx0.9$. Nevertheless, we present our results even at lower densities, $\rho_{0,7}\gtrsim0.47$, as long as we were able to integrate with high accuracy. Figure~\ref{fig:CJProof} shows that the deviation between $D_{\textrm{CJ}}$ and $D_{*}$ decreases exponentially with $1/\rho_{0}$, which suggests that the detonation remains pathological even at lower upstream densities. There could be a change in this behaviour at lower densities (maybe connected with the maximum of $D_{\textrm{CJ}}$ at $\rho_{0,7}\approx0.35$), but we are unable to find evidence for CJ detonations at low upstream densities. 

The claim that at low upstream densities, $\rho_{0,7}\lesssim1$, the detonation is CJ was made by \citet{Imshennik1984} for a pure $^{12}$C initial composition and an upstream temperature of $T_{0,9}=0.2$. Their claim is based on inspecting whether $q(x)$ monotonically increases during CJ detonations. However, it is not clear at which point they stopped the integration, and whether the accuracy of the integration is sufficient for meaningful results close to the NSE state. We find that in order to determine the position at which $q(x)$ begins to decrease, higher and higher numerical accuracy is required for lower and lower upstream densities. This is demonstrated in Figure~\ref{fig:CJProof}, which shows that the difference between the maximal value of $q$ along the integration and $q_{01}$ at the end of the integration, $\Delta q$, normalized by $q_{01}$, decreases exponentially with $1/\rho_{0}$. It is, therefore, likely that the numerical accuracy of \citet{Imshennik1984} did not reach the level needed to identify the point at which $q(x)$ begins to decrease for $\rho_{0,7}\lesssim1$. \citet{Sharpe1999} states that he finds CO detonations to be CJ-type below about $\rho_{0,7}<2$, but do not explore these densities in detail, constraining the study to higher densities at which the pathological nature is more clear. \citet{Gamezo99} claim that for CO the detonation is CJ at low densities, based on inspecting whether the flow hits a sonic point and is subsonic downstream and upstream of that point. From their demonstration of this method (bottom panel of their figure 3), it is clear that in their integration they actually did not hit the sonic point, as $u$ deviates from $c_{s}$ by $\myapprox1.5\%$. This procedure depends on numerical accuracy as well, and it seems that \citet{Gamezo99} did not have the required numerical accuracy to detect pathological detonations at low densities (compare their $1.5\%$ accuracy with the red line in Figure~\ref{fig:CJProof}). 

A few key parameters of these pathological detonations are given in Table~\ref{tbl:CO CJ} for $T_{0,9}=0.2$. The results are similar to the CJ results, demonstrating that the final CJ NSE conditions provide a good approximation of the pathological NSE conditions. 

Similarly to the CJ case, these results do not depend much on the initial upstream temperature. For $T_{0,9}=0.04$, we are able to integrate within the same range of upstream densities with high enough accuracy. Within this range, the $D_{\textrm{*}}$ values for $T_{0,9}=0.04$ deviate from the results for $T_{0,9}=0.2$ by less than $8\times10^{-4}$, and the key parameters of Table~\ref{tbl:CO CJ} deviate by less than $\myapprox0.6\%$, with the largest deviation obtained for $q_{01,*}$ at $\rho_{0,7}=300$.

The results calculated with the NSE$4$ (NSE$5$) isotope list deviate from the results presented above by less than $7.6\times10^{-3}$ ($1.2\times10^{-3}$), which suggests that our isotope list is converged to at least $\mysim10^{-3}$. For a given $D_{*}$, the uncertainty of the results is similar to the CJ case (dominated by the uncertainty of the Coulomb correction terms for the EOS and the Coulomb correction terms for the NSE state), and we estimate it to be on the sub-percent level (see detailed discussion in Section~\ref{sec:screening}). However, the values of $D_{*}$ itself depend also on the reaction rates and are influenced by uncertainties in these rates. The study of this uncertainty is beyond the scope of this paper, but because of the slight deviation ($\lesssim1.4\%$) of $D_{*}$ from $D_{\textrm{CJ}}$ (that does not depend on the reaction rates), we speculate that this uncertainty is smaller than a few percent. 

\subsubsection{The dependence of the burning scales on the upstream density for CO}
\label{sec:CO scan scale}

\begin{figure}
\includegraphics[width=0.48\textwidth]{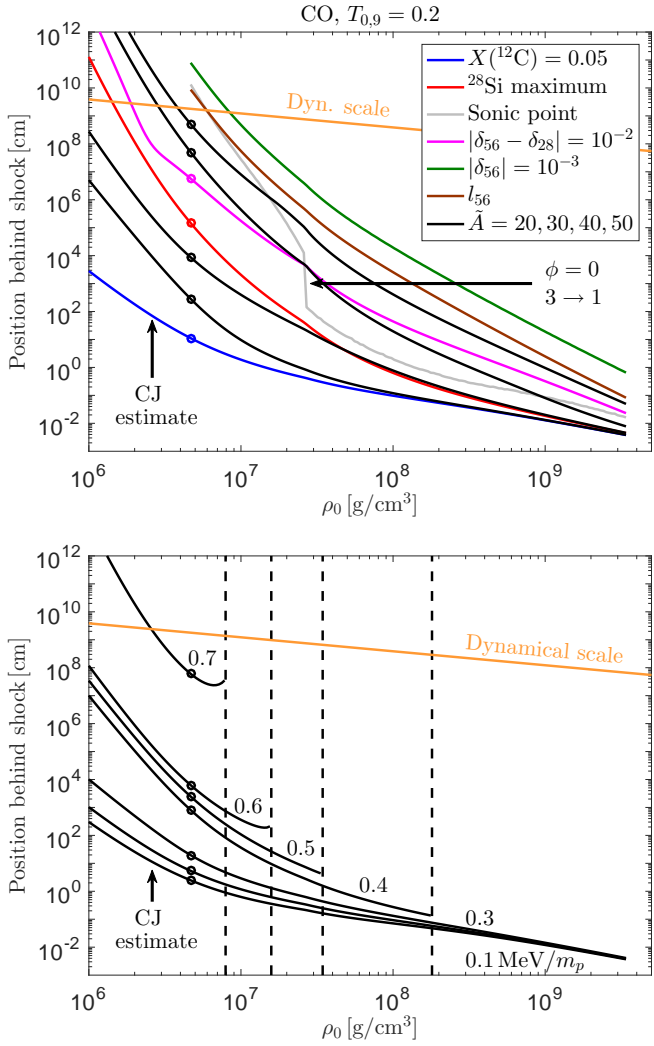}
\caption{Different scales of the CO pathological detonation in comparison with a typical dynamical scale of $v/\sqrt{G\rho_{0}}$ with $v=10^{4}\,\textrm{km}/\textrm{s}$ (orange). Top panel: the $^{12}$C consumption scale (blue), $^{28}$Si maximum (red), the location where $\tilde{A}=20,30,40$, and $50$ (bottom to top, black), the location where $|\delta_{56}-\delta_{28}|=10^{-2}$ (magenta), and the location where $|\delta_{56}|=10^{-3}$ (green) and $l_{\textrm{56}}$ (brown). Bottom panel: the locations where the energy release is $0.1,0.2,...,0.7\,\textrm{MeV}/m_{p}$ (bottom to top, black). These scales are determined from the profiles with the lowest detonation speed that is still larger than $D_{*}$ (\textit{slightly overdriven}). The sonic point location (grey, top panel) is determined from the profiles with the highest detonation speed that is still lower than $D_{*}$. For low densities, $\rho_{0,7}\lesssim0.47$ (indicated by points), where we are unable to determine $D_{*}$ we estimate the position of the scales (except the location of the sonic point, $|\delta_{56}|=10^{-3}$ and $l_{56}$) by integrating with $D=D_{CJ}$. Dashed lines in the bottom panel mark the upstream densities at which $q_{01}$ obtained at the NSE state matches the indicated energy release. A discontinuous behaviour of the sonic point location, from $x\sim100\,\textrm{cm}$ to $x\sim10^{4}\,\textrm{cm}$, is obtained around $\rho_{0,7}\approx2.7$.
\label{fig:CO_DetonationPropQAt}}
\end{figure}

In Figure~\ref{fig:CO_DetonationPropQAt}, different scales of the CO pathological detonation are compared with a typical dynamical scale of $v/\sqrt{G\rho_{0}}$ with $v=10^{4}\,\textrm{km}/\textrm{s}$. All scales, except for the sonic point location, are determined from the profiles with the lowest detonation speed that is still larger than $D_{*}$ (\textit{slightly overdriven}). For low densities, $\rho_{0,7}\lesssim0.47$, where we are unable to determine $D_{*}$ we estimate the scales by integrating with $D=D_{CJ}$. Since at these densities $D_{*}$ (if exists) probably deviates from $D_{\textrm{CJ}}$ by less than $10^{-4}$ and we are able to integrate with high accuracy up to the presented scales, our results should be an excellent estimate. The location where $|\delta_{56}|=10^{-3}$ and $l_{\textrm{56}}$ are shown as well, which allows to estimate the position of a smaller deviation from NSE. Note that many works present a finite position for NSE that does not have a clear meaning \citep{Khokhlov89,Townsley2016,Dunkley2013}, since the NSE is only obtained asymptotically at infinity. The numerical accuracy of all the scales in Figure~\ref{fig:CO_DetonationPropQAt} is $\mylesssim10^{-3}$, except for the sonic point location with a numerical accuracy of the order of a few tens of percent (see discussion above), which is also the reason for the noisy appearance of this curve. 

The ordering of the different scales as a function of the upstream density is similar to the case $\rho_{0,7}=1$, described in detail in Section~\ref{sec:CO example}. Following some induction time, the $^{12}$C is consumed and $\myapprox0.3\,\textrm{MeV}/m_{p}$ are released. This is followed by $^{16}$O burning that synthesizes heavier elements, $\tilde{A}\approx30$, roughly when the mass fraction of $^{28}$Si is maximal. Slightly later, the material is in NSQE ($|\delta_{56}(x)-\delta_{28}(x)|=0.01$), and it approaches NSE while heavier elements are synthesized with $\tilde{A}\gtrsim50$ without releasing much energy. 

\begin{figure}
\includegraphics[width=0.48\textwidth]{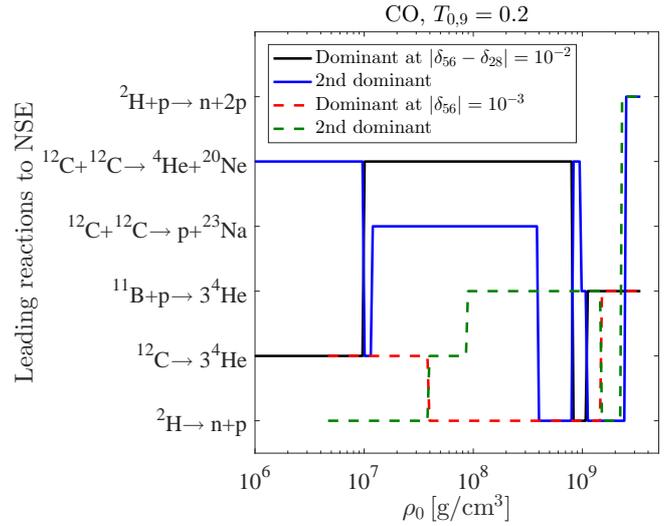}
\caption{The reactions that dominate the net change of $\tilde{Y}$ at $|\delta_{56}-\delta_{28}|=10^{-2}$ (black and blue) and at $|\delta_{56}|=10^{-3}$ (red and green), as a function of the upstream density for CO.
\label{fig:COLeadingNSE}}
\end{figure}

In order to determine which reactions control the approach to NSE (where $\tilde{Y}$ approaches its NSE value), we inspect at the location where $|\delta_{56}-\delta_{28}|=10^{-2}$ and at the location where $|\delta_{56}|=10^{-3}$ all the reactions that can change the value of $\tilde{Y}$. Of those reactions, the ones that are not in a detailed balance with their reverse reactions dominate the net change in $\tilde{Y}$, so we sort the reactions according to the absolute value of the difference between them and their reverse reactions. The reactions with the largest differences, which control the approach to NSE, are shown in Figure~\ref{fig:COLeadingNSE}. The approach to NSE at the location where $|\delta_{56}|=10^{-3}$ is controlled at low upstream densities, $\rho_{0,7}\lesssim10$, by the inverse triple-$\alpha$ reaction, $^{12}$C$\rightarrow3^{4}$He, and to some extent by $^{2}$H$\rightarrow n+p$, while at high densities, $^{2}$H$\rightarrow n+p$ is the dominant process with an a additional contribution from $^{11}$B$+p\rightarrow3^{4}$He. At very high densities, $\rho_{0,7}\gtrsim200$,  $^{11}$B$+p\rightarrow3^{4}$He and $p+^{2}$H$\rightarrow n+2p$ are dominant and comparable. Earlier in the process, where $|\delta_{56}-\delta_{28}|=10^{-2}$, the reactions $^{12}$C$+^{12}$C$\rightarrow^{4}$He$+^{20}$Ne and $^{12}$C$+^{12}$C$\rightarrow p+^{23}$Na are important as well. Except for the inverse triple-$\alpha$ reaction that was known to determine the length-scale of the detonation wave at low densities, the importance of the other reactions was not identified in the past.

The scales themselves shorten significantly as the upstream density increases, due to the increase in the post-shock temperature. Furthermore, the temperature at the NSE state increases monotonically with $\rho_{0}$, which decreases both $\bar{A}$ and $q_{01}$ at these states (see Table~\ref{tbl:CO CJ}). At large upstream densities, the released energy is not much larger than the contribution from carbon burning. Usually the detonated material will later cool and $^{4}$He will recombine to release more energy without a large change in $\tilde{A}$. The upstream densities in which some values of $q_{01}$ are obtained at the NSE state are marked with dashed lines at the bottom panel of Figure~\ref{fig:CO_DetonationPropQAt}. Note that for CJ detonations the scale at which these $q_{01}$ values are obtained should diverge as the upstream density decreases. However, since for pathological detonations the energy release is not monotonic, these $q_{01}$ values are obtained after a finite distance behind the shock wave. Figure~\ref{fig:CO_DetonationPropQAt} allows to estimate for a given upstream density and physical scale the amount of guaranteed energy release and the obtained value of $\tilde{A}$ (for example, whether iron group elements can be synthesized). 

The sonic point location is always above the locations where the mass fraction of $^{28}$Si is maximal and where $\tilde{A}=30$. This observation differs from the claims of \citet{Gamezo99}.

\begin{figure}
\includegraphics[width=0.48\textwidth]{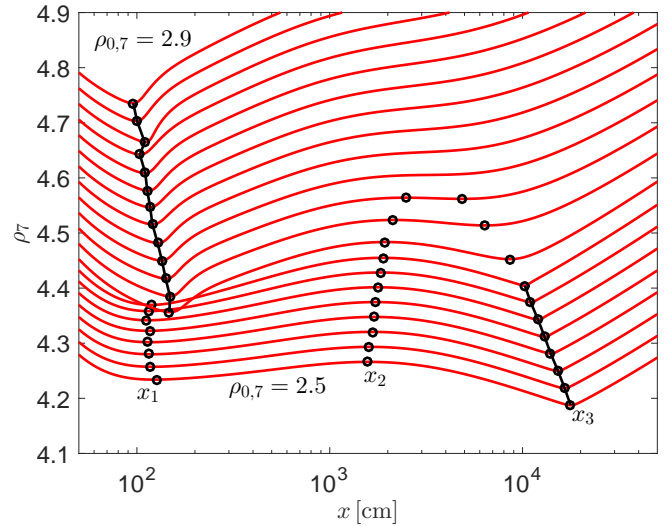}
\caption{The slightly overdriven density profiles for CO, $T_{0,9}=0.2$ and an upstream density in the range of $[2.5,2.9]\times10^{7}\,\textrm{g}/\textrm{cm}^{3}$ (the spacing between the densities used for the red lines is $2\times10^{5}\,\textrm{g}/\textrm{cm}^{3}$). For the low upstream densities, there are three locations where $\phi=0$ ($x_{1}$, $x_{2}$, and $x_{3}$, black points). Each of these points is an extremum point of the density. The sonic point location (black line) for these upstream densities is near $x_{3}\sim10^{4}\,\textrm{cm}$. As the upstream density increases, there remains only a single location where $\phi=0$, which is close to $x_{1}\sim10^{2}\,\textrm{cm}$. Around this transition, the sonic point changes location to $x_{1}$. 
\label{fig:PhiTransition}}
\end{figure}

A discontinuous behaviour of the sonic point location, from $x\sim100\,\textrm{cm}$ to $x\sim10^{4}\,\textrm{cm}$ around $\rho_{0,7}\approx2.7$, is seen in the bottom panel of Figure~\ref{fig:CO_DetonationPropQAt} (it was observed by \citet{Dunkley2013} but assumed, without investigation, to be related to the transition between CJ and pathological behavior based on the previous work of \citet{Gamezo99}. There is also a hint for this transition in figure 1 of \citet{Townsley2016}). This is also seen as a minimum of $D_{*}$ at this upstream density in the upper panel of Figure~\ref{fig:DetonationSpeed_CJAbu}. The reason for this behaviour is explained in Figure~\ref{fig:PhiTransition}, which shows the slightly overdriven density profiles for $T_{0,9}=0.2$ and an upstream density in the range of $[2.5,2.9]\times10^{7}\,\textrm{g}/\textrm{cm}^{3}$. For the low upstream densities, there are three locations where $\phi=0$ ($x_{1}$, $x_{2}$, and $x_{3}$). Each of those points is an extremum point of the density (there is another such point at infinity). The sonic point location for these upstream densities is near $x_{3}\sim10^{4}\,\textrm{cm}$. As the upstream density increases, there remains only a single location where $\phi=0$, which is close to $x_{1}\sim10^{2}\,\textrm{cm}$. Around this transition, the sonic point changes location to $x_{1}$. We mark this transition as $\phi=0,\,3\rightarrow1$ in the upper panels of Figures~\ref{fig:DetonationSpeed_CJAbu} and~\ref{fig:CO_DetonationPropQAt}. The slight jittering of the sonic point location as the density changes is a consequence of the rapid shift in the sonic point location as $D$ approaches $D_{*}$. 

\begin{figure}
\includegraphics[width=0.48\textwidth]{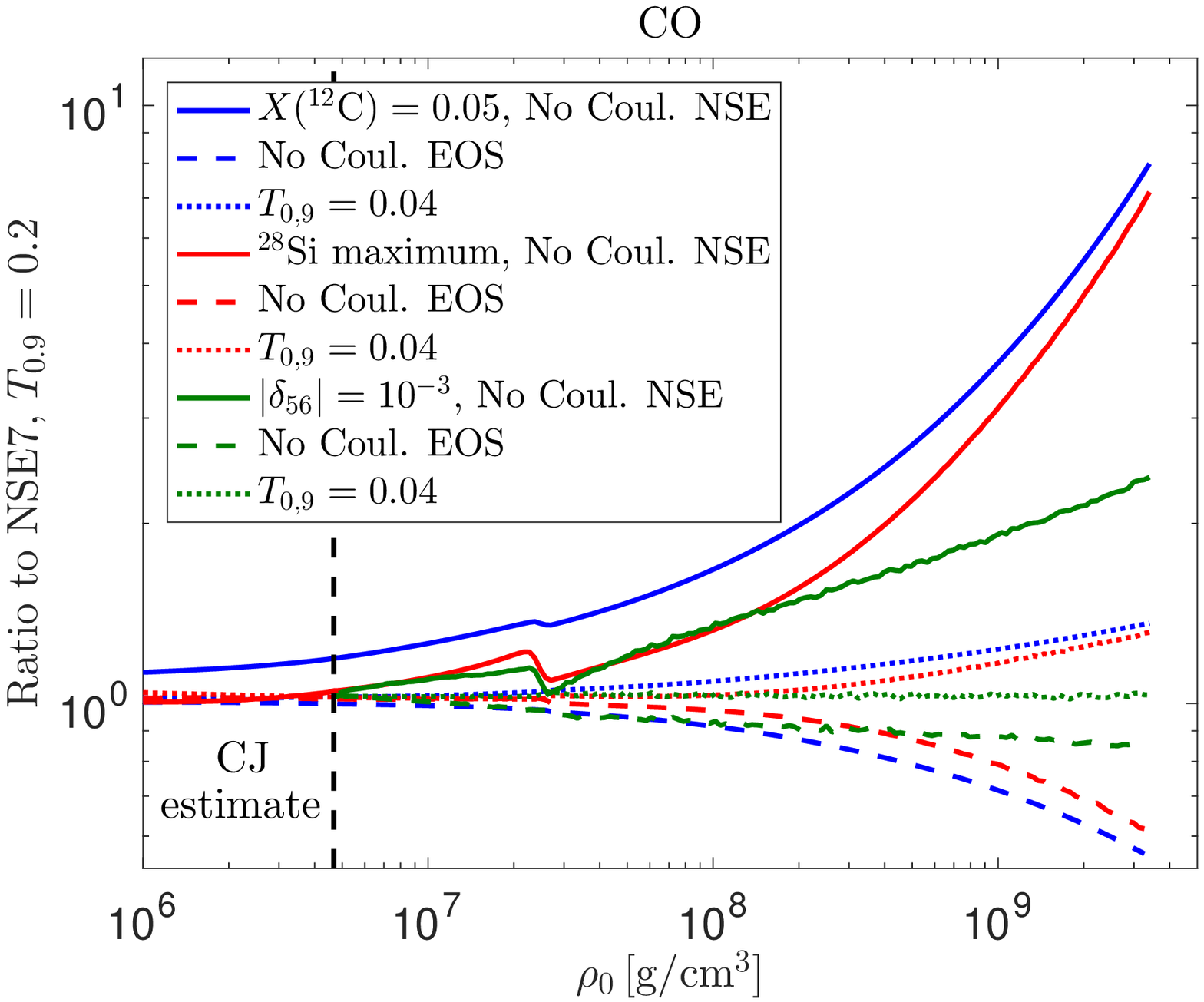}
\caption{The ratio between the carbon-burning scale (blue), the positions where $^{28}$Si is maximal (red) and where $|\delta_{56}|=10^{-3}$ (green) obtained under various assumptions and these scales obtained with our default input physics and $T_{0,9}=0.2$. The solid lines are without the addition of the  Coulomb correction term of the EOS, dashed lines are without the addition of the Coulomb correction term to the NSE state and the dotted lines are for $T_{0,9}=0.04$. For low densities, $\rho_{0,7}\lesssim0.47$ (left to the black dashed line), we integrated with $D=D_{CJ}$.
\label{fig:CO_DetonationProp_Uncertain}}
\end{figure}

\begin{figure}
\includegraphics[width=0.48\textwidth]{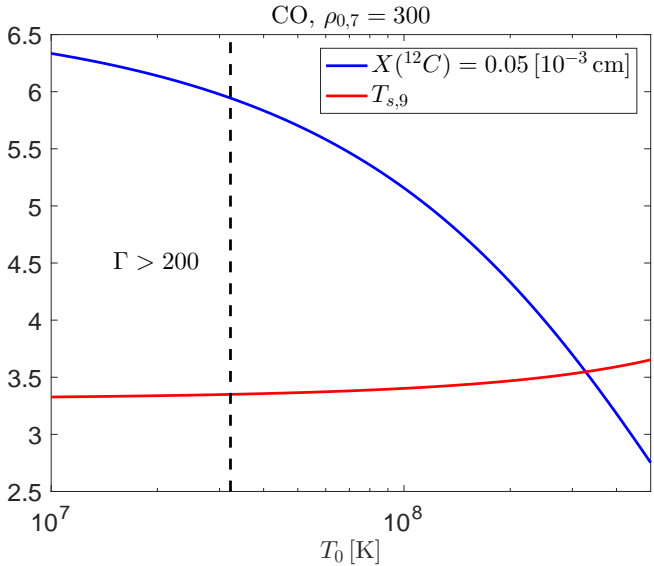}
\caption{The carbon-burning scale (blue) and the post-shock temperature (red) as a function of the upstream temperature for CO and $\rho_{0,7}=300$. The ion coupling parameter, $\Gamma$, of the upstream plasma is larger than $200$ for $T_{0,9}\lesssim0.032$ (left to the black dashed line), where the fit for $f(\Gamma)$ is not valid. 
\label{fig:Tdep}}
\end{figure}

Our analysis indicates some minor dependence of the scales on the upstream temperature (see dotted lines in Figure~\ref{fig:CO_DetonationProp_Uncertain}). The largest one is for the carbon-burning scale at high densities. The carbon-burning scale is shown as a function of the upstream temperature in Figure~\ref{fig:Tdep} for $\rho_{0,7}=300$. The burning scale decreases as the upstream temperature increases. This is because the post-shock temperature, $T_{s}$, depends slightly on the upstream temperature. This effect is obtained at high densities, where the post-shock plasma is slightly degenerate, making the temperature a sensitive function of the pressure. We note that the ion coupling parameter, $\Gamma$, of the upstream plasma is larger than $200$ for $T_{0,9}\lesssim0.032$, where the fit for $f(\Gamma)$ is not valid. This is the reason that we choose $T_{0,9}=0.04$ for the temperature sensitivity tests in the previous CO sections. 

\subsubsection{The uncertainty of the CO results}
\label{sec:CO uncertain}

\begin{figure}
\includegraphics[width=0.48\textwidth]{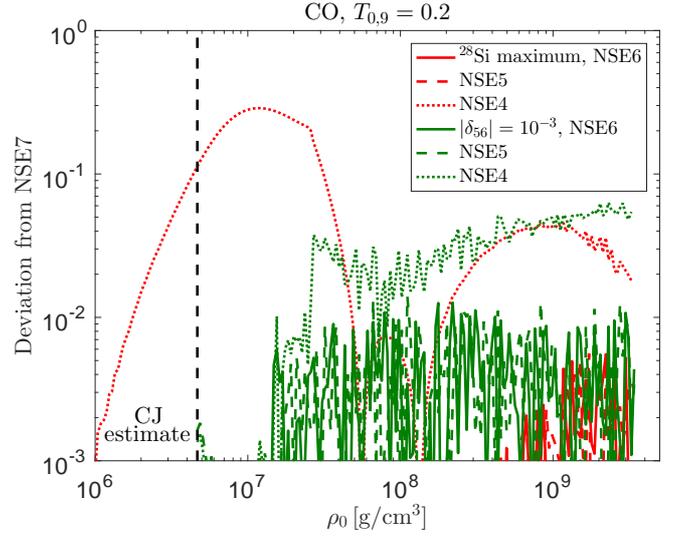}
\caption{The deviations of the positions where $^{28}$Si is maximal (red) and where $|\delta_{56}|=10^{-3}$ (green), calculated with the NSE$4$ (dotted lines), NSE$5$ (dashed lines) and NSE$6$ (solid lines) isotope lists, from the results calculated with the NSE$7$ isotope list, for CO at $T_{0,9}=0.2$ as a function of the upstream density. For low densities, $\rho_{0,7}\lesssim0.47$ (left to the black dashed line), we integrated with $D=D_{CJ}$.
\label{fig:CO_DetonationProp_nets}}
\end{figure}

The deviations of the positions where $^{28}$Si is maximal and where $|\delta_{56}|=10^{-3}$, calculated with the NSE$4$, NSE$5$ and NSE$6$ isotope lists, from the results calculated with the NSE$7$ isotope list are presented in Figure~\ref{fig:CO_DetonationProp_nets}. Deviations as high as $\myapprox30\%$ are obtained for NSE$4$, while the deviations of NSE$5$ and NSE$6$ are smaller than the percent level. The other scales shown in Figure~\ref{fig:CO_DetonationPropQAt} have smaller deviations. We verified that the deviations of the results obtained with the NSE$7$Si list deviate by less than a percent from the results obtained with the NSE$7$ list. This suggests that our calculation of the length-scales is converged to the percent level. The effect of the Coulomb correction is examined in Figure~\ref{fig:CO_DetonationProp_Uncertain}. The Coulomb corrections to the EOS are only important at high densities, and they change at most the carbon-burning scale by a factor of $\mysim2$. The Coulomb correction terms to the NSE have a significant effect at high densities, where they can decrease the length-scales by up to one order of magnitude, as they increase the reaction rates. Uncertainty in the reaction rates can be at the same level or even higher, making the length-scales uncertain to a factor of a few. However, a detailed study of the sensitivity to uncertainty in the reaction rates is beyond the scope of this paper. 

\subsubsection{Comparing the detonation wave structure in CO to \citet{Khokhlov89}}
\label{sec:Khokhlov89}

\citet{Khokhlov89} calculated the detonation wave structure for CO, an upstream temperature of $T_{0,9}=0.2$ and a few values of the upstream density in the range of $[10^{7},3\times10^{9}]\,\textrm{g}/\textrm{cm}^{3}$. The EOS used by \citet{Khokhlov89} is the same as the EOS used by \citet{Khokhlov88}\footnote{Although the nuclear level excitations are missing from the description of the EOS in \citet{Khokhlov89}.}. Since \citet{Khokhlov89} does not mention the apparent numerical bug in the EOS used by \citet{Khokhlov88}, as we showed in Section~\ref{sec:Khokhlov88}, and is citing the same $D_{\textrm{CJ}}$ values from \citet{Khokhlov88}, we assume that the EOS used by \citet{Khokhlov89} suffers from the same shortcomings as the EOS used by \citet{Khokhlov88}. The list of isotopes included $114$ isotopes, and we used the modern values of the binding energies and partition functions.

\begin{figure}
\includegraphics[width=0.48\textwidth]{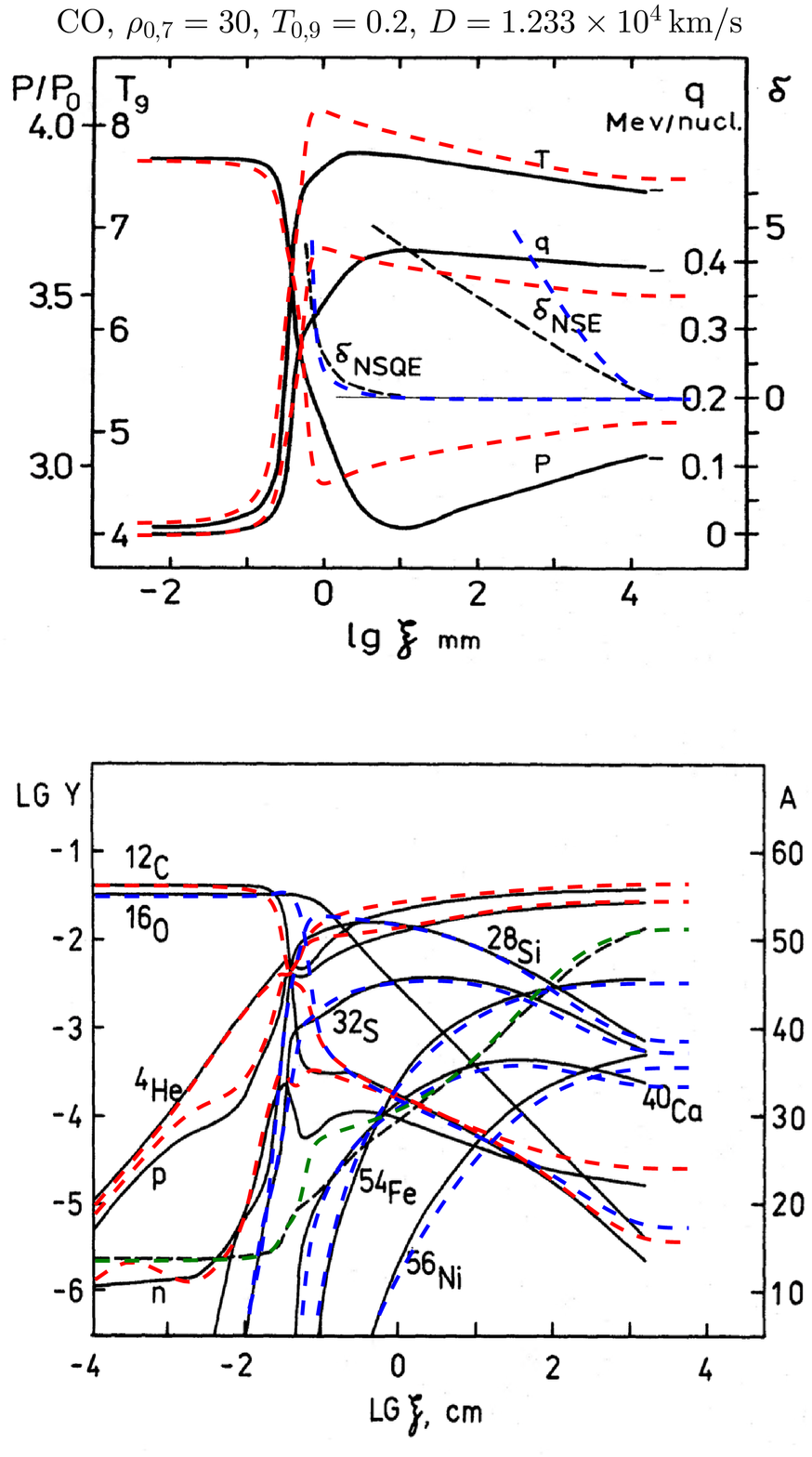}
\caption{Figures 7 and 8 from \citet{Khokhlov89}. The structure of an overdriven detonation wave for CO, $T_{0,9}=0.2$, $\rho_{0,7}=30$ and $D=1.233\times10^{4}\,\textrm{km}/\textrm{s}$, as a function of the distance behind the shock. Black lines are the results of \citet{Khokhlov89}, while the coloured lines are our results with the input physics of \citet{Khokhlov89}. Here, $\delta_{\textrm{NSQE}}=\delta_{56}-\delta_{28}$ (but we actually plot $\delta_{28}-\delta_{56}$, since it seems that \citet{Khokhlov89} plotted this as well) and $\delta_{\textrm{NSE}}=\delta_{56}$. The green dashed line in the bottom panel is $\tilde{A}$ \citep[note that the right y-axis label, $A$, is probably a typo, and should be $\langle A\rangle$ with the definitions of][]{Khokhlov89}. Note that the $x$-axes units in the two panels are different. 
\label{fig:KhoComp}}
\end{figure}

We concentrate on the $\rho_{0,7}=30$, for which \citet{Khokhlov89} provides detailed results. \citet{Khokhlov89} reports that $D_{*}=1.218\times10^{4}\,\textrm{km}/\textrm{s}$, while we find that $D_{*}=1.2107\times10^{4}\,\textrm{km}/\textrm{s}$ for the same input physics (similar deviation was found in Section~\ref{sec:Khokhlov88} for $D_{\textrm{CJ}}$). It is apparent from our comparison of the structure of an overdriven detonation with $D=1.233\times10^{4}\,\textrm{km}/\textrm{s}$ (Figure~\ref{fig:KhoComp}, note the different units of the $x$-axes of the two panels) that the NSE state is different between the two calculations (especially in the upper panel). This difference is similar in magnitude to the one we found in Section~\ref{sec:Khokhlov88} for the CJ state, suggesting that it is connected with the apparent numerical bug in the EOS used by \citet{Khokhlov89}. This could also be the reason for the higher (lower) temperatures (pressures) that we get around $1\,\textrm{mm}$. For the $\delta_{\textrm{NSQE}}$, it seems that \citet{Khokhlov89} plotted $\delta_{28}-\delta_{56}$ (and not $\delta_{56}-\delta_{28}$, as claimed by \citet{Khokhlov89}), so we plot this as well. Note that the scale of $\delta_{\textrm{NSE}}$ and $\delta_{\textrm{NSQE}}$ is linear. The abundance of the isotopes, shown in the bottom panel of Figure~\ref{fig:KhoComp}, is similar in the two calculations, except for the much faster consumption of $^{16}$O around $1\,\textrm{mm}$ in our calculation, which is because of the higher temperatures we get there. 

\begin{figure}
\includegraphics[width=0.48\textwidth]{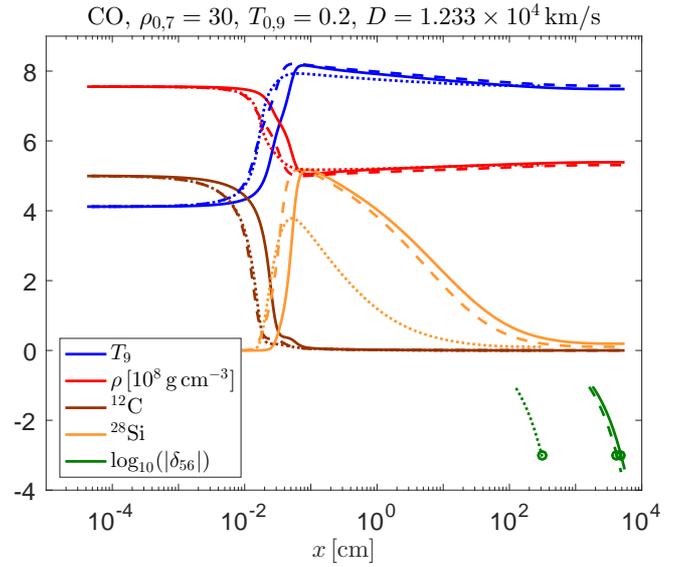}
\caption{The structure of an overdriven detonation wave for CO, $T_{0,9}=0.2$, $\rho_{0,7}=30$ and $D=1.233\times10^{4}\,\textrm{km}/\textrm{s}$, as a function of the distance behind the shock. We show the temperature (blue), density (red), $^{12}$C mass fraction (brown), $^{28}$Si mass fraction (orange), and $\delta_{56}$ (green). The solid lines present the results with the input physics of \citet{Khokhlov89}, the dashed lines are with the addition of the Coulomb correction terms to the NSE, and the dotted lines are the results with the default input physics. The green points mark the locations where $|\delta_{56}|=10^{-3}$.
\label{fig:KhoComp3}}
\end{figure}

We next compare in Figure~\ref{fig:KhoComp3} our results with the input physics of \citet[][solid lines]{Khokhlov89} to the results with the default input physics (dotted lines). The carbon- and silicon-burning length-scales are smaller by a factor of $\mysim2$ in the default case, and $|\delta(56)|=10^{-3}$ at a distance that is smaller by a factor of $\mysim10$. The inclusion of the Coulomb correction term for the NSE (dashed lines) decreases the carbon-burning length-scale to the default value (see also Figure~\ref{fig:CO_DetonationProp_Uncertain}). The remaining discrepancies are because of the isotope list used by \citet{Khokhlov89}. We verified that the default results are reproduced by adding the missing isotopes from NSE$7$ with $Z\le14$ and from the $\alpha$-ext lists to the list used by \citet{Khokhlov89}, which increases the number of isotopes to $161$. In fact, the results from NSE$4$ deviate by less than $10\%$ for this upstream density (see Figure~\ref{fig:CO_DetonationProp_nets}), which shows that with $137$ isotopes (although somewhat different from the $114$ used by \citet{Khokhlov89}) better results can be
obtained. 

\subsubsection{Comparison to \citet{Townsley2016}}
\label{sec:Townsley2016}

\citet{Townsley2016} calculated the detonation wave structure for an initial composition of $X(^{12}\textrm{C})=0.5$, $X(^{16}\textrm{O})=0.48$, $X(^{22}\textrm{Ne})=0.02$ (CONe), an upstream temperature of $T_{0,9}=0.4$ and a few values of the upstream density in the range of $[5\times10^{6},2\times10^{8}]\,\textrm{g}/\textrm{cm}^{3}$. The calculation were performed with TORCH \citep{Timmes1999TORCH}, which uses the \textit{Helmholtz} EOS. The list of isotopes included $200$ isotopes\footnote{\citet{Townsley2016} probably used the \textsc{torch200} net, which actually contains 205 isotopes.}, and screening was applied for the reaction rates. We concentrate on the overdriven detonation in which $\rho_{0,7}=1$ and $D=1.166\times10^{4}\,\textrm{km}/\textrm{s}$, for which \citet{Townsley2016} provide detailed results. We calculate the detonation wave structure for this case by using the \textit{Helmholtz} EOS, the \textsc{torch200} isotope list (without $^{8}$Be and $^{9}$B, see Section~\ref{sec:net}, which sums up to $203$ isotopes), and using the option of the \textsc{extended screening} of {\sc MESA}. As the other input physics, we use our usual default values. Since neither the \textsc{extended screening} nor the screening used by TORCH respect a detailed balance, the integration does not terminate at NSE, but rather at some other steady-state configuration. We integrate up to $t=10\,\textrm{s}$, at which point this steady state was obtained.  

\begin{figure}
\includegraphics[width=0.48\textwidth]{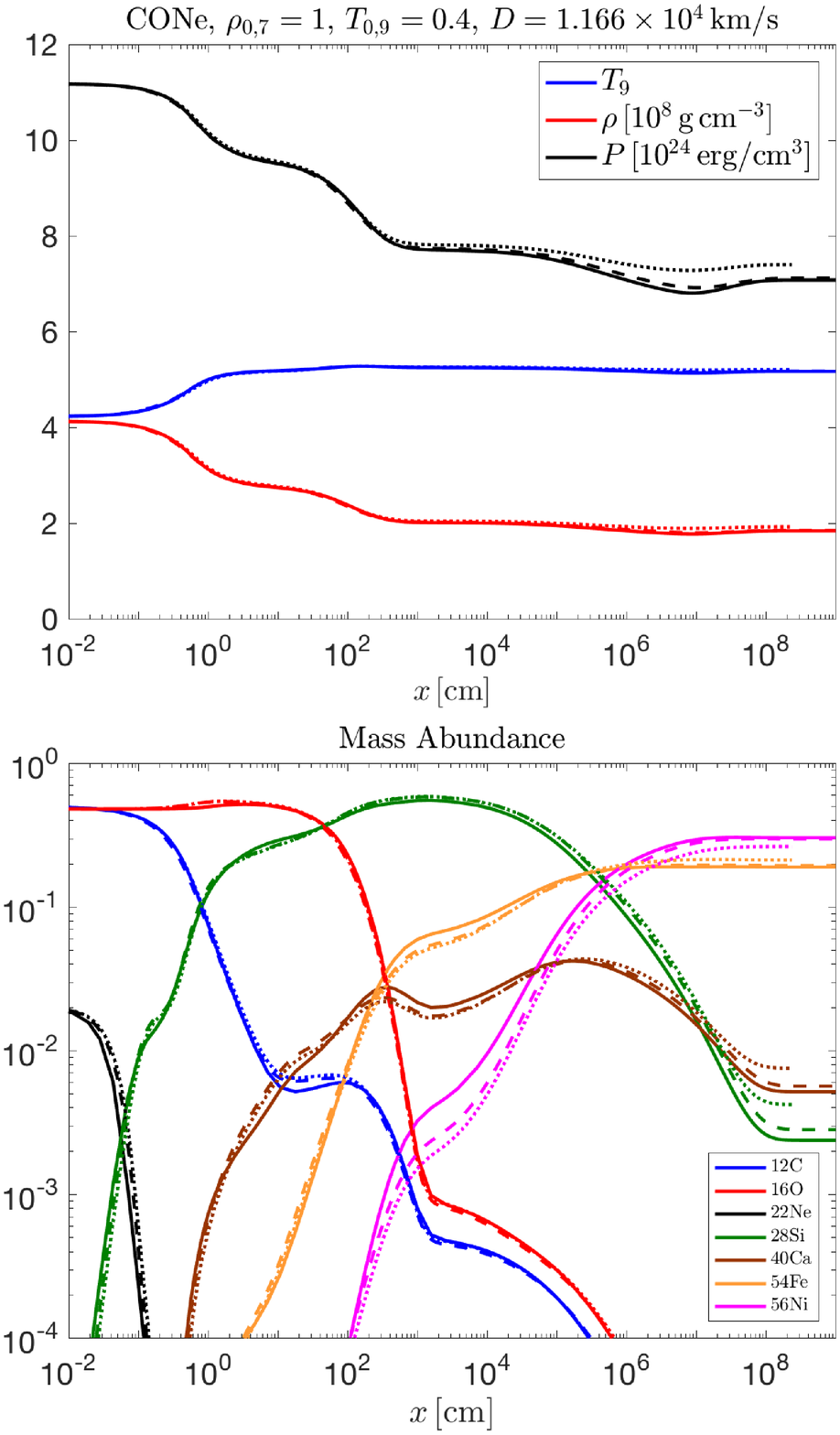}
\caption{The structure of an overdriven detonation wave for $X(^{12}\textrm{C})=0.5$, $X(^{16}\textrm{O})=0.48$, $X(^{22}\textrm{Ne})=0.02$, $T_{0,9}=0.4$, $\rho_{0,7}=1$ and $D=1.166\times10^{4}\,\textrm{km}/\textrm{s}$, as a function of the distance behind the shock. Solid lines are the results from \citet{Townsley2016}, dashed lines are our results with the input physics of \citet{Townsley2016} and the \textsc{extended screening} option of {\sc MESA}, and dotted lines are the results with our default input physics. Upper panel: temperature (blue), density (red) and pressure (black). Bottom panel: the mass fraction of key isotopes. Note that since the \textsc{extended screening} option does not respect a detailed balance, the integration does not terminate at NSE, so we integrate up to $t=10\,\textrm{s}$. 
\label{fig:TownsleyComp}}
\end{figure}

The deviation of our results (Figure~\ref{fig:TownsleyComp}, dashed lines) from those of \citet[][solid lines]{Townsley2016} is quite small\footnote{We thank Dean Townsley for sharing their results with us.}. For example, the deviation in the pressures is smaller than $2\%$. This difference is probably because of the somewhat different reaction rates and screening factors incorporated into each of the two calculations. A calculation with our default input physics is presented as well in Figure~\ref{fig:TownsleyComp} (dotted lines). As usual, the integration is performed up to $\delta_{\max}=10^{-3}$. Larger deviations are obtained between the default calculation and the results of \citet{Townsley2016}. For example, a deviation of $\myapprox7\%$ is obtained in the pressure at a distance of $x\sim10^{7}\,\textrm{cm}$. It is evident that the NSE values obtained with our default input physics deviate by a few percent from the steady-state configuration obtained by \citet{Townsley2016}. The easiest way to analyse these differences is to compare their NSE states (which are independent of reaction rates), but as explained above, such a state does not exist for the input physics of \citet{Townsley2016}.

With respect to figure 1 of \citet{Townsley2016}, since there the pathological detonation speed was not calibrated to high accuracy and the sonic point location was determined by the location of the density minimum\footnote{Townsley (private communication).}, the position of the $^{28}$Si abundance maximum and the sonic point location are not adequate for an accurate comparison.


\subsection{The structure of the detonation wave in He}
\label{sec:He structure}

In this section, we present the structure of the detonation wave in He. In Section~\ref{sec:Helium example}, we present an example of the structure of a detonation wave for some specific initial conditions. In Section~\ref{sec:He scan scale}, we calculate the structure of the detonation wave as a function of the upstream density. We then comment on the uncertainty of the results in Section~\ref{sec:He uncertain}. Finally, we compare out results to \citet{Khokhlov89} in Section~\ref{sec:Khokhlov89_He}.
 
\subsubsection{An example for He: $\rho_{0,7}=1$ and $T_{0,9}=0.2$}
\label{sec:Helium example}

\begin{figure}
\includegraphics[width=0.45\textwidth]{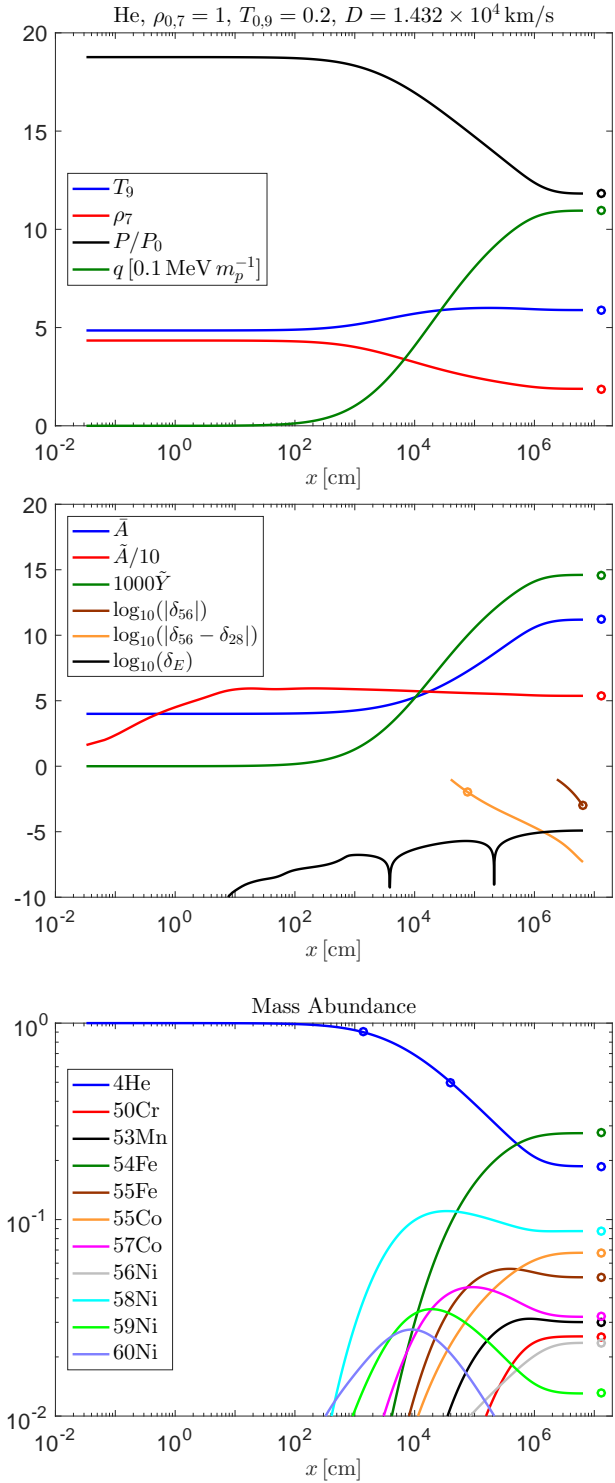}
\caption{The structure of an overdriven detonation wave as a function of the distance behind the shock. Upper panel: temperature (blue), density (red), pressure (black) and thermonuclear energy release (green). Middle panel: $\bar{A}$ (blue), $\tilde{A}$ (red), $\tilde{Y}$ (green), $\delta_{56}$ (see the text, brown), $\delta_{56}(x)-\delta_{28}(x)$ (which monitors the NSQE state, orange) and $\delta_{E}$ (which monitors energy conservation, black). The orange point marks the location where $|\delta_{56}-\delta_{28}|=10^{-2}$, and the brown point marks the location where $|\delta_{56}|=10^{-3}$. Bottom panel: mass fractions of a few key isotopes. The blue points mark the locations where the mass fraction of $^{4}$He reaches $0.9$, $0.5$. The points at the right edges of the panels represent the NSE values.
\label{fig:He_1e7_2e8}}
\end{figure}

We present in Figure~\ref{fig:He_1e7_2e8}, as an example, the structure of a detonation wave for He, $\rho_{0,7}=1$, $T_{0,9}=0.2$ and a detonation speed of $D=1.432\times10^{4}\,\textrm{km}/\textrm{s}$ ($>D_{\textrm{CJ}}\approx1.4304\times10^{4}\,\textrm{km}/\textrm{s}$ for these upstream conditions, see Table~\ref{tbl:He CJ}). The structure of this detonation wave is very different from the structure of a detonation wave in CO. The burning of $^{4}$He immediately synthesizes heavy elements with $\tilde{A}\approx55$ (see detailed discussion in \citet{Khokhlov1984} and a somewhat more accurate description in \citet{Khokhlov1985}). This mode of burning depletes the $^{4}$He by $10(50)\%$ at $x\approx1.4\times10^{3}(4.0\times10^{4})\,\textrm{cm}$ (blue points in the lower panel), while increasing $\tilde{Y}$ and $\bar{A}$, almost without changing $\tilde{A}$, and releasing $\myapprox1.1\,\textrm{MeV}/m_{p}$. Most of the energy is being release with the plasma not in NSQE, as $|\delta_{56}(x)-\delta_{28}(x)|=0.01$ at $x\approx7.8\times10^{4}\,\textrm{cm}$ (orange point in the middle panel), where already $\myapprox0.77\,\textrm{MeV}/m_{p}$ have been released. 

The middle panel shows that $\tilde{Y}$ increases towards the NSE value (compare with Figure~\ref{fig:CO_1e7_2e8}, in which $\tilde{Y}$ decreases towards the NSE value), and we verified that the increase is controlled by the triple-$\alpha$ reaction, $3^{4}$He$\rightarrow^{12}$C. The middle panel shows that $|\delta_{56}|=0.1$ at $x\approx2.4\times10^{6}\,\textrm{cm}$. From that position, $|\delta_{56}|$ decreases exponentially with an $e$-folding distance of $l_{\textrm{56}}\approx 2.6\times10^{6}\,\textrm{cm}$. The brown point marks the location where $|\delta_{56}|=10^{-3}$. As usual, we stop the integration when $\delta_{\max}=10^{-3}$. The deviation of the solution parameters at the end of the integration from the NSE values (points at the right edges of the panels), which are calculated only from conservation laws, is smaller than $10^{-3}$. The middle panel shows that the value of $\delta_{E}$ increases towards NSE and is $\myapprox10^{-5}$ at the end of the integration. This demonstrates the high accuracy of our integration. 

\subsubsection{The dependence of the burning scales on the upstream density for He}
\label{sec:He scan scale}

\begin{figure}
\includegraphics[width=0.48\textwidth]{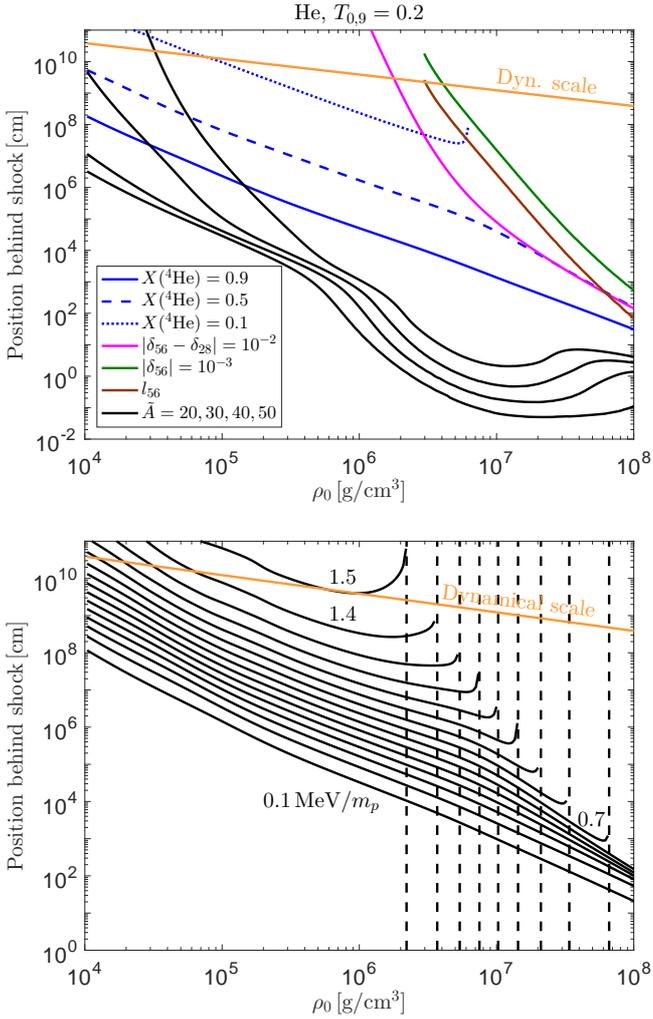}
\caption{Different scales of the He CJ detonation in comparison with a typical dynamical scale of $v/\sqrt{G\rho_{0}}$ with $v=10^{4}\,\textrm{km}/\textrm{s}$ (orange). Top panel: the $^{4}$He consumption scale (blue, $X(^{4}\textrm{He})=0.9,0.5,0.1$ solid, dashed, dotted, respectively), the location where $\tilde{A}=20,30,40,50$ (bottom to top, black), the location where $|\delta_{56}-\delta_{28}|=10^{-2}$ (magenta) the location where $|\delta_{56}|=10^{-3}$ (green) and $l_{\textrm{56}}$ (brown). Bottom panel: the locations where the energy release is $0.1,0.2,...,1.5\,\textrm{MeV}/m_{p}$ (bottom to top, black). Dashed lines in the bottom panel mark the upstream densities at which $q_{01,\textrm{CJ}}$ obtained at the NSE state matches the indicated energy release.
\label{fig:He_DetonationPropQAt}}
\end{figure}

For He, the detonation is of the CJ type (see detailed discussion in Section~\ref{sec:CJ condition}). Different scales of the He CJ detonation are shown in Figure~\ref{fig:He_DetonationPropQAt}. For low densities, $\rho_{0,7}\lesssim0.30$, we are unable to integrate with high accuracy up to the location where $|\delta_{56}|=10^{-3}$, so this location and $l_{\textrm{56}}$ are not shown for these densities. The numerical accuracy of all scales in Figure~\ref{fig:He_DetonationPropQAt} is $\mylesssim10^{-3}$. 

\begin{figure}
\includegraphics[width=0.48\textwidth]{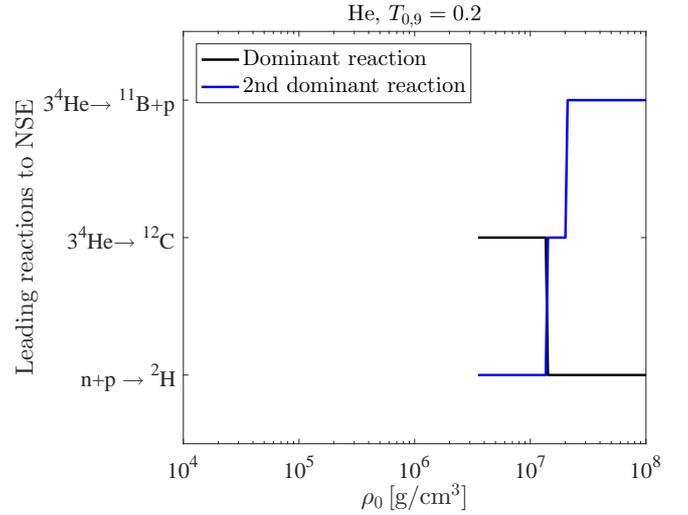}
\caption{The reactions that dominate the net change of $\tilde{Y}$ at $|\delta_{56}-\delta_{28}|=10^{-2}$ and at $|\delta_{56}|=10^{-3}$ (in this case the same reactions dominate at both positions), as a function of the upstream density for He. 
\label{fig:HeLeadingNSE}}
\end{figure}

For high upstream densities, $\rho_{0,7}\gtrsim0.015$, the ordering of the different scales as a function of the upstream density is similar to the case of $\rho_{0,7}=1$ that was described in detail in Section~\ref{sec:Helium example}. The burning of $^{4}$He synthesizes heavy elements with $\tilde{A}\approx55$ much faster than the rate in which $^{4}$He is depleted.  At lower upstream densities, the depletion rate of $^{4}$He is faster than the rate at which heavy elements are synthesized. The energy release roughly follows the $^{4}$He depletion, and most of the energy is being release with the plasma not in NSQE. The reactions that dominate the approach to NSE are shown in Figure~\ref{fig:HeLeadingNSE} (in this case the same reactions dominate both at $|\delta_{56}-\delta_{28}|=10^{-2}$ and at $|\delta_{56}|=10^{-3}$). The approach to NSE is controlled at low upstream densities, $0.3\lesssim\rho_{0,7}\lesssim10$, by the triple-$\alpha$ reaction, $3^{4}$He$\rightarrow^{12}$C, and to some extent by $n+p\rightarrow^{2}$H, while at high densities $n+p\rightarrow^{2}$H is the dominant process with an additional contribution from $3^{4}$He$\rightarrow^{11}$B$+p$. 

The scales themselves shorten significantly as the upstream density increases, due to the rise in the post-shock temperature. Furthermore, the temperature at the CJ NSE state increases monotonically with $\rho_{0}$, which decreases both $\bar{A}$ and $q_{01}$ at these states (see Table~\ref{tbl:He CJ}). We mark with dashed lines in the bottom panel of Figure~\ref{fig:He_DetonationPropQAt} the upstream densities in which some values of $q_{01}$ are obtained at the CJ NSE state.

\begin{figure}
\includegraphics[width=0.48\textwidth]{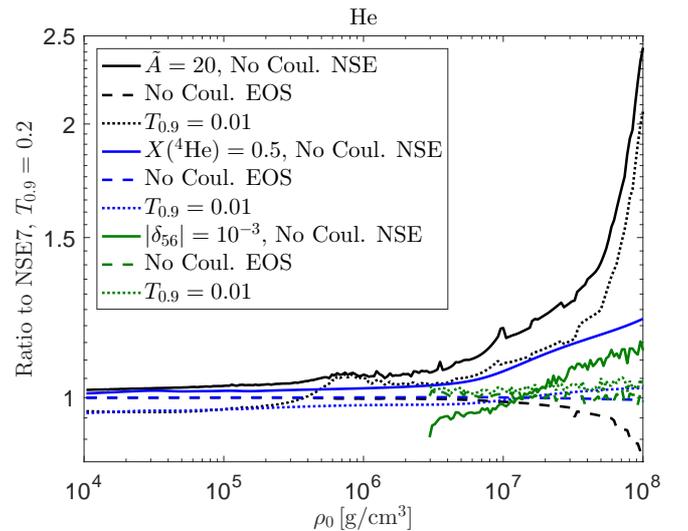}
\caption{The ratio between the positions at which $\tilde{A}=20$ (black), where half of the $^{4}$He is consumed (blue) and where $|\delta_{56}|=10^{-3}$ (green) is obtained under various assumptions and the scales that are obtained with our default input physics and $T_{0,9}=0.2$. The solid lines are without the Coulomb correction term of the EOS, dashed lines are without the Coulomb correction term to the NSE state and dotted lines are for $T_{0,9}=0.01$. The electron--electron term is neglected here, and it is a few percent correction for $\rho_{0,7}\lesssim0.027$ and $T_{0,9}=0.01$.
\label{fig:He_DetonationProp_Uncertain_Helium}}
\end{figure}

\begin{figure}
\includegraphics[width=0.48\textwidth]{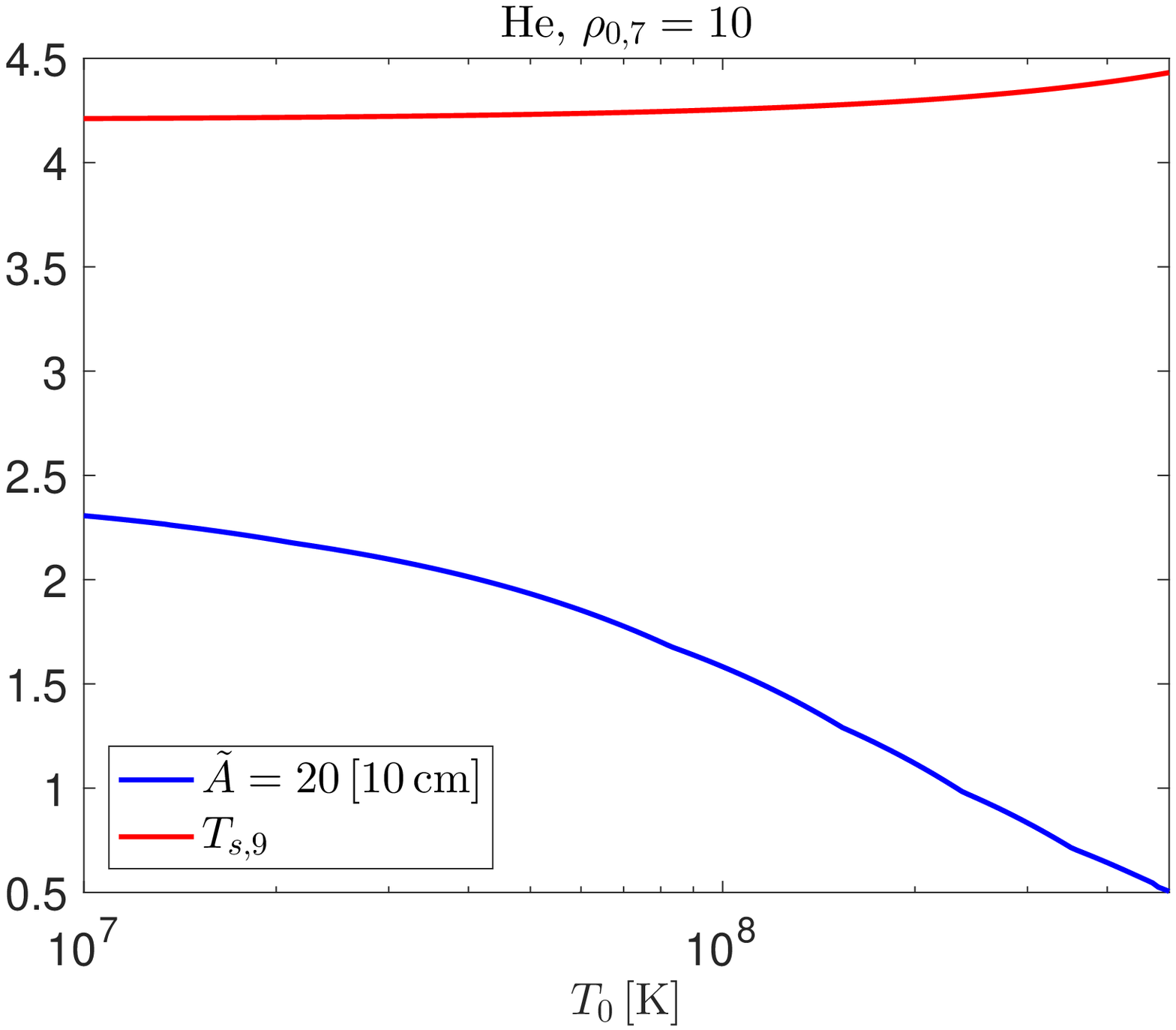}
\caption{The scale at which $\tilde{A}=20$ (blue) and the post-shock temperature (red) as a function of the upstream temperature for He and $\rho_{0,7}=10$.
\label{fig:Tdep_Helium}}
\end{figure}

Some minor dependence of the scales on the upstream temperature are obtained (see dotted lines in Figure~\ref{fig:He_DetonationProp_Uncertain_Helium}, the electron--electron term is neglected here, and it is a few percent correction for $\rho_{0,7}\lesssim0.027$ and $T_{0,9}=0.01$). The largest one is for the scale at which $\tilde{A}=20$ at high densities. This scale is shown as a function of the upstream temperature in Figure~\ref{fig:Tdep_Helium} for $\rho_{0,7}=10$. The scale decreases as the upstream temperature increases, because the post-shock temperature, $T_{s}$, depends slightly on the upstream temperature. This effect is obtained at high densities, where the post-shock plasma is slightly degenerate, making the temperature a sensitive function of the pressure. 

\subsubsection{The uncertainty of the He results}
\label{sec:He uncertain}

\begin{figure}
\includegraphics[width=0.48\textwidth]{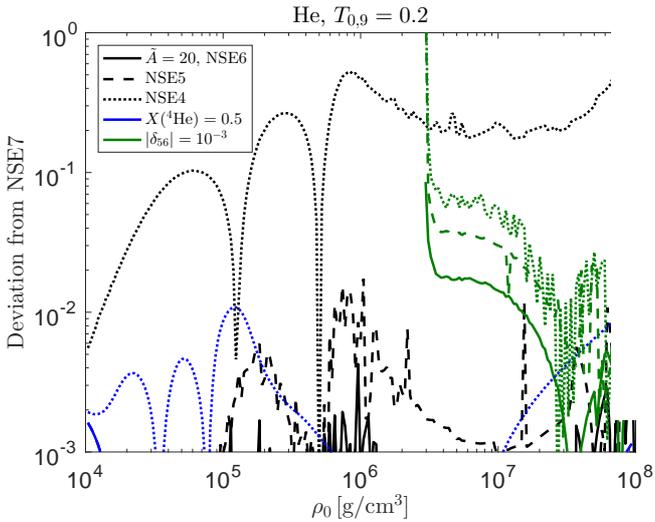}
\caption{The deviations of the positions where $\tilde{A}=20$ (black), where half of the $^{4}$He is consumed (blue) and where $|\delta_{56}|=10^{-3}$ (green), calculated with the NSE$4$ (dotted lines), NSE$5$ (dashed lines) and NSE$6$ (solid lines) isotope lists, from the results calculated with the NSE$7$ isotope list, for He with $T_{0,9}=0.2$, as a function of the upstream density.
\label{fig:He_DetonationProp_nets}}
\end{figure}

The deviations of the positions where $\tilde{A}=20$, where half of the $^{4}$He is consumed and where $|\delta_{56}|=10^{-3}$, calculated with the NSE$4$, NSE$5$ and NSE$6$ isotope lists, from the results calculated with the NSE$7$ isotope list are presented in Figure~\ref{fig:He_DetonationProp_nets}. Deviations as high as an order unity are obtained for NSE$4$, while the deviations of NSE$5$ and NSE$6$ are smaller than a few percent (not including $|\delta_{56}|=10^{-3}$ near $\rho_{0,7}\lesssim0.30$, where we are unable to integrate with high accuracy up to this location). The other scales shown in Figure~\ref{fig:He_DetonationPropQAt} have smaller deviations. We verified that the deviations of the results obtained with the NSE$7$Si list deviate by less than a few percent from the results obtained with the NSE$7$ list. This suggests that our calculation of the length-scales is converged to a few percent. The effect of the Coulomb correction is examined in Figure~\ref{fig:He_DetonationProp_Uncertain_Helium}. The Coulomb correction terms to the EOS are only important at high densities, and they change at most the $\tilde{A}=20$ scale by $\myapprox15\%$. The Coulomb correction terms to the NSE have an effect at high densities, where they can decrease the length-scales by up to a factor of $2$, as they increase the reaction rates. Uncertainty in the reaction rates can be at the same magnitude or even higher, making the length-scales uncertain to a factor of a few. However, a detailed study of the sensitivity to uncertain reaction rates is beyond the scope of this paper. 

\subsubsection{Comparing the detonation wave structure in He to \citet{Khokhlov89}}
\label{sec:Khokhlov89_He}

\begin{figure}
\includegraphics[width=0.48\textwidth]{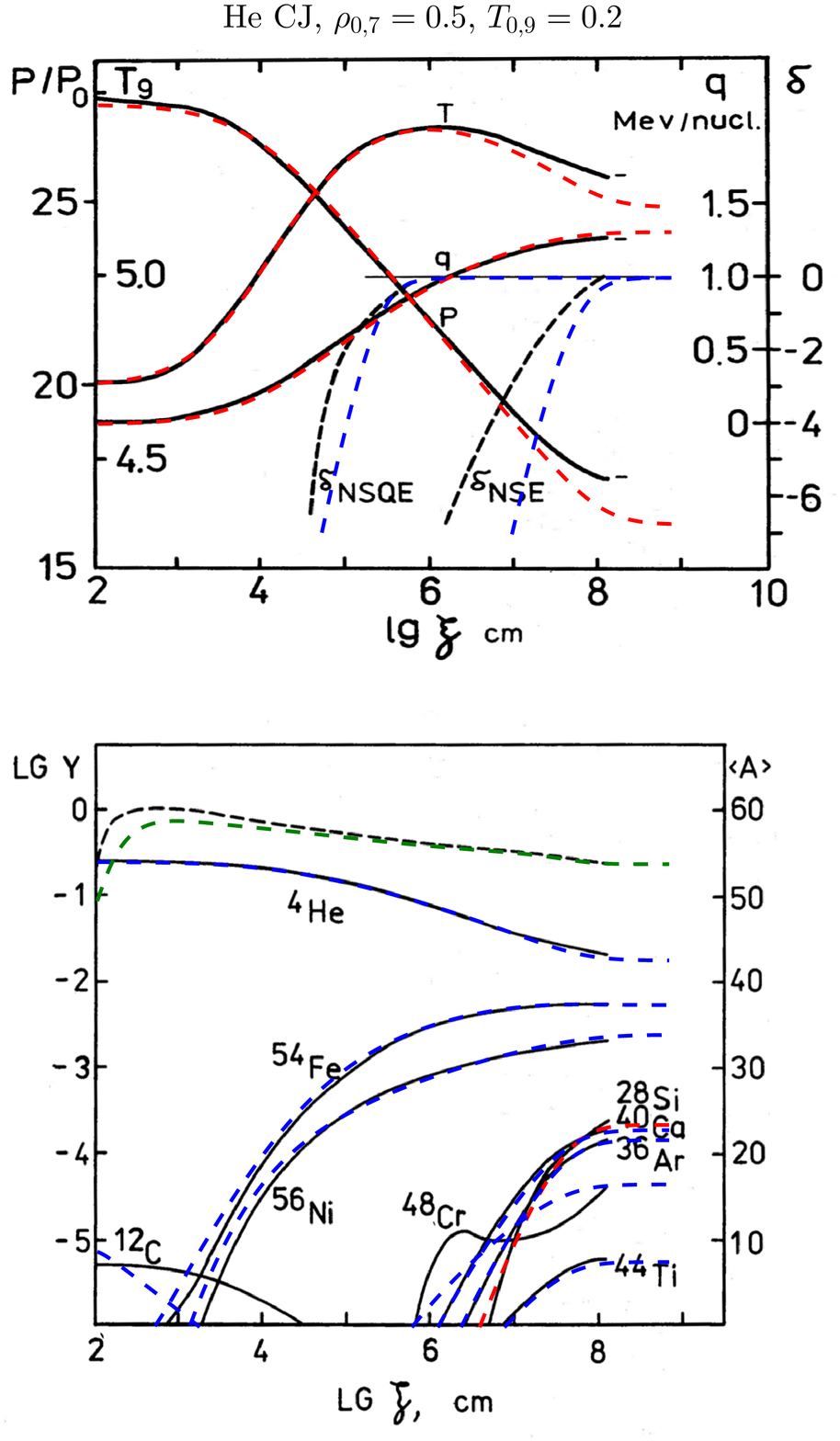}
\caption{Figures 10 and 11 of \citet{Khokhlov89}. The structure of a CJ detonation wave for He, $T_{0,9}=0.2$, $\rho_{0,7}=0.5$ and $D=1.233\times10^{4}\,\textrm{km}/\textrm{s}$, as a function of the distance behind the shock. Black lines are the results of \citet{Khokhlov89}, while the coloured lines are our results with the input physics of \citet{Khokhlov89}. The value of $D_{\textrm{CJ}}$ used by \citet{Khokhlov89} is probably different from our value of $D_{\textrm{CJ}}\approx1.4906\times10^{4}\,\textrm{km}/\textrm{s}$ due to the apparent numerical bug in the EOS used by \citet{Khokhlov89}. The green dashed line in the bottom panel is $\tilde{A}$. 
\label{fig:KhoComp_Helium}}
\end{figure}

\citet{Khokhlov89} calculated the CJ detonation wave structure for He, an upstream temperature of (probably) $T_{0,9}=0.2$ and a few values of the upstream density in the range of $[\textrm{few}\times10^{5},\textrm{few}\times10^{9}]\,\textrm{g}/\textrm{cm}^{3}$. The value of $D_{\textrm{CJ}}$ used by \citet{Khokhlov89} is probably different from our value of $D_{\textrm{CJ}}\approx1.4906\times10^{4}\,\textrm{km}/\textrm{s}$, calculated with the input physics of \citet{Khokhlov89}, due to the apparent numerical bug in the EOS used by \citet{Khokhlov89}. We compare in Figure~\ref{fig:KhoComp_Helium} the structure of the CJ detonation. It is apparent from the upper panel of Figure~\ref{fig:KhoComp_Helium} that the NSE state is different in the two calculations. This difference is similar in magnitude to the one we found in Section~\ref{sec:Khokhlov88_He}, suggesting that it is due to the shortcomings of the EOS used by \citet{Khokhlov89}. Note, however, that in Table~\ref{tbl:K88_Helium} we consistently get for CJ detonations a higher $T_{\textrm{CJ}}$ and lower $q_{01,\textrm{CJ}}$ than the results of \citet{Khokhlov88}, which is not the case for the NSE state in Figure~\ref{fig:KhoComp_Helium}. This could suggest that the results of \citet{Khokhlov88} are inconsistent with the results of \citet{Khokhlov89}. The abundance of the isotopes shown in the bottom panel of Figure~\ref{fig:KhoComp_Helium} are similar in the two calculations. 

\begin{figure}
\includegraphics[width=0.48\textwidth]{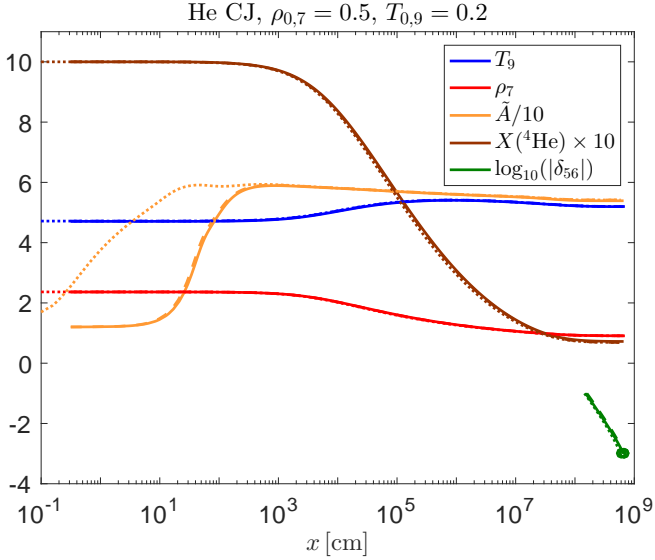}
\caption{The structure of a CJ detonation wave for He, $T_{0,9}=0.2$ and $\rho_{0,7}=0.5$ as a function of the distance behind the shock. We show the temperature (blue), density (red), $\tilde{A}$ (orange), $^{4}$He mass fraction (brown) and $\delta_{56}$ (green). The solid lines are the results with the input physics of \citet{Khokhlov89}, the dashed lines are with the addition of Coulomb correction terms to the NSE, and the dotted lines are the results with the default input physics. Green points mark the location where $|\delta_{56}|=10^{-3}$.
\label{fig:KhoComp3_He}}
\end{figure}

We next compare in Figure~\ref{fig:KhoComp3_He} our results with the input physics of \citet[][solid lines]{Khokhlov89} to the results with our default input physics (dotted lines). The synthesis of heavy elements is significantly faster in the default case (compare the profile of $\tilde{A}$). The inclusion of the Coulomb correction term for the NSE (dashed lines) changes the profiles by $\mylesssim10\%$ (see also Figure~\ref{fig:He_DetonationProp_Uncertain_Helium}). The main discrepancy is because of the isotope list used by \citet{Khokhlov89}. We verified that the default results are reproduced by adding the missing isotopes from NSE$7$ with $Z\le14$ and from the $\alpha$-ext lists to the list used by \citet{Khokhlov89}, which increases the number of isotopes to $161$. In fact, the results from NSE$4$ deviate by less than $30\%$ for this upstream density (see Figure~\ref{fig:He_DetonationProp_nets}), which shows that with $137$ isotopes (although somewhat different than the $114$ used by \citet{Khokhlov89}) better results can be obtained. 

\section{An approximate condition for CJ detonations}
\label{sec:CJ condition}

In Section~\ref{sec:CO scan D}, we found that CO detonations are pathological for all upstream densities values, as far as our numerical accuracy allowed us to test this. In Section~\ref{sec:He scan scale}, we claimed, without justifying it, that He detonations are of the CJ type. In this section, we show that He detonations are indeed of the CJ type, and we further provide an approximate condition, independent of reaction rates, that allows to estimate whether arbitrary upstream values (including composition) will support a detonation of the CJ type. 

For each upstream value, we can calculate the $\tilde{Y}_0$ of the initial conditions and the $\tilde{Y}_\textrm{CJ}$ of the NSE state for a CJ detonation. The assumption we make now is that along the detonation wave, $\tilde{Y}$ is monotonic between $\tilde{Y}_0$ and $\tilde{Y}_\textrm{CJ}$. This behaviour holds for CO and He (see e.g. Figures~\ref{fig:CO_1e7_2e8} and~\ref{fig:He_1e7_2e8}), but certainly breaks down when $\tilde{Y}_0\approx\tilde{Y}_\textrm{CJ}$. Our analysis is, therefore, approximate in the sense that it applies only when $\tilde{Y}_0$ and $\tilde{Y}_\textrm{CJ}$ are significantly different. Under our assumption, there are two cases -- either $\tilde{Y}$ is monotonically decreasing (as in CO detonations) or it is monotonically increasing (as in He detonations). We can, therefore, inspect the solution of the CJ detonation wave near the NSE state by solving for NSQE with $\tilde{Y}$ slightly larger or smaller than $\tilde{Y}_\textrm{CJ}$. It should be realized that for NSQE, the value of $\tilde{Y}$ completely defines the state of the plasma for a given $D_{\textrm{CJ}}$. This allows us to calculate $\delta q=q(\tilde{Y}_\textrm{CJ})-q(\tilde{Y}_\textrm{CJ}+\delta\tilde{Y})$ near the NSE state (with $\delta\tilde{Y}>0$ for decreasing $\tilde{Y}$ and with $\delta\tilde{Y}<0$ for increasing $\tilde{Y}$). In the case that $\delta q>0(<0)$, the energy release increases (decreases) towards the NSE state, which is the signature of a CJ (pathological) detonation. For all the cases that we examined, we find that
\begin{eqnarray}\label{eq:dqdy}
\left(\frac{dq}{d\tilde{Y}}\right)_{\textrm{CJ,NSE}}>0,
\end{eqnarray}
but we are unable to provide a proof for it. If Equation~\eqref{eq:dqdy} always holds, then we get the following simple condition for a CJ detonation:
\begin{eqnarray}\label{eq:CJ cond}
\tilde{Y}_0<\tilde{Y}_{\textrm{CJ}}.
\end{eqnarray}

\begin{figure}
\includegraphics[width=0.48\textwidth]{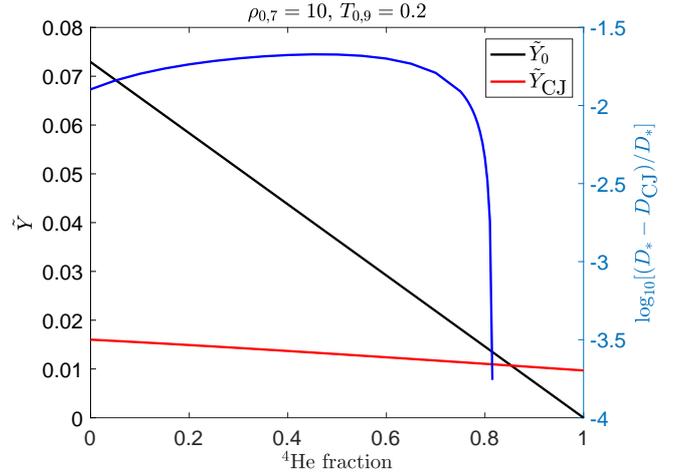}
\caption{$\tilde{Y}_{0}$ (black) and $\tilde{Y}_{\textrm{CJ}}$ (red) for $\rho_{0,7}=10$, $T_{0,9}=0.2$, and a $^{4}$He, $^{12}$C, and $^{16}$O mixture with $X(^{12}\textrm{C})=X(^{16}\textrm{O})$  as a function of $X(^{4}\textrm{He})$. The right $y$-axis shows the deviation between $D_{\textrm{CJ}}$ and $D_{*}$ (blue). For $X(^{4}\textrm{He})\lesssim0.81$ we are able to resolve $D_{*}>D_{\textrm{CJ}}$. However, the deviation between $D_{*}$ and $D_{\textrm{CJ}}$ decreases abruptly with higher mass fractions of $^{4}$He, which our numerical accuracy does not allow us to resolve. The abrupt decrease suggests that for $X(^{4}\textrm{He})\gtrsim0.81$ the detonation is of the CJ type, which supports the claim that He detonations are of the CJ type. Furthermore, the approximate condition of Equation~\eqref{eq:CJ cond} predicts the transition to happen at $X(^{4}\textrm{He})\approx0.85$, which is in agreement with the detailed calculations.
\label{fig:HeCO}}
\end{figure}

To test the approximate condition~\eqref{eq:CJ cond}, we calculate $D_{\textrm{CJ}}$ and $D_{*}$ for $\rho_{0,7}=10$, $T_{0,9}=0.2$ and for a $^{4}$He, $^{12}$C, and $^{16}$O mixture with $X(^{12}\textrm{C})=X(^{16}\textrm{O})$ (and varying amounts of $X(^{4}\textrm{He})$). The results are presented in Figure~\ref{fig:HeCO}. For $X(^{4}\textrm{He})\lesssim0.81$, we are able to resolve $D_{*}>D_{\textrm{CJ}}$. However, the deviation between $D_{*}$ and $D_{\textrm{CJ}}$ decreases abruptly with higher mass fractions of $^{4}$He, which our numerical accuracy does not allow us to resolve. The abrupt decrease suggests that for $X(^{4}\textrm{He})\gtrsim0.81$ the detonation is of the CJ type, which supports the claim that He detonations are of the CJ type. Furthermore, the approximate condition~\eqref{eq:CJ cond} predicts the transition to happen at $X(^{4}\textrm{He})\approx0.85$, which is in agreement with the detailed calculations. Similar results were obtained for different values of $\rho_{0}$ as well. We, therefore, conclude that the approximate condition of Equation~\eqref{eq:CJ cond} is valid. 

\section{The effect of weak reaction on the results}
\label{sec:weak}

In this section, we justify the assumption of the absence of weak reactions throughout the paper. Physically, since neutrinos are lost from the system, energy constantly leaves the system and a steady-state solution cannot be obtained. However, this effect can be smaller than the numerical accuracy of the integration, allowing, for example, the condition $\delta_{\max}=10^{-3}$ to be fulfilled. We test the effects of weak reactions separately for thermal neutrino emission (\textsc{neu} module of {\sc MESA}) and for weak nuclear reactions (\textsc{weaklib} module of {\sc MESA}). We calculate overdriven detonations for the cases in Tables~\ref{tbl:CO CJ} with $D=D_{*}+10\,\textrm{km}/\textrm{s}$ (and for the cases in Table~\ref{tbl:He CJ} with $D=D_{\textrm{CJ}}+10\,\textrm{km}/\textrm{s}$) with and without weak reactions. For CO, the deviation in the carbon-burning length-scale is completely negligible. The deviation in the position of the $^{28}$Si maximum is not negligible only for $\rho_{0}=10^{6}\,\textrm{g}/\textrm{cm}^{3}$ (where there is enough time for the neutrino losses to be significant); however, in this case the maximum position is much larger than the dynamical scale. For He, the deviation in the positions where $\tilde{A}=20$ and where half of the $^{4}$He is consumed is negligible. We further compare the position in which the density profile deviates by more than $1\%$ from the default case. It is either that the condition $\delta_{\max}=10^{-3}$ is fulfilled and there is no deviation larger than $1\%$, or that the deviation happens at scales comparable to (or much larger than) the dynamical scale. We, therefore, conclude that the assumption of absence of weak reactions is justified.


\section{Summary}
\label{sec:discussion}

In this work, we revisited the problem of thermonuclear detonation waves. We constructed lists of isotopes that allow the calculation of a thermonuclear detonation wave (Section~\ref{sec:net}) with some prescribed accuracy. For all isotopes, we used the most updated (measured) values of their mass and ground-state spin, and we provide fit parameters to the nuclear partition functions for all isotopes (Section~\ref{sec:nse}). We examined in detail the EOS and constructed an EOS with an uncertainty in the range of one percent (Section~\ref{sec:eos}). For this level of uncertainty, the nuclear level excitations (Section~\ref{sec:nuclear level excitations}) and the ion--ion interaction terms (Section~\ref{sec:screening}) must be included. It seems possible to construct an EOS with a $\mysim0.1\%$ level of uncertainty \citep{Potekhin2010}, but this accuracy is not required for current applications of supernovae. The EOS we constructed allows us to calculate CJ detonations with a degree of uncertainty in the percent level. We further provide the parameters of CJ detonations for initial compositions of CO (Section~\ref{sec:CO CJ}) and He (Section~\ref{sec:He CJ}) over a wide range of upstream plasma conditions that are relevant for supernovae. By comparing to previous works, we demonstrate that this is the first time that such a level of accuracy is obtained for the calculation of CJ detonations. Our results have a numerical accuracy of $\mysim0.1\%$, which allows an efficient benchmarking for future studies. We provide all the relevant information needed to fully reproduce our results. 

Our calculation of the structure of a detonation wave for both CO (Section~\ref{sec:CO structure}) and He (Section~\ref{sec:He structure}) over a wide range of upstream plasma conditions, demonstrates that we are able to perform such a calculation to a numerical accuracy of $\mysim0.1\%$. Our determination of the pathological detonation speed for CO, as well as the NSE state for these detonations, is with a degree of uncertainty in the percent level.  By comparing to previous works, we demonstrate that this is the first time that such a degree of accuracy has been reached. The uncertainty of different physical scales within the detonation waves is uncertain to a factor of a few, because the uncertainty is dominated by uncertain reaction rates. A detailed study of this uncertainty is beyond the scope of this work. The calculation of the physical scales is done with a numerical accuracy that is in the percent level, except for the location of the sonic point for pathological detonations, which is calculated with a numerical accuracy of a few tens of percent.  

Besides providing accurate results and highlighting a few shortcomings of previous works, we report here a few new insights into the structure of thermonuclear detonation waves. We show that CO detonations are pathological for all upstream density values, as far as our numerical accuracy allowed us to test this (Section~\ref{sec:CO scan D}). This is different from previous studies, which concluded that for low upstream densities CO detonations are of the CJ type. These claims were probably due to low numerical accuracy. We further provide an approximate condition, independent of reaction rates, that allows to estimate whether arbitrary upstream values (including composition) will support a detonation of the CJ type (Section~\ref{sec:CJ condition}). Using this argument, we were able to show that CO detonations are pathological for all upstream densities and to verify that He detonations are of the CJ type, as was previously claimed for He. We also show that for CO detonations the location of the sonic point changes position in a discontinuous manner from $x\sim100\,\textrm{cm}$ to $x\sim10^{4}\,\textrm{cm}$ around $\rho_{0,7}\approx2.7$.

Our analysis of the reactions that control the approach to NSE, which determines the length-scale of this stage, revealed that at high densities, the reaction $^{11}$B$+p\leftrightarrow3^{4}$He plays a significant role, which was previously unknown. This will help to focus the effort of improving reaction-rate measurements. 

The implications of the various improvements introduced in this work to supernova modelling will be studied in the future. 

\section*{Acknowledgements}
We thank Boaz Katz, Eli Waxman, Dean Townsley and Frank Timmes for useful discussions.  DK is supported by the Israel Atomic Energy Commission -- The Council for Higher Education -- Pazi Foundation -- and by a research grant from The Abramson Family Center for Young Scientists.









\begin{appendix}

\section{Differences between \textsc{winvn\_v2.0.dat} and ENSDF}
\label{sec:mass and spin}

For some isotopes, the values of the nuclear masses, $m_{i}$, included in \textsc{winvn\_v2.0.dat} differ from the most updated values given in the ENSDF data base, $\tilde{m}_{i}$. The list of isotopes for which $m_{i}$ and $\tilde{m}_{i}$ differ is given in Table~\ref{tbl:mass}, together with their mass (excess) values. For some isotopes, the values of $J_{i,0}$ included in \textsc{winvn\_v2.0.dat} differ from the most updated values given in the ENSDF data base, $\tilde{J}_{i,0}$. The list of isotopes for which $J_{i,0}$ and $\tilde{J}_{i,0}$ differ is given in Table~\ref{tbl:spin}, together with their spin values.

\begin{table*}
\begin{minipage}{110mm}
\caption{The list of isotopes for which the values of the nuclear masses, $m_{i}$, included in \textsc{winvn\_v2.0.dat} differ from the most updated values given in the ENSDF data base, $\tilde{m}_{i}$. For each isotope, we provide the mass excess value, $\Delta m_{i}$, included in \textsc{winvn\_v2.0.dat} and the  mass excess value, $\Delta\tilde{m}_{i}$, given in the ENSDF data base.}
\begin{tabular}{|c||c||c||c||c||c||c||c||c|}
\hline
$\textrm{Isotope}$  & $\Delta m_{i}$ & $\Delta\tilde{m}_{i}$ & $\textrm{Isotope}$  & $\Delta m_{i}$ & $\Delta\tilde{m}_{i}$ & $\textrm{Isotope}$  & $\Delta m_{i}$ & $\Delta\tilde{m}_{i}$  \\
 & $[\textrm{MeV}]$ & $[\textrm{MeV}]$ & & $[\textrm{MeV}]$ & $[\textrm{MeV}]$ &  & $[\textrm{MeV}]$ & $[\textrm{MeV}]$ \\ \hline
$^{13}\textrm{Be}$	&	$33.208$	&	$33.659$	&	$^{19}\textrm{O}$	&	$3.334$	&	$3.333$	&	$^{22}\textrm{O}$	&	$9.282$	&	$9.283$	\\
$^{15}\textrm{F}$	&	$16.813$	&	$16.567$	&	$^{23}\textrm{F}$	&	$3.310$	&	$3.285$	&	$^{24}\textrm{F}$	&	$7.560$	&	$7.545$	\\
$^{25}\textrm{F}$	&	$11.364$	&	$11.334$	&	$^{26}\textrm{F}$	&	$18.665$	&	$18.649$	&	$^{27}\textrm{F}$	&	$24.630$	&	$25.450$	\\
$^{25}\textrm{Ne}$	&	$-2.060$	&	$-2.036$	&	$^{26}\textrm{Ne}$	&	$0.479$	&	$0.481$	&	$^{27}\textrm{Ne}$	&	$7.036$	&	$7.051$	\\
$^{28}\textrm{Ne}$	&	$11.292$	&	$11.300$	&	$^{30}\textrm{Ne}$	&	$23.040$	&	$23.280$	&	$^{31}\textrm{Ne}$	&	$30.820$	&	$31.182$	\\
$^{32}\textrm{Ne}$	&	$37.278$	&	$36.999$	&	$^{34}\textrm{Ne}$	&	$53.121$	&	$52.842$	&	$^{19}\textrm{Na}$	&	$12.928$	&	$12.929$	\\
$^{29}\textrm{Na}$	&	$2.670$	&	$2.680$	&	$^{30}\textrm{Na}$	&	$8.374$	&	$8.475$	&	$^{31}\textrm{Na}$	&	$12.540$	&	$12.246$	\\
$^{32}\textrm{Na}$	&	$18.810$	&	$18.640$	&	$^{33}\textrm{Na}$	&	$24.889$	&	$23.780$	&	$^{34}\textrm{Na}$	&	$32.761$	&	$31.680$	\\
$^{35}\textrm{Na}$	&	$39.582$	&	$38.231$	&	$^{36}\textrm{Na}$	&	$47.953$	&	$46.303$	&	$^{37}\textrm{Na}$	&	$55.275$	&	$53.534$	\\
$^{20}\textrm{Mg}$	&	$17.559$	&	$17.478$	&	$^{21}\textrm{Mg}$	&	$10.913$	&	$10.904$	&	$^{30}\textrm{Mg}$	&	$-8.892$	&	$-8.884$	\\
$^{31}\textrm{Mg}$	&	$-3.190$	&	$-3.122$	&	$^{32}\textrm{Mg}$	&	$-0.912$	&	$-0.829$	&	$^{33}\textrm{Mg}$	&	$4.947$	&	$4.962$	\\
$^{34}\textrm{Mg}$	&	$8.560$	&	$8.323$	&	$^{37}\textrm{Mg}$	&	$29.249$	&	$28.211$	&	$^{38}\textrm{Mg}$	&	$34.996$	&	$34.074$	\\
$^{39}\textrm{Mg}$	&	$43.568$	&	$42.275$	&	$^{40}\textrm{Mg}$	&	$50.235$	&	$48.350$	&	$^{22}\textrm{Al}$	&	$18.183$	&	$18.201$	\\
$^{29}\textrm{Al}$	&	$-18.215$	&	$-18.208$	&	$^{30}\textrm{Al}$	&	$-15.872$	&	$-15.865$	&	$^{31}\textrm{Al}$	&	$-14.955$	&	$-14.951$	\\
$^{32}\textrm{Al}$	&	$-11.062$	&	$-11.099$	&	$^{33}\textrm{Al}$	&	$-8.437$	&	$-8.497$	&	$^{34}\textrm{Al}$	&	$-3.047$	&	$-3.000$	\\
$^{35}\textrm{Al}$	&	$-0.220$	&	$-0.224$	&	$^{39}\textrm{Al}$	&	$21.396$	&	$20.650$	&	$^{40}\textrm{Al}$	&	$29.295$	&	$27.590$	\\
$^{41}\textrm{Al}$	&	$35.704$	&	$33.420$	&	$^{42}\textrm{Al}$	&	$43.678$	&	$40.100$	&	$^{43}\textrm{Al}$	&	$48.428$	&	$47.020$	\\
$^{23}\textrm{Si}$	&	$23.772$	&	$23.697$	&	$^{24}\textrm{Si}$	&	$10.755$	&	$10.745$	&	$^{35}\textrm{Si}$	&	$-14.360$	&	$-14.391$	\\
$^{36}\textrm{Si}$	&	$-12.418$	&	$-12.436$	&	$^{37}\textrm{Si}$	&	$-6.594$	&	$-6.571$	&	$^{42}\textrm{Si}$	&	$18.434$	&	$16.470$	\\
$^{43}\textrm{Si}$	&	$26.697$	&	$23.101$	&	$^{44}\textrm{Si}$	&	$32.844$	&	$28.513$	&	$^{27}\textrm{P}$	&	$-0.716$	&	$-0.722$	\\
$^{28}\textrm{P}$	&	$-7.149$	&	$-7.148$	&	$^{38}\textrm{P}$	&	$-14.643$	&	$-14.622$	&	$^{39}\textrm{P}$	&	$-12.795$	&	$-12.775$	\\
$^{40}\textrm{P}$	&	$-8.074$	&	$-8.114$	&	$^{29}\textrm{S}$	&	$-3.157$	&	$-3.156$	&	$^{30}\textrm{S}$	&	$-14.062$	&	$-14.059$	\\
$^{40}\textrm{S}$	&	$-22.930$	&	$-22.838$	&	$^{41}\textrm{S}$	&	$-19.089$	&	$-19.009$	&	$^{42}\textrm{S}$	&	$-17.678$	&	$-17.638$	\\
$^{43}\textrm{S}$	&	$-12.070$	&	$-12.195$	&	$^{44}\textrm{S}$	&	$-9.100$	&	$-9.204$	&	$^{31}\textrm{Cl}$	&	$-7.066$	&	$-7.035$	\\
$^{42}\textrm{Cl}$	&	$-24.913$	&	$-24.832$	&	$^{43}\textrm{Cl}$	&	$-24.408$	&	$-24.159$	&	$^{44}\textrm{Cl}$	&	$-20.605$	&	$-20.384$	\\
$^{45}\textrm{Cl}$	&	$-18.360$	&	$-18.262$	&	$^{46}\textrm{Cl}$	&	$-13.810$	&	$-13.859$	&	$^{46}\textrm{Ar}$	&	$-29.729$	&	$-29.773$	\\
$^{47}\textrm{Ar}$	&	$-25.210$	&	$-25.366$	&	$^{48}\textrm{Ar}$	&	$-23.716$	&	$-23.281$	&	$^{49}\textrm{Ar}$	&	$-18.146$	&	$-17.190$	\\
$^{50}\textrm{K}$	&	$-25.736$	&	$-25.728$	&	$^{51}\textrm{K}$	&	$-22.002$	&	$-22.516$	&	$^{51}\textrm{Ca}$	&	$-35.873$	&	$-36.332$	\\
$^{52}\textrm{Ca}$	&	$-32.509$	&	$-34.266$	&	$^{53}\textrm{Ca}$	&	$-27.898$	&	$-29.388$	&	$^{54}\textrm{Ca}$	&	$-23.893$	&	$-25.161$	\\
$^{52}\textrm{Sc}$	&	$-40.357$	&	$-40.443$	&	$^{53}\textrm{Sc}$	&	$-37.623$	&	$-38.907$	&	$^{54}\textrm{Sc}$	&	$-34.219$	&	$-33.891$	\\
$^{55}\textrm{Sc}$	&	$-29.581$	&	$-30.159$	&	$^{56}\textrm{Sc}$	&	$-25.271$	&	$-24.852$	&	$^{41}\textrm{Ti}$	&	$-15.090$	&	$-15.697$	\\
$^{54}\textrm{Ti}$	&	$-45.594$	&	$-45.622$	&	$^{56}\textrm{Ti}$	&	$-38.937$	&	$-39.320$	&	$^{57}\textrm{Ti}$	&	$-33.544$	&	$-33.916$	\\
$^{58}\textrm{Ti}$	&	$-30.767$	&	$-31.110$	&	$^{42}\textrm{V}$	&	$-8.169$	&	$-7.620$	&	$^{43}\textrm{V}$	&	$-17.814$	&	$-17.916$	\\
$^{45}\textrm{V}$	&	$-31.880$	&	$-31.886$	&	$^{55}\textrm{V}$	&	$-49.153$	&	$-49.147$	&	$^{56}\textrm{V}$	&	$-46.080$	&	$-46.155$	\\
$^{57}\textrm{V}$	&	$-44.189$	&	$-44.413$	&	$^{58}\textrm{V}$	&	$-40.209$	&	$-40.402$	&	$^{44}\textrm{Cr}$	&	$-13.461$	&	$-13.360$	\\
$^{45}\textrm{Cr}$	&	$-19.436$	&	$-19.515$	&	$^{47}\textrm{Cr}$	&	$-34.559$	&	$-34.563$	&	$^{58}\textrm{Cr}$	&	$-51.835$	&	$-51.992$	\\
$^{59}\textrm{Cr}$	&	$-47.891$	&	$-48.086$	&	$^{46}\textrm{Mn}$	&	$-12.512$	&	$-12.570$	&	$^{47}\textrm{Mn}$	&	$-22.661$	&	$-22.566$	\\
$^{48}\textrm{Mn}$	&	$-29.323$	&	$-29.296$	&	$^{49}\textrm{Mn}$	&	$-37.615$	&	$-37.621$	&	$^{48}\textrm{Fe}$	&	$-18.160$	&	$-18.000$	\\
$^{49}\textrm{Fe}$	&	$-24.766$	&	$-24.751$	&	$^{50}\textrm{Fe}$	&	$-34.489$	&	$-34.476$	&	$^{51}\textrm{Fe}$	&	$-40.221$	&	$-40.203$	\\
$^{50}\textrm{Co}$	&	$-17.832$	&	$-17.630$	&	$^{51}\textrm{Co}$	&	$-27.542$	&	$-27.342$	&	$^{52}\textrm{Co}$	&	$-33.916$	&	$-34.361$	\\
$^{62}\textrm{Co}$	&	$-61.431$	&	$-61.424$	&	$^{63}\textrm{Co}$	&	$-61.840$	&	$-61.851$	&	$^{52}\textrm{Ni}$	&	$-22.654$	&	$-22.330$	\\
$^{53}\textrm{Ni}$	&	$-29.851$	&	$-29.631$	&	$^{54}\textrm{Ni}$	&	$-39.223$	&	$-39.278$	&	$^{54}\textrm{Cu}$	&	$-22.062$	&	$-21.410$	\\
$^{55}\textrm{Cu}$	&	$-31.994$	&	$-31.635$	&	$^{56}\textrm{Cu}$	&	$-38.694$	&	$-38.643$	&	$^{56}\textrm{Zn}$	&	$-26.137$	&	$-25.390$	\\
$^{57}\textrm{Zn}$	&	$-32.945$	&	$-32.550$	&	$^{61}\textrm{Zn}$	&	$-56.343$	&	$-56.349$	&	$^{59}\textrm{Ga}$	&	$-34.087$	&	$-33.760$	\\
$^{60}\textrm{Ga}$	&	$-40.004$	&	$-39.590$	&	$^{61}\textrm{Ga}$	&	$-47.088$	&	$-47.135$	&	$^{60}\textrm{Ge}$	&	$-27.858$	&	$-27.090$	\\
$^{61}\textrm{Ge}$	&	$-34.065$	&	$-33.360$	&	$^{62}\textrm{Ge}$	&	$-42.377$	&	$-41.740$	&	$^{63}\textrm{As}$	&	$-33.687$	&	$-33.500$	\\
$^{64}\textrm{As}$	&	$-39.518$	&	$-39.532$	&	$^{69}\textrm{As}$	&	$-63.086$	&	$-63.112$	&	$^{80}\textrm{As}$	&	$-72.172$	&	$-72.214$	\\
$^{64}\textrm{Se}$	&	$-27.504$	&	$-26.700$	&	$^{65}\textrm{Se}$	&	$-33.325$	&	$-33.020$	&	$^{66}\textrm{Se}$	&	$-41.832$	&	$-41.660$	\\
$^{69}\textrm{Se}$	&	$-56.301$	&	$-56.435$	&	$^{69}\textrm{Br}$	&	$-46.265$	&	$-46.260$	&	& & \\

 \hline
\end{tabular}
\centering
\label{tbl:mass}
\end{minipage}
\end{table*}

\begin{table}
\caption{The list of isotopes for which the values of $J_{i,0}$ included in \textsc{winvn\_v2.0.dat} differ from the most updated values given in the ENSDF data base, $\tilde{J}_{i,0}$.}
\begin{tabular}{|c||c||c||c||c||c|}
\hline
$\textrm{Isotope}$  & $J_{i,0}$ & $\tilde{J}_{i,0}$ & $\textrm{Isotope}$  & $J_{i,0}$ & $\tilde{J}_{i,0}$   \\ \hline
$^{13}\textrm{Be}$	&	$3/2$	&	$1/2$	&	$^{18}\textrm{N}$	&	$2$	&	$1$	\\
$^{21}\textrm{O}$	&	$1/2$	&	$5/2$	&	$^{23}\textrm{O}$	&	$3/2$	&	$1/2$	\\
$^{14}\textrm{F}$	&	$0$	&	$2$	&	$^{23}\textrm{F}$	&	$3/2$	&	$5/2$	\\
$^{24}\textrm{F}$	&	$0$	&	$3$	&	$^{25}\textrm{F}$	&	$1/2$	&	$5/2$	\\
$^{26}\textrm{F}$	&	$2$	&	$1$	&	$^{27}\textrm{F}$	&	$3/2$	&	$5/2$	\\
$^{29}\textrm{Ne}$	&	$1/2$	&	$3/2$	&	$^{19}\textrm{Na}$	&	$3/2$	&	$5/2$	\\
$^{31}\textrm{Na}$	&	$5/2$	&	$3/2$	&	$^{32}\textrm{Na}$	&	$0$	&	$3$	\\
$^{21}\textrm{Mg}$	&	$3/2$	&	$5/2$	&	$^{31}\textrm{Mg}$	&	$3/2$	&	$1/2$	\\
$^{33}\textrm{Mg}$	&	$5/2$	&	$3/2$	&	$^{35}\textrm{Mg}$	&	$3/2$	&	$5/2$	\\
$^{22}\textrm{Al}$	&	$3$	&	$4$	&	$^{31}\textrm{Al}$	&	$3/2$	&	$5/2$	\\
$^{33}\textrm{Al}$	&	$3/2$	&	$5/2$	&	$^{34}\textrm{Al}$	&	$2$	&	$4$	\\
$^{35}\textrm{Si}$	&	$5/2$	&	$7/2$	&	$^{37}\textrm{Si}$	&	$3/2$	&	$5/2$	\\
$^{36}\textrm{P}$	&	$2$	&	$4$	&	$^{38}\textrm{P}$	&	$2$	&	$0$	\\
$^{39}\textrm{S}$	&	$3/2$	&	$7/2$	&	$^{43}\textrm{S}$	&	$7/2$	&	$3/2$	\\
$^{44}\textrm{Cl}$	&	$4$	&	$2$	&	$^{45}\textrm{Cl}$	&	$3/2$	&	$1/2$	\\
$^{46}\textrm{Cl}$	&	$0$	&	$2$	&	$^{43}\textrm{Ar}$	&	$3/2$	&	$5/2$	\\
$^{45}\textrm{Ar}$	&	$1/2$	&	$5/2$	&	$^{49}\textrm{K}$	&	$3/2$	&	$1/2$	\\
$^{51}\textrm{K}$	&	$1/2$	&	$3/2$	&	$^{53}\textrm{Ca}$	&	$3/2$	&	$1/2$	\\
$^{54}\textrm{Sc}$	&	$1$	&	$3$	&	$^{56}\textrm{Sc}$	&	$3$	&	$1$	\\
$^{57}\textrm{Ti}$	&	$3/2$	&	$5/2$	&	$^{44}\textrm{V}$	&	$3$	&	$2$	\\
$^{56}\textrm{V}$	&	$2$	&	$1$	&	$^{57}\textrm{V}$	&	$3/2$	&	$7/2$	\\
$^{58}\textrm{V}$	&	$2$	&	$1$	&	$^{45}\textrm{Cr}$	&	$5/2$	&	$7/2$	\\
$^{59}\textrm{Cr}$	&	$3/2$	&	$1/2$	&	$^{58}\textrm{Mn}$	&	$3$	&	$1$	\\
$^{59}\textrm{Mn}$	&	$3/2$	&	$5/2$	&	$^{60}\textrm{Mn}$	&	$3$	&	$1$	\\
$^{50}\textrm{Co}$	&	$4$	&	$6$	&	$^{52}\textrm{Co}$	&	$1$	&	$6$	\\
$^{71}\textrm{Ni}$	&	$1/2$	&	$9/2$	&	$^{55}\textrm{Cu}$	&	$1/2$	&	$3/2$	\\
$^{56}\textrm{Cu}$	&	$3$	&	$4$	&	$^{70}\textrm{Cu}$	&	$1$	&	$6$	\\
$^{72}\textrm{Cu}$	&	$1$	&	$2$	&	$^{60}\textrm{Ga}$	&	$1$	&	$2$	\\
$^{63}\textrm{Ge}$	&	$1/2$	&	$3/2$	&	$^{66}\textrm{As}$	&	$2$	&	$0$	\\
$^{70}\textrm{As}$	&	$0$	&	$4$	&	$^{65}\textrm{Se}$	&	$1/2$	&	$3/2$	\\
$^{69}\textrm{Se}$	&	$3/2$	&	$1/2$	&	$^{71}\textrm{Se}$	&	$3/2$	&	$5/2$	\\
$^{69}\textrm{Br}$	&	$9/2$	&	$5/2$	&	$^{70}\textrm{Br}$	&	$5$	&	$0$	\\
$^{72}\textrm{Br}$	&	$3$	&	$1$	&	$^{86}\textrm{Br}$	&	$2$	&	$1$	\\
$^{71}\textrm{Kr}$	&	$9/2$	&	$5/2$	&	$^{73}\textrm{Kr}$	&	$5/2$	&	$3/2$	\\

 \hline
\end{tabular}
\centering
\label{tbl:spin}
\end{table}

\section{The inconsistency of the \textit{Helmholtz} EOS}
\label{sec:helm}

\begin{figure}
\includegraphics[width=0.48\textwidth]{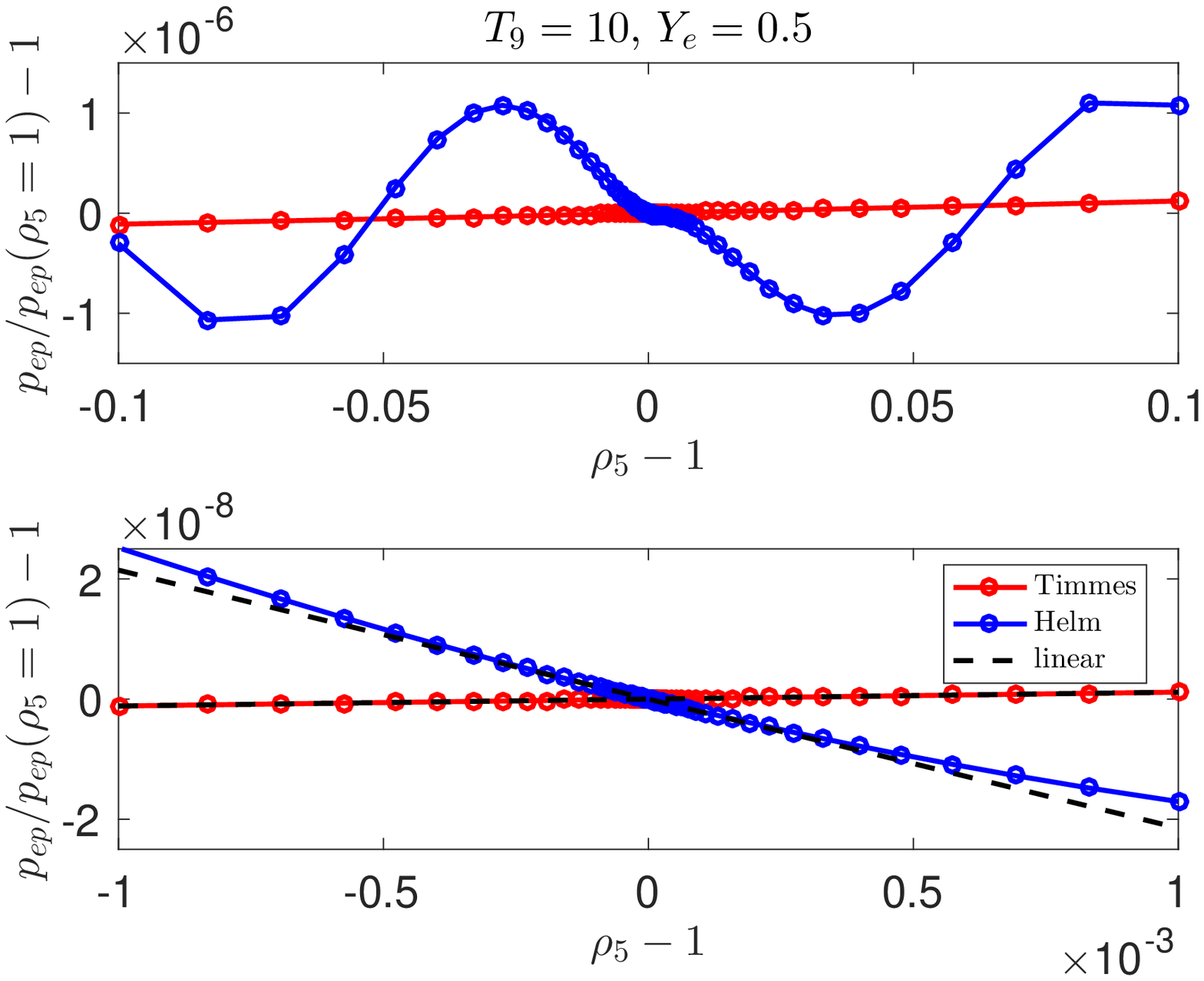}
\caption{The electron--positron pressure, $p_{ep}$, as a function of density (around $\rho_5\equiv\rho[\textrm{g}/\textrm{cm}^{3}]/10^{5}=1$) for $T_9=10$ and $Y_e=0.5$, as calculated by the Timmes EOS (red) and by the \textit{Helmholtz} EOS (blue). The lower panel is a zoomed version of the upper panel, and includes linear approximations to the EOSs (black), taken with a finite differencing.
\label{fig:HelmVsFXT}}
\end{figure}

Integrating Equations~\eqref{eq:ZND t} in a highly accurate manner requires a high degree of accuracy for the partial derivatives of the pressure and the internal energy with respect to the independent variables. We have found that the \textit{Helmholtz} EOS does not provide consistent values for $\partial p/\partial \rho$ at high temperatures and low densities. In order to demonstrate this inconsistency, we use the version of  \textit{Helmholtz} EOS with the densest grid (20 entries per decade; 'four times nominal grid' of \citet{Timmes00})\footnote{Note that the tables provided by {\sc MESA} and {\sc FLASH} are with 10 entries per decade.}, available through Frank Timmes website\footnote{http://cococubed.asu.edu/}. We consider the parameters $\rho_7=0.01$, $T_9=10$ and $Y_e=0.5$, and we compare the electron--positron pressure, $p_{ep}$, and the derivative of this pressure with respect to the density, $\partial p_{ep}/\partial\rho$, as calculate by the \textit{Helmholtz} EOS to the (accurate) values provided by the Timmes EOS. We find that, as reported by \citet{Timmes00}, $p_{ep}$ and $\partial p_{ep}/\partial\rho$ are calculated accurately by \textit{Helmholtz} EOS to better than $10^{-7}$. However, as demonstrated in Figure~\ref{fig:HelmVsFXT}, the behaviour of $p_{ep}$ as a function of the density, as provided by  the \textit{Helmholtz} EOS, is inconsistent with the provided $\partial p_{ep}/\partial\rho$. While the values of $p_{ep}$ and $\partial p_{ep}/\partial\rho$ are always accurate to better than $\mysim10^{-6}$, the pressure can actually decrease with increasing density while $\partial p_{ep}/\partial\rho$ is positive. This inconsistency precludes the accurate integration of Equations~\eqref{eq:ZND t}, and may be problematic for other applications as well.

In order to estimate the level of this inconsistency, we may compare the value provided by \textit{Helmholtz} EOS for $\partial p_{ep}/\partial\rho$ and the value calculated by directly differencing the pressure provided by \textit{Helmholtz} EOS with respect to the density (the relative difference of the density was $10^{-7}$ for the direct differencing), $(\partial p/\partial \rho)_d$. As demonstrated in Figure~\ref{fig:HelmVsFXT}, the pressure is well behaved, so a simple direct differencing is sufficient (compare the black dashed lines, which are linear approximations, taken with a finite differencing, to the actual values of the EOS). 

\begin{figure}
\includegraphics[width=0.48\textwidth]{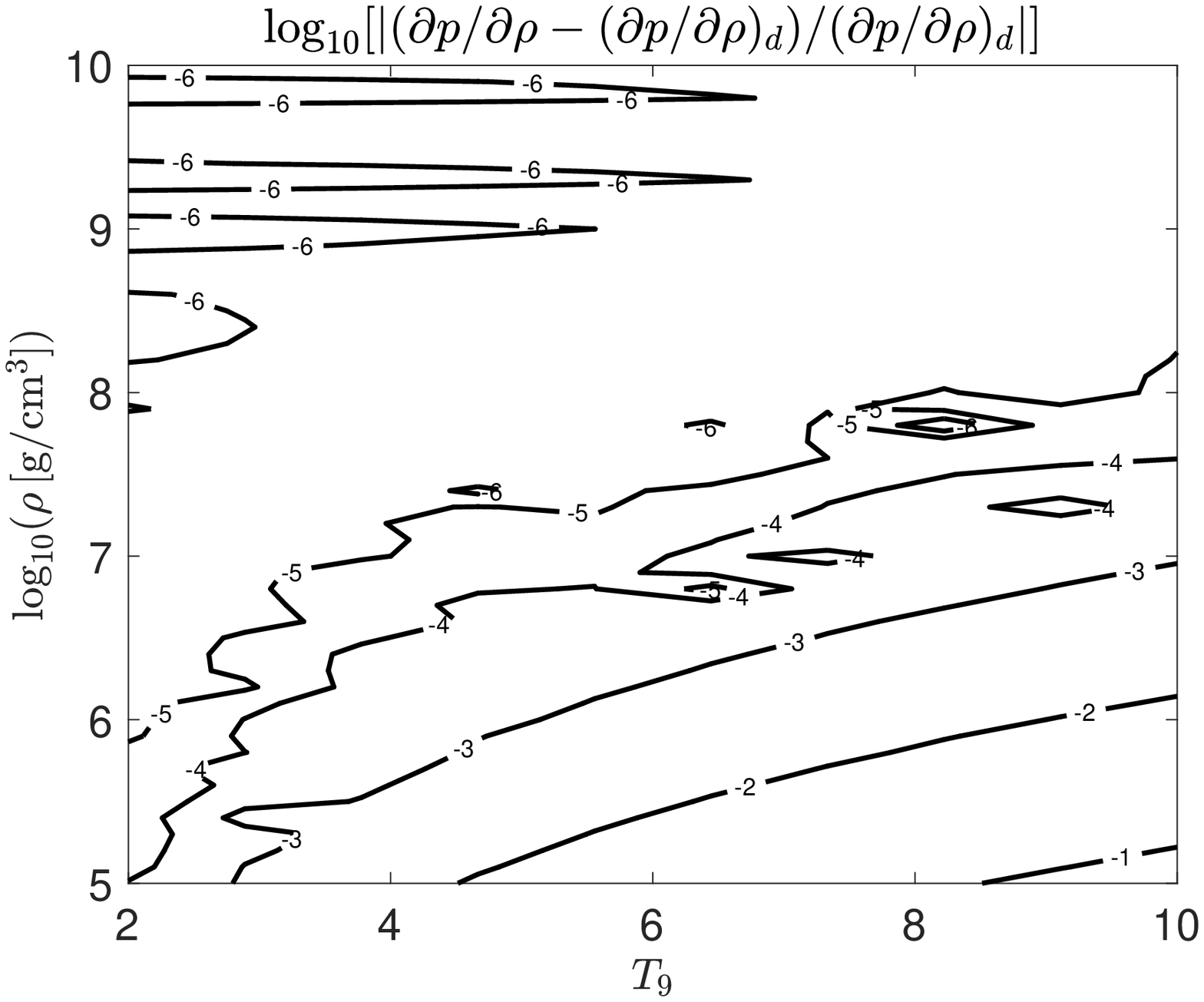}
\caption{The relative difference between $\partial p/\partial \rho$, as provided by \textit{Helmholtz} EOS, and the value calculated by the direct differencing of the pressure provided by \textit{Helmholtz} EOS with respect to the density (the relative difference of the density was $10^{-7}$ for the direct differencing), $(\partial p/\partial \rho)_d$, for CO. Inconsistencies that exceed $10^{-3}$ are obtained at high temperatures and low densities. In fact, the inconsistency of the electron--positron part of the pressure is much larger (and exceeds unity), but the other (analytical) parts of the pressure dominate at high temperatures and low densities.
\label{fig:HelmAcu}}
\end{figure}

Figure~\ref{fig:HelmAcu} presents the relative difference between $\partial p/\partial \rho$ and $(\partial p/\partial \rho)_d$, for CO. Inconsistencies that exceed $10^{-3}$ are obtained at high temperatures and low densities. In fact, the inconsistency of the electron--positron part of the pressure is much larger (and exceeds unity), but the other (analytical) parts of the pressure dominate at high temperatures and low densities. We, therefore, use the Timmes EOS, for which $\partial p/\partial \rho$ is consistent to better than $\mysim10^{-5}$ (and to better than $\mysim10^{-3}$ just for the electron--positron part of the pressure).

\section{Corrections to the exponential mass formula of Cameron \& Elkin (1965)}
\label{sec:CE bugs}

It seems that the exponential mass formula of \citet{Cameron1965} contains possible errors and that the following correction are required:
\begin{enumerate}
\item The pre-factors for $E_{c}$ and $E_{ex}$ (p. 1291) should be $Z^{2}/A^{1/3}$ and $Z^{4/3}/A^{1/3}$ and not $Z^{2}/r_{0}A^{1/3}$ and $Z^{4/3}/r_{0}A^{1/3}$, respectively.  
\item The fourth term inside the parentheses in the $E_{ex}$ expression should include the factor $r_{0}^{3}$ and not $r_{0}$.
\item The value for $\beta$ (p. 1292) should be $-35.939$ (given for $\gamma$ by \citet{Cameron1965}).
\item The value for $\gamma$ should be $-26.587$ (given for $-\beta$ by \citet{Cameron1965}).
\item The mass excess is actually given in the $^{16}$O scale (and not in the $^{12}$C scale, as claimed by \citet{Cameron1965}).
\end{enumerate}

\end{appendix}

\bsp	
\label{lastpage}
\end{document}